\documentclass[review]{jfm}
\usepackage{lineno}
\usepackage{graphicx}
\usepackage{natbib}
\usepackage{hyperref}
\hypersetup{
    colorlinks = true,
    urlcolor   = blue,
    citecolor  = black,
}

\newcommand{\RomanNumeralCaps}[1]
\linenumbers

\usepackage{nomencl}
\usepackage{amsmath}
\usepackage{amsfonts}
\usepackage{multirow}
\usepackage{subcaption}

\makenomenclature
\usepackage{xpatch}
\xpatchcmd{\thenomenclature}{\section*{\nomname}
}{}{\typeout{Success}}{\typeout{Failure}}
\makenomenclature
\RequirePackage{ifthen}
\printnomenclature[0.8cm]

\usepackage{soul}
\usepackage{xcolor}
\usepackage{bm}
\usepackage[colorinlistoftodos,disable]{todonotes}

\usepackage{url}

\graphicspath{}

\let\oldequation\equation
\let\oldendequation\endequation
\renewenvironment{equation}
  {\linenomathNonumbers\oldequation}
  {\oldendequation\endlinenomath}

\shorttitle{Ensemble learning of turbulence model}
\shortauthor{X.-L. Zhang, H. Xiao, X. Luo and G. He}

\title{Ensemble Kalman method for learning turbulence models from indirect observation data}

\author{Xin-Lei Zhang\aff{1,2},
Heng Xiao\aff{3}\corresp{\email{hengxiao@vt.edu}},
Xiaodong Luo\aff{4}    
\and
Guowei He\aff{1,2}\corresp{\email{hgw@lnm.imech.ac.cn}}}

\affiliation{
  \aff{1}The State Key Laboratory of Nonlinear Mechanics, Institute of Mechanics, Chinese Academy of Sciences, Beijing 100049, China
  \aff{2}School of Engineering Sciences, University of Chinese Academy of Sciences, Beijing 100049, China
\aff{3}Kevin T. Crofton Department of Aerospace and Ocean Engineering, Virginia Tech, Blacksburg, VA 24060, USA
\aff{4}Norwegian Research Centre (NORCE), Bergen, Norway}

\begin{document}
\maketitle

\begin{abstract}
In this work, we propose using an ensemble Kalman method to learn a nonlinear eddy viscosity model, represented as a tensor basis neural network, from velocity data. Data-driven turbulence models have emerged as a promising alternative to traditional models for providing closure mapping from the mean velocities to Reynolds stresses. Most data-driven models in this category need full-field Reynolds stress data for training, which not only places stringent demand on the data generation but also makes the trained model ill-conditioned and lacks robustness. This difficulty can be alleviated by incorporating the Reynolds-averaged Navier-Stokes (RANS) solver in the training process. However, this would necessitate developing adjoint solvers of the RANS model, which requires extra effort in code development and maintenance. Given this difficulty, we present an ensemble Kalman method with an adaptive step size to train a neural network-based turbulence model by using indirect observation data. To our knowledge, this is the first such attempt in turbulence modelling. 
The ensemble method is first verified on the flow in a square duct, where it correctly learns the underlying turbulence models from velocity data. Then, the generalizability of the learned model is evaluated on a family of separated flows over periodic hills. It is demonstrated that the turbulence model learned in one flow can predict flows in similar configurations with varying slopes.
\end{abstract}

\section{Introduction}
Despite the growth of available computational resources and the development of high-fidelity methods, industrial computational fluid dynamics (CFD) simulations still predominantly rely on Reynolds-averaged Navier-Stokes (RANS) solvers with turbulence models. This is expected to remain so in the decades to come, particularly for outer loop applications such as design optimization and uncertainty quantification~\citep{pi2014cfd}. Therefore, it is still of practical interest to develop more accurate and robust turbulence models.

Most of the currently used models are linear eddy viscosity models such as $k$--$\varepsilon$ model~\citep{launder1974application} and Spalart--Allmaras model~\citep{spalart1992one-equation}, which are based on two major assumptions~\citep{pope2001turbulent}: (1) weak equilibrium assumption, i.e., only the non-equilibrium in the magnitude of the Reynolds stress is accounted for through the transport equations, while its anisotropy is modelled based on local strain rate, and (2) Boussinesq assumption, i.e., the Reynolds stress anisotropy is assumed to be aligned with the strain rate tensor. 
Reynolds stress transport models 
(also referred to as differential stress models) have been developed in the past few decades to address the shortcomings caused by the weak equilibrium assumption~\citep{launder1975progress, speziale1991modelling, eisfeld2016verification}.
As to the second assumption, various nonlinear eddy viscosity and explicit algebraic stress models have been developed~\citep{spalart2000strategies, wallin2000explicit}, and some have even achieved dramatic successes in specialized flows (e.g., those with secondary flows or rotation). 
However, these complex models face challenges from the lack of robustness, increased computational costs and implementation complexity, and the difficulty of generalizing to a broader range of flows.
Consequently, turbulence modellers and CFD practitioners often face a compromise between the predictive performance and practical usability~\citep{xiao2019quantification}.

In the past few years, data-driven methods have emerged as a promising alternative for developing more generalizable and robust turbulence models. For example, nonlocal models based on vector-cloud neural networks have been proposed to emulate Reynolds stress transport equations~\citep{han2022vcnn, zhou2022frame}.
While this line of research is still in an early stage, it has the potential of leading to more robust and flexible non-equilibrium Reynolds stress models without solving the tensorial transport equations.
Alternatively, data-driven nonlinear eddy viscosity models have achieved much more success. Researchers have used machine learning to discover data-driven turbulence models or corrections thereto, which are nonlinear mappings from the strain rate and rotation rate to Reynolds stresses learned from data. Such functional mappings can be in the form of symbolic expressions~\citep{weatheritt2016novel,schmelzer2020discovery}, tensor basis neural networks~\citep{ling2016reynolds}, and random forests~\citep{wang2017physics, wu2019physics}, among others.
The data-driven nonlinear eddy viscosity models are a major improvement over their traditional counterparts in that they can leverage calibration data more systematically and explore a much larger functional space of stress--strain-rate mappings. 
However, they have some major shortcomings.  
First, as with their traditional counterparts, these data-driven models only addressed the Boussinesq assumption of the linear models as their strain--stress relations are still local, and thus they cannot address the weak equilibrium assumption described above. This is in contrast to the data-driven nonlocal Reynolds stress models~\citep{han2022vcnn, zhou2022frame}, which emulates the Reynolds stress transport equations and fully non-equilibrium models.
Second, the training of such models often requires full-field Reynolds stresses (referred to as \emph{direct data} hereafter), which are rarely available except from high fidelity simulations such as direct numerical simulations (DNS) and wall-resolved large eddy simulations (LES)~\citep{yang2021grid}. This would inevitably constrain the training flows to those accessible for DNS and LES, i.e., flows with simple configurations at low Reynolds numbers. 
It is not clear whether the data-driven models trained with such data would be applicable to practical industrial flows.  
Finally, the training of data-driven models is often performed in an \textit{a priori} manner, i.e., without involving RANS solvers in the training process. Consequently, the trained model may have poor predictions of the mean velocity in \textit{a posteriori} tests where the trained turbulence model is coupled with the RANS solvers. 
This is caused by the inconsistency between the training and prediction environments~\citep{duraisamy2021perspectives}. Specifically, even small errors in the Reynolds stress can be dramatically amplified in the predicted velocities due to the intrinsic ill-conditioning of the RANS operator~\citep{wu2019reynolds,brener2021conditioning}. Such ill-conditioning is particularly prominent in high Reynolds number flows; even an apparently simple flow such as a plane channel flow can be extremely ill-conditioned~\citep{wu2019reynolds}.
Such an inconsistency can be more severe in  learning dynamic models such as subgrid-scale models of LES, since the training environment is static while the prediction environment is dynamic.
Moreover, the model with the best \textit{a posterior}  performance may not necessarily excel in \textit{a priori} evaluations~\citep{park2021toward}.
In view of the drawbacks in \textit{a priori} training of turbulence models with direct data (Reynolds stress), it is desirable to leverage indirect observation data (e.g., sparse velocities and drag) to train data-driven turbulence models in the prediction environments by involving the RANS solvers in the training process. 
These indirect data are often available from experiments at high Reynolds numbers. Such a strategy is referred to as ``model-consistent learning'' in the literature~\citep{duraisamy2021perspectives}.

Model-consistent learning amounts to finding the turbulence model that, when embedded in the RANS solvers, produces outputs in the best agreement with the training data. Specifically, in incompressible flows these outputs include the velocity and pressure as well as their post-processed or sparsely observed quantities. Assuming the turbulence model is represented by a neural network to be trained with the stochastic gradient descent method, every iteration in the training process involves solving the RANS equations and finding the sensitivity of the discrepancy between the observed and predicted velocities with respect to the neural network weights. This is in stark contrast to the traditional method of training neural networks that learns from direct data (output of the neural network, i.e., Reynolds stresses in this case), where the gradients can be directly obtained from back-propagation. 
In model-consistent training, one typically uses adjoint solvers to obtain the RANS solver-contributed gradient (i.e., the sensitivity of velocity with respect to the Reynolds stress), as the full model consists of both the neural network and the RANS solver~\citep{holland2019field,strofer2021end}. The adjoint sensitivity is then multiplied by the neural network gradient according to the chain rule to yield the full gradient.
Similar efforts of combining adjoint solvers and neural network gradients have been made in learning subgrid-scale models in LES~\citep{macart2021embedded}.
These adjoint-based methods have been demonstrated to learn models with good posterior velocity predictions.
Moreover, for turbulence models represented as symbolic expressions, model-consistent learning is similarly performed by combining the model with the RANS solver in the learning processes~\citep{zhao2020rans,saidi2021cfddriven}, although the chain-rule based gradient evaluation is no longer needed in gradient-free optimizations such as genetic optimization.

In view of the extra efforts in developing adjoint solvers, particularly for legacy codes and multi-physics coupled solvers, \citet{strofer2021ensemble} explored ensemble-based gradient approximation as an alternative to the adjoint solver used in \citet{strofer2021end} to learn turbulence model from indirect data. 
Such a gradient is combined with that from the neural network via chain rule and then used in an explicit gradient-descent training. 
They found that the learned model was less accurate than that learned by using adjoint solvers
in the prediction of Reynolds stress and velocity.
This is not surprising, because the ensemble-based gradient approximation is less accurate than the analytic gradient from the adjoint solvers~\citep{evensen2018analysis}. 
Therefore, instead of using an ensemble to approximate gradients in optimization, it can be advantageous to directly use ensemble Kalman methods for training neural networks~\citep{chen2019ensemble,kovachki2019ensemble}. 
This is because such ensemble methods do not merely perform explicit, first-order gradient-descent optimization as is typically done in neural network training (deep learning). Rather, they implicitly use the Hessian matrix (second-order gradient) along with the Jacobian (first-order gradient) to accelerate convergence. 
Indeed, ensemble-based learning has gained significant success recently~\citep{schneider2020ensemble,schneider2020imposing}, but the applications focused mostly on learning from \emph{direct} data. 
They have not been used to learn from indirect data, where physical models such as RANS solvers become an integral part of the learning process.

In this work, we propose using an iterative ensemble Kalman method to train a neural network-based turbulence model by using indirect observation data. To the authors' knowledge, this is the first such attempt in turbulence modelling. Moreover, in view of the strong nonlinearity of the problem, we adjust the step size adaptively in the learning process~\citep{luo2015iterative}, which serves a similar purpose to that of the learning-rate scheduling in deep learning. 
Such an algorithmic modification is crucial for accelerating convergence and improving robustness of the learning, which can make an otherwise intractable learning problem with the adjoint method computationally feasible with the ensemble method.
A comparison is performed between the present method and the continuous adjoint method based on our particular implementation~\citep{strofer2021end}.
We show that, by incorporating Hessian information with adaptive stepping, the ensemble Kalman method exceeds the performance of the adjoint-based learning in both accuracy and robustness. Specifically, the present method successfully learned a generalizable nonlinear eddy viscosity model for the separated flows over periodic hills (Section~\ref{sec:results}), which the adjoint method was not able to achieve due to the lack of robustness. 
We emphasize that all these improvements are achieved at a much lower computational cost (measured in wall-time) and with a significantly lower implementation effort compared to the adjoint method. Both methods used the same representation of Reynolds stresses based on the tensor basis neural network~\citep{ling2016reynolds}.

In summary, the present framework of ensemble-based learning from indirect data has three key advantages. First, compared to methods that learn from direct data, the present framework relaxes the data requirements and only needs the measurable flow quantities, e.g., sparse measurements of the mean velocities or integral quantities such as drag and lift, rather than full-field Reynolds stresses.
Second, the model is trained in the prediction environment, thereby alleviating the ill-condition of the explicit data-driven RANS equation and avoiding the inconsistency between training and prediction.
Finally, the ensemble method is non-intrusive and thus very straightforward to implement for any solvers. In particular, it does not require adjoint solvers, which allows different quantities to be used in the objective function without additional code re-developments. 

The rest of this paper is organized as follows.
The architecture of the neural network and the model-consistent training algorithm are presented in Section~\ref{sec:method}.
The case setup for testing the performance of the proposed non-intrusive model-consistent training workflow is detailed in Section~\ref{sec:case_setup}.
The training results are presented and analyzed in Section~\ref{sec:results}.
The parallelization and the flexibility of the proposed method are discussed in Section~\ref{sec:discuss}.
Finally, conclusions are provided in Section~\ref{sec:conclusion}.

\section{Reynolds stress representation and model-consistent training}
\label{sec:method}

The objective is to develop a data-driven turbulence modelling framework that meets the following requirements: 
\begin{enumerate}
    \item The Reynolds stress representation shall be frame invariant and sufficiently flexible in expressive power to represent a wide range of flows.
    \item The model shall be trained in the prediction environment for robustness.
    \item It shall be able to incorporate sparse and potentially noisy observation data as well as Reynolds stress data.
\end{enumerate}
To this end, we choose the tensor basis neural networks~\citep{ling2016reynolds} to represent the mapping from the mean velocities to the Reynolds stresses. 
This representation has the merits of the embedded Galilean invariance and the flexibility to model complicated nonlinear relationships.
Furthermore, we use the ensemble Kalman method to learn the neural network-based model in a non-intrusive, model-consistent manner. 

\begin{figure}
    \centering
    \includegraphics[width=0.9\textwidth]{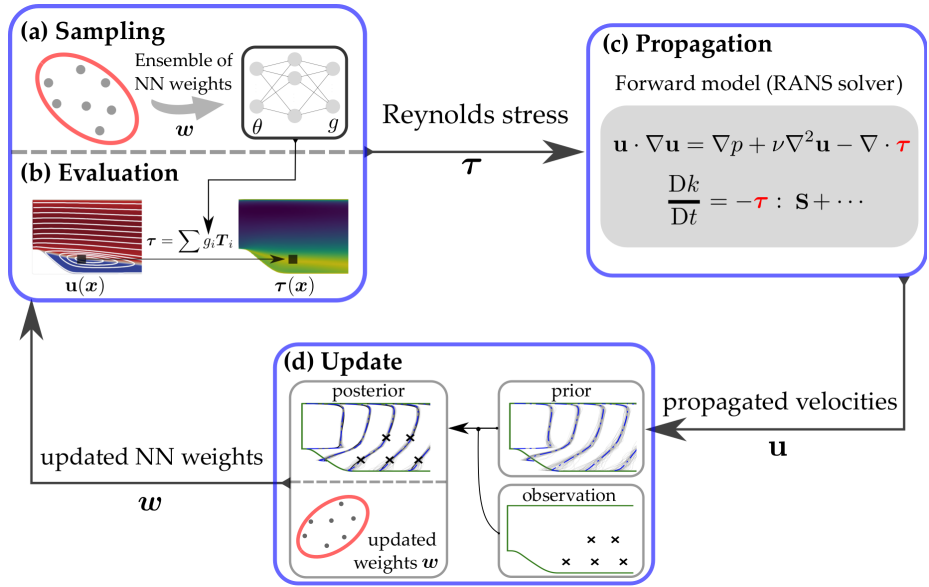}
    \caption{Schematic of the ensemble-based learning with sparse velocity data, consisting of the following four steps: (a) sampling the weights of the tensor basis neural network; (b) construct the Reynolds stress by evaluating the neural network-based turbulence model; (c) propagate the constructed Reynolds stress tensor to velocity by solving RANS equations; (d) update the neural network weights by incorporating observation data.}
    \label{fig:daml_scheme}
\end{figure}

The proposed workflow for training the tensor basis neural networks with indirect observation data is schematically illustrated in Figure~\ref{fig:daml_scheme}.
Traditionally, ensemble Kalman methods have been used in data assimilation applications to infer the \emph{state} of the system (e.g., velocities and pressures of a flow field). However, in our application, we aim to learn a turbulence model represented by a neural network. Therefore, the parameters (weight vector $\bm{w}$) of the network are the quantities to be inferred. The iterative ensemble Kalman method adopted for model learning consists of the following steps:
\begin{enumerate}[(1)]
    \item Sample the parameters (neural network weight vector $\bm{w}$) based on the initial prior distribution (Fig.~\ref{fig:daml_scheme}a). 
    The initial parameters are obtained by pretraining based on a baseline model.
    \item Construct the Reynolds stress field from the mean velocity field by evaluating the neural network-based turbulence model (Fig.~\ref{fig:daml_scheme}b).  
    The initial velocity field is obtained from the prediction with the baseline model.
    For a given mean velocity field $\bm{u}(\bm{x})$, each of the sample $\bm{w}_j$ (with $j$ being the sample index) implies a different turbulence model and thus a different Reynolds stress field, leading to an ensemble of Reynolds stress field in the whole computational domain;
    \item Propagate each Reynolds stress field in the ensemble to velocity field by solving the RANS equations (Fig.~\ref{fig:daml_scheme}c), based on which the observations can be obtained via post-processing (e.g., extracting velocities at specific points or integrating surface pressure to obtain drag);
    \item Update the parameters (network weights $\bm{w}$) through statistical analysis of the predicted observable quantities (e.g., velocities or drag) and comparison to observation data (Fig.~\ref{fig:daml_scheme}d).
\end{enumerate}
Steps (ii)--(iv) are repeated until convergence is achieved. The implementation  details are provided in Appendix~\ref{sec:implementation}. 

In this section, we introduce the Reynolds stress representation based on the tensor basis neural network and the ensemble-based learning algorithm. The latter is compared to other learning algorithms in the literature.

\subsection{Embedded neural network for Reynolds stress representation}

For constant-density, incompressible turbulent flows, the mean flow can be described by the RANS equation:
\begin{equation}
\begin{aligned}
    \nabla \cdot \boldsymbol{u} &= 0  \\
    \boldsymbol{u} \cdot \nabla \boldsymbol{u}  &= - \nabla p + \nu \nabla^2 \boldsymbol{u} - \nabla \cdot \boldsymbol{\tau} \text{,}
\end{aligned}
\end{equation}
where $p$ denotes mean pressure normalized by the constant flow density, and the Reynolds stress $\boldsymbol{\tau}$ indicates the effects of the small-scale turbulence on the mean flow quantities, which are required to be modelled.
The Reynolds stress can be decomposed into a deviatoric part~$\boldsymbol{a}$ and a spherical part as
\begin{equation}
 \boldsymbol{\tau}  = \boldsymbol{a} + \frac{2}{3} k \mathbf{I},
\end{equation}
where $k$ is the turbulence kinetic energy, and $\mathbf{I}$ is the second order identity tensor.
Different strategies have been developed to represent the deviatoric part of the Reynolds stress, and here we use the tensor basis neural network~\citep{ling2016reynolds}.

The neural network represents the deviatoric part of Reynolds stress with the scalar invariants and the tensor bases of turbulence field.
Specifically, the neural network is used to represent the mapping between the scalar invariants and coefficients of tensor bases.
Further, the output of the neural network is combined with the tensor bases to construct the Reynolds stress field such that the framework has the embedded Galilean invariance.
The deviatoric part of the Reynolds stress~$\boldsymbol{a}$ can be constructed as~\citep{pope1975more}:
\begin{align}
    \boldsymbol{a} &= 2 k \sum_{i=1}^{10} g^{(i)} \mathbf{T}^{(i)}, \\
    \textrm{with} \quad g^{(i)} &= g^{(i)}(\theta_1, \dots, \theta_5) \text{,}
\end{align}
where $\mathbf{T}$ and $\mathbf{\theta}$ are the tensor basis and scalar invariant of the input tensors, and $g$ is the scalar coefficient functions to be learned.
There are $10$ independent tensors that give the most general form of eddy viscosity.
The first four tensors are given as
\begin{equation}
\begin{aligned}
    \mathbf{T}^{(1)} &= \mathbf{S}, \qquad \mathbf{T}^{(2)} = \mathbf{S W} - \mathbf{W S} \\ 
    \qquad \mathbf{T}^{(3)} &= \mathbf{S}^2 - \frac{1}{3}\{\mathbf{S}^2\} \mathbf{I}, \qquad \mathbf{T}^{(4)} = \mathbf{W}^2-\frac{1}{3}\{\mathbf{W}^2\}\mathbf{I}
\end{aligned}
\label{eq:tensor_basis}
\end{equation}
where the curly bracket~$\{ \cdot \}$ indicates the trace of a matrix.
The first two scalar invariants are
\begin{equation}
        \theta_1 =\{\mathbf{S}^2\} \quad \textrm{and} \quad \theta_2 = \{\mathbf{W}^2\} \text{.}
\label{eq:scalar_invariant}
\end{equation}
Both the symmetric tensor $\mathbf{S}$ and the anti-symmetric tensor~$\mathbf{W}$ are normalized by the turbulence time scale~$\frac{k}{\varepsilon}$ as $\mathbf{S}=\frac{1}{2} \frac{k}{\varepsilon} \left[\nabla \boldsymbol{u} + (\nabla \boldsymbol{u})^\top\right]$ and $\mathbf{W}=\frac{1}{2} \frac{k}{\varepsilon} \left[\nabla \boldsymbol{u} - (\nabla \boldsymbol{u})^\top\right]$.
The time scale~$\frac{k}{\varepsilon}$ is obtained from the turbulent quantities solved from the transport equations for turbulence kinetic energy~$k$ and dissipation rate~$\varepsilon$.
For a two-dimensional flow, only two scalar invariants are nonzero, and the first three tensor bases are linearly independent~\citep{pope1975more}.
Further for incompressible flows, the components of the third tensor have $\mathbf{T}^{(3)}_{11} = \mathbf{T}^{(3)}_{22}$ and $\mathbf{T}^{(3)}_{12}=\mathbf{T}^{(3)}_{21}=0$. Hence,
the third tensor basis can be incorporated into the pressure term in the RANS equation, leaving only two tensor functions and two scalar invariants. 
In the turbulence transport equation, the turbulence production term is modified to account for the expanded formulation of Reynolds stress $ \mathcal{P} = -\boldsymbol{\tau} : \mathbf{S}$, where $:$ denotes double contraction of tensors.
For details of the implementation, readers are referred to \citet{strofer2021end}.
Note that the representation of the Reynolds stress is based on the following three hypotheses: (1) the Reynolds stress can be locally described with the scalar invariants and the independent tensors; (2) the coefficients of tensor bases can be represented by a neural network with the scalar invariants as input features; (3) a universal model form exists for flows having similar distributions of scalar invariants in the feature space.
Admittedly, the nonlinear eddy viscosity model is essentially still under the weak equilibrium assumption. 
Here we choose the nonlinear eddy viscosity model as a base model, mainly due to the following considerations. 
First, it is a more general representation of Reynolds stress tensor compared to the linear eddy viscosity model.
It utilizes ten tensor bases formed by the strain-rate tensor and rotation-rate tensor to represent the Reynolds stress, while the linear eddy viscosity model only uses the first tensor basis.
Second, the nonlinear eddy viscosity model uses uniform model inputs with Galilean invariance, i.e., scalar invariants, without requiring feature selections based on physical knowledge of specific flow applications.
Finally, the model expresses the Reynolds stress in an algebraic form, and no additional transport equation is solved.
From a practical perspective, the model is straightforward to implement and computationally more efficient than the Reynolds stress transport models.

In this work the tensor basis neural network is embedded into the RANS equation during the training process.
Specifically, the RANS equation is solved to propagate the Reynolds stress to the velocity by coupling with the neural network-based model, and the propagated velocity and the indirect observations are analyzed to train the neural network through model learning algorithms.
We use an ensemble Kalman method to train the neural network-based turbulence model embedded in the RANS equations, which is elaborated in Section~\ref{sec:enkf} below.
More detailed comparisons between the proposed method and other related schemes are presented in Section~\ref{sec:comparison_scheme}.

\subsection{Ensemble-based model-consistent training}
\label{sec:enkf}
The goal of the model-consistent training is to reduce the model prediction error by optimizing the weights $\bm{w}$ of the neural network.
The corresponding cost function can be formulated as~\citep{zhang2020regularized}
\begin{equation}
    J = \| \bm{w} - \bm{w}^0 \|_\mathsf{P}^2 + \| \mathsf{y} - \mathcal{H}[\bm{w}]  \|_{ \mathsf{R}}^2 \text{,}
    \label{eq:cost}
\end{equation}
where $\|\cdot\|_\mathsf{A}$ indicates weighted norm (defined as $\| \bm{v} \|_\mathsf{A}^2 = \bm{v}^\top \mathsf{A}^{-1} \bm{v}$ for a vector~$\bm{v}$ with the weight matrix~$\mathsf{A}$),  $\mathsf{P}$ is the model error covariance matrix indicating the uncertainties of the initial weights, 
$\mathsf{R}$ is the observation error covariance matrix, and $\mathsf{y}$ is the training data which is subjected to the Gaussian noise  $\epsilon \sim \mathcal{N}(0, \mathsf{R})$.
For simplicity we introduce the operator $\mathcal{H}$, which is a composition of RANS solver and the associated post-processing (observation). It maps the weights~$\bm{w}$ to the observation space (e.g., velocity or drag coefficient).
The first term in Equation~\eqref{eq:cost} is introduced to regularize the updated weights $\bm{w}$ by penalizing large deviations from their initial values~$\bm{w}^0$.
The second term describes the discrepancy between the model prediction~$\mathcal{H}[\bm{w}]$ and the observation~$\mathsf{y}$.
The training of the neural network is equivalent to minimization of the cost function~\eqref{eq:cost} by optimizing the weights~$\bm{w}$.
Note that the cost function can be modified to include other observation quantities such as friction coefficient and transition location.

In this work, we use the iterative ensemble Kalman method with adaptive stepping~\citep{luo2015iterative} to train the neural network framework.
This algorithm is a variant of the ensemble-based method where the observation error covariance matrix $\mathsf{R}$ is inflated such that the step size is adjusted adaptively at each iteration step. 
The corresponding cost function involves the regularization based on the difference from the last iteration, i.e.,
\begin{equation}
    J = \| \bm{w}_j^{l+1} - \bm{w}_j^l \|_\mathsf{P}^2 + \| \mathsf{y}_j - \mathcal{H}[\bm{w}_j^l]  \|_{\gamma \mathsf{R}}^2 \text{,}
    \label{eq:cost_iter}
\end{equation}
where~$l$ is the iteration index, $j$ is the sample index, and~$\gamma$ is a scaling parameter.
The weight update scheme of the iterative ensemble Kalman method is formulated as
\begin{subequations}
\label{eq:ies_weight} 
\begin{align} 
    \bm{w}_j^{l+1} & = \bm{w}_j^l + \mathsf{K} \left(\mathsf{y}_j - \mathcal{H}[\bm{w}_j^l]\right) \label{eq:ies_weight_update} \\
    \textrm{with} \quad \mathsf{K} & = \mathsf{S}_w \mathsf{S}_y^\top \left(\mathsf{S}_y \mathsf{S}_y^\top + \gamma^l \mathsf{R} \right)^{-1} \text{.} \label{eq:ies_weight_kalman}
\end{align}
\end{subequations}
The square root matrices $\mathsf{S}_w$ and $\mathsf{S}_y$ can be estimated from the ensemble at each iteration. See step (vi) and Equation~\eqref{eq:sqrt_root} of the detailed implementation in Appendix~\ref{sec:implementation}.

Note that the Kalman gain matrix above has a slightly different form than the more common formulation $\mathsf{K} = \mathsf{P}\mathsf{H}^\top\left(\mathsf{H}\mathsf{P}\mathsf{H}^\top + \gamma^l \mathsf{R} \right)^{-1}$. This is because we have written the terms  associated with the model error covariance matrix $\mathsf{P}$ by using the square root matrix $\mathsf{S}_w$ and its projection $\mathsf{S}_y$ to the observation space, i.e.,
\begin{equation}
\mathsf{P} = \mathsf{S}_w \mathsf{S}_w^\top
\quad \textrm{and} \quad 
\mathsf{S}_y = \mathsf{H} \mathsf{S}_w \end{equation}
where $\mathsf{H}$ is the local gradient of the observation operator $\mathcal{H}$ with respect to the parameter~$\bm{w}$. The equivalence between the two formulations is illustrated in Appendix~\ref{sec:theory}. 

The Kalman gain matrix in Equation~\eqref{eq:ies_weight_kalman} implicitly contains the inverse of the approximated second-order derivatives (Hessian matrix) as well as the gradient (Jacobian) of the cost function (both with respect to the weights $\bm{w})$. This can be seen from the derivations presented in Appendix~\ref{sec:theory}. Including both the gradient and the Hessian information significantly accelerate the convergence of the iteration process and thus improves the learning efficiency. This is in stark contrast to using only the gradient in typical training procedures of deep learning. Moreover, this is done in ensemble Kalman methods economically without significant overhead in computational costs or memory footprint.

The inflation parameter~$\gamma^l$ in Equation~\eqref{eq:ies_weight_kalman} can be considered a coefficient for adjusting the relative weight between the prediction discrepancies and the regularization terms. As such, we let 
\[
\gamma^l = \beta^l \{ \mathsf{S}_y^l (\mathsf{S}_y^l)^\top \} / \{ \mathsf{R} \},
\]
where $\beta^l$ is a scalar coefficient whose value also changes over the iteration process. The detailed algorithm for scheduling $\beta^l$ (and thus  $\gamma^l$) is presented in step (vii) of the detailed implementation in Appendix~\ref{sec:implementation}.

The ensemble-based method has the following three practical advantages.
First, it produces an ensemble of weights of the neural network, based on which uncertainty quantification can be conducted for the model prediction similarly to the Bayesian neural network~\citep{sun2020physics}.
Second, unlike the adjoint-based method, the ensemble-based method is non-intrusive and derivative-free, which means that it can be applied to black-box systems without the need for modifying the underlying source code. This feature makes it convenient to implement the ensemble-based method in practice and promotes the generalizability of the implemented ensemble method to different problems.
Finally, to reduce the consumption of computer memory, commonly used training algorithms, such as stochastic gradient descent, typically only involve the use of gradients of an objective function to update the weights of a neural network, while the ensemble-based method incorporates the information of low-rank approximated Hessian without a substantial increment of computer memory. 
Utilizing the Hessian information significantly improves convergence as discussed above.
In addition, the method can be used to train the model jointly with data from different flow configurations.
In such scenarios, the observation vector and the corresponding error covariance matrix would contain different quantities, e.g., the velocity and drag coefficient. 
The ensemble-based learning method interacts with the prediction environment during the training process, which is similar to the reinforcement learning in this sense.
However, the reinforcement learning usually learns a control policy for a dynamic scenario~\citep[e.g.,][]{novati2021automating,bae2022scientific}, while the present work learns a closure model with supervised learning.
Moreover, the reinforcement learning approach uses particular policy gradient algorithms to update the policy, while the ensemble Kalman method uses the ensemble-based gradient and Hessian to find the minimum of the underlying objective function.

An open-source platform OpenFOAM~\citep{opencfd21openfoam} is used in this work to solve the RANS equations with turbulence models.
Specifically, the built-in solver \textit{simpleFoam} is applied to solve the RANS equation coupling with the specialized neural network model.
Moreover, the DAFI code~\citep{strofer2021dafi} is used to implement the ensemble-based training algorithm.
A fully connected neural network is used in this work, and the detailed architecture for each case will be explained later. The rectified linear unit (ReLU) activation function is used for the hidden layers, and the linear activation function is used for the output layer. The machine learning library TensorFlow~\citep{abadi2015tensorflow} is employed to construct the neural network.
The code developed for this work is publicly available on Github~\citep{zhang2022dafi}.

\subsection{Comparison to other learning methods}
\label{sec:comparison_scheme}

Various approaches have been proposed for data-driven turbulence modelling, such as the direct training method~\citep{ling2016reynolds}, the adjoint-based differentiable method~
\citep{holland2019field,macart2021embedded,strofer2021end},
the ensemble gradient method~\citep{strofer2021ensemble}, 
and the ensemble Kalman inversion~\citep{kovachki2019ensemble}.
Here we present an algorithmic comparison of the proposed method with other model learning strategies in a unified perspective.

Conventional methods use the Reynolds stress of DNS to train the model in the \textit{a priori} manner, with the goal to minimize the discrepancy between the output of a neural network and the training data based on the backpropagation technique.
This concept can be formulated as a corresponding minimization problem (with the proposed solution), as follows:
\begin{equation}
\begin{aligned}
   \mathop{\arg \min}_{\bm{w}} J &= \| \boldsymbol{\tau}(\bm{w}, \widetilde{\mathbf{S}}, \widetilde{\mathbf{W}}) - \boldsymbol{\tau}^\text{DNS} \|^2 \text{,} \\
   \bm{w}^{l+1} &= \bm{w}^{l} - \beta \frac{\partial \boldsymbol{\tau}(\bm{w}, \widetilde{\mathbf{S}}, \widetilde{\mathbf{W}})}{\partial \bm{w}}\left[\boldsymbol{\tau}(\bm{w}, \widetilde{\mathbf{S}}, \widetilde{\mathbf{W}}) - \boldsymbol{\tau}^\text{DNS} \right], 
\end{aligned}
\end{equation}
where the input features $\widetilde{\mathbf{S}}$ and  $\widetilde{\mathbf{W}}$ are processed from the DNS results.
Further the trained neural network is coupled with the RANS solver for the posterior tests in similar configurations.
It is obvious that inconsistency exists between the training and prediction environments.
Specifically, during the training process, the model inputs are post-processed from the DNS data, while the learned model uses the RANS prediction to construct the input features.
Besides, the training process aims to minimize the cost function associated with the Reynolds stress, while the prediction aims to achieve the least discrepancies in the velocity.
This inconsistency would lead to unsatisfactory prediction due to the ill-conditioning issue of the RANS equation~\citep{wu2019reynolds}.
To tackle this problem, model-consistent training is required to construct the input features and the cost function with respect to more appropriate predicted quantities, e.g., the velocity.

For model-consistent training, the corresponding minimization problem (together with its solution) is changed to
\begin{equation}
\begin{aligned}
   \mathop{\arg \min}_{\bm{w}} J &= \| \boldsymbol{u}^\text{DNS} - \boldsymbol{u}(\boldsymbol{\tau}(\bm{w}, \mathbf{S}, \mathbf{W})) \|^2 \text{,} \\
   \bm{w}^{l+1} &= \bm{w}^{l} - \beta \frac{\partial J}{\partial \bm{w}}, 
\end{aligned}
\end{equation}
where the input feature~$\mathbf{S}$ and $\mathbf{W}$ are processed from the RANS prediction.
Both the input feature and the objective function used for training are consistent with the prediction environment.
Different approaches can be used to train the model, such as the adjoint-based differentiable method, the ensemble-based gradient method, and the ensemble Kalman inversion method.
Specifically, the adjoint-based differentiable framework~\citep{strofer2021ensemble} decomposes the gradient of the cost function into~$\frac{\partial J}{\partial \boldsymbol{\tau} }$ and $\frac{\partial \boldsymbol{\tau}}{\partial \bm{w}}$ by using the chain rule.
The weight-update scheme can be written as
\begin{equation}
     \bm{w}^{l+1} = \bm{w}^{l} - \beta \frac{\partial J}{\partial  \boldsymbol{\tau}} \frac{\partial \boldsymbol{\tau}}{\partial \bm{w}} \text{.}
\end{equation}
The gradient~$\frac{\partial J}{\partial \boldsymbol{\tau}}$ is computed using the adjoint method, and the gradient~$\frac{\partial \boldsymbol{\tau}}{\partial \bm{w}}$ is computed based on the backpropagation method.
The ensemble-based gradient method applies the Monte Carlo technique to draw samples from a Gaussian distribution.
Moreover, the data noise is taken into account by weighting the cost function with the observation error covariance matrix~$\mathsf{R}$. 
Further, the cross-covariance matrix computed by the ensemble method can be used to approximate the adjoint-based gradient as
\begin{equation}
    \frac{\partial J}{\partial \boldsymbol{\tau}} \approx \mathsf{S}_\tau \mathsf{S}_y^\top \mathsf{R}^{-1} \left(\mathcal{H}[\bm{w}]-\mathsf{y} \right) \text{.}
\end{equation}
The above-mentioned training approach employs the readily available analytic gradient of the neural network based on the backpropagation method.
Further the gradient of the cost function can be constructed by coupling with adjoint- or ensemble-based sensitivity of the RANS equation.

The ensemble Kalman inversion method~\citep{kovachki2019ensemble} adds a regularization term into the cost function and approximates the gradient of the cost function with respect to the weights of the neural network based on implicit linearization.
The minimization problem and the corresponding weight update scheme are
\begin{equation}
\begin{aligned}
    \mathop{\arg \min}_{\bm{w}} J &= \| \bm{w}^{l+1} - \bm{w}^l \|^2_\mathsf{P} + \| \boldsymbol{u}^\text{DNS} - \boldsymbol{u} \|^2_\mathsf{R} \\
    \bm{w}_j^{l+1} &= \bm{w}_j^l +  \mathsf{S}_w^l \left(\mathsf{S}_y^l \right)^\top \left(\mathsf{S}_y^l \left(\mathsf{S}_y^l \right)^\top + \mathsf{R} \right)^{-1} \left(\mathsf{y}_j - \mathcal{H}[\bm{w}^l]\right) \text{.}
\end{aligned}
\end{equation}
Note that this method involves the Hessian of the cost function~\citep{evensen2018analysis, luo2021novel} and provides quantified uncertainties based on Bayesian analysis~\citep{zhang2020evaluation}.
Similar to the ensemble gradient method, the ensemble Kalman inversion method also approximates the sensitivity of velocity to neural-network weights based on the ensemble cross-covariance matrix, without involving the analytic gradient of the neural network. 
However, the ensemble Kalman inversion method includes approximated Hessian in the weight-update scheme, which is missing in the ensemble gradient method.
The present algorithm can be considered a variant of the ensemble Kalman inversion method, which inherits the advantages of ensemble-based methods in terms of the non-intrusiveness and quantified uncertainty,
Moreover, the present method adjusts the relative weight of the prediction discrepancy and the regularization terms at each iteration step, which helps to speed up the convergence of the iteration process and enhance the robustness of the weight-update scheme.
For convenience of comparison, the training algorithms of different model-consistent data-driven turbulence modelling frameworks are summarized in Table~\ref{tab:compare_algorithm}.

\begin{table} 
    \centering
    \begin{tabular}{c c c}
        Method & Cost function & Update scheme \\
        \\
        \shortstack{Learning from \\ direct data} & $J = \| \boldsymbol{\tau}^\text{DNS}
          - \boldsymbol{\tau} \|^2$ &  $\bm{w}^{l+1} = \bm{w}^{l} + \beta \frac{\partial \boldsymbol{\tau}}{\partial \bm{w}} \left( \boldsymbol{\tau}^\text{DNS}
          - \boldsymbol{\tau}
          \right)$ \\
        \\
        \shortstack{Adjoint-based \\ learning} & $J = \| \boldsymbol{u}^\text{DNS} - \boldsymbol{u} \|^2$ & $\bm{w}^{l+1} = \bm{w}^{l} - \beta \frac{\partial J}{\partial \boldsymbol{\tau}} \frac{\partial \boldsymbol{\tau}}{\partial \bm{w}}$ \\
        \\
         \shortstack{Ensemble gradient \\ learning}  & $J = \| \boldsymbol{u}^\text{DNS} - \boldsymbol{u} \|_\mathsf{R}^2$ &
        \shortstack{
        \begin{math}
        \begin{aligned}
           \bm{w}_j^{l+1} &= \bm{w}_j^l + \mathsf{K}(\mathsf{y}_j - \mathcal{H}[\bm{w}_j^l]) \frac{\partial \boldsymbol{\tau}}{\partial \boldsymbol{w}}\\ 
        \textrm{with} \; \mathsf{K} &=  \mathsf{S}_\tau \mathsf{S}_y^\top \mathsf{R}^{-1}
        \end{aligned}
        \end{math}
        } \\
        \\
        \shortstack{
        Ensemble Kalman method \\ 
         with adaptive stepping \\
         (present framework)
        }  & \shortstack[r]{
        \begin{math}
        \begin{aligned}[c] 
        J = & \| \bm{w}_j^{l+1} - \bm{w}_j^l \|_\mathsf{P}^2 \\
        & + \| \boldsymbol{u}^\text{DNS} - \boldsymbol{u} \|_{\gamma \mathsf{R}}^2
        \end{aligned}
        \end{math}}
        & 
        \shortstack[r]{
        \begin{math}
        \begin{aligned}[c] 
        \bm{w}_j^{l+1} &= \bm{w}_j^l + \mathsf{K} (\mathsf{y}_j - \mathcal{H}[\bm{w}_j^l]) \; \textrm{with} \\ 
        \mathsf{K} &=  \mathsf{S}_w \mathsf{S}_y ^\top \left(\mathsf{S}_y \mathsf{S}_y^\top + \gamma \mathsf{R}\right)^{-1}
        \end{aligned}
        \end{math}
        }
    \end{tabular}
    \caption{ 
     Summary of different approaches for learning turbulence models in terms of the cost function and update schemes.
    We compared the ensemble Kalman method with adaptive stepping~\citep{kovachki2019ensemble,luo2015iterative} 
    with other related methods, including learning from direct data, i.e., the Reynolds stresses~\citep{ling2016reynolds}, adjoint-based learning~ \citep{holland2019field,macart2021embedded,strofer2021end},
    and ensemble gradient learning~\citep{strofer2021ensemble}.
    The DNS mean velocities are used as example indirect data.} 
    \label{tab:compare_algorithm}
\end{table}

The performance of the aforementioned methods in two applications, i.e., flow in a square duct and flow over periodic hills, is summarized in Table~\ref{tab:performance_sum}.
The square duct case is a synthetic case to assess the capability of the methods in learning underlying model functions, where the prediction with Shih's quadratic model~\citep{shih1993realizable} is used as the training data. 
For this reason, learning from direct data (referred to as ``direct learning method'' hereafter) can construct the synthetic model function accurately, and the results are omitted for brevity.
We present the results of the direct learning method for the periodic hill case in Section~\ref{sec:result_pehills}, where the DNS data are used as the training data.
The direct learning method is able to learn a model that improves the estimation of both velocity and Reynolds stress. 
However, when generalized to configurations with varying slopes, the learned model lacks robustness and leads to large discrepancies as shown in Figure~\ref{fig:pehill_generalizability}.
The adjoint-based learning method accurately reconstructs both the velocity and Reynolds stress fields in the square duct case as shown in Section~\ref{sec:result_squareduct}.
However, the method failed to learn a nonlinear eddy viscosity model in the periodic hill case -- it diverged during the training as reported in~\citet{strofer2021end}.
The ensemble gradient method was not able to recover the underlying model function in the synthetic square duct case~\citep{strofer2021ensemble} and also diverged in the periodic hill case.
In contrast, the present method is capable of learning the functional mapping in both cases.
Moreover, the learned model is generalized well to similar configurations with varying slopes as shown in Figure~\ref{fig:pehill_generalizability}.

\begin{table}
    \centering
    \begin{tabular}{c c c}
        Method & \shortstack{Square duct \\ (Learn synthetic model)} & \shortstack{Periodic hill \\ (Learn general nonlinear model)} \\ \\
        \shortstack{Learning from \\ direct data} & -- &  \shortstack{\emph{Poor} (see Fig.~\ref{fig:pehill_generalizability}) \\ (Learned nonlinear model; \\ poor generalization)} \\ \\
        \shortstack{Adjoint-based \\ learning} & \shortstack{\emph{Good} (see Fig.~\ref{fig:duct_g}) \\
        (Learned functional mapping)} & \shortstack{\emph{Diverged} \\ (Only learned linear model)} \\ \\
        \shortstack{Ensemble gradient \\
        learning} & \shortstack{\emph{Poor} \\ (Failed to learn functional mapping)} & \emph{Diverged} \\ \\
        \shortstack{
        Ensemble Kalman method \\ 
         with adaptive stepping \\
         (present framework)
        } & \shortstack{\emph{Good} (see Fig.~\ref{fig:duct_g}) \\ (Learned functional mapping)} & \shortstack{\emph{Good} (see Fig.~\ref{fig:pehill_generalizability}) \\ (Learned nonlinear model; \\ generalized well)} \\
    \end{tabular}
    \caption{ Summary of the performance of different approaches for learning turbulence models in two different cases, i.e., flow in a square duct and flow over periodic hills. 
    We compare the present method~\citep{kovachki2019ensemble,luo2015iterative} with other related methods, including learning from direct data~\citep{ling2016reynolds}, adjoint-based learning~ \citep{holland2019field,macart2021embedded,strofer2021end},
     and ensemble gradient learning~\citep{strofer2021ensemble}. The square duct case uses the prediction from Shih's quadratic model as training data, while the periodic hill case uses the DNS results as training data.}
    \label{tab:performance_sum}
\end{table}

\section{Case setup}
\label{sec:case_setup}

We use two test cases to show the performance of the proposed method for learning turbulence models: (1) flow in a square duct and (2) separated flows over periodic hills.
Both are classical test cases that are well-known to be challenging for linear eddy viscosity models\citep{xiao2019quantification}.
We aim to learn neural network-represented nonlinear eddy viscosity models from velocity data by using the ensemble method.  The learned models are evaluated by comparing to the ground truth for the square duct case and assessing its generalization performance in the separated flows over periodic hills. The results are also compared to those of the adjoint-based method. Details of the case setup are discussed below.

\subsection{Secondary flows in a square duct}
The first case is the flow in a square duct, where the linear eddy viscosity model is not able to capture the in-plane secondary flow.
The nonlinear eddy viscosity model, e.g., Shih's quadratic model~\citep{shih1993realizable}, is able to simulate the secondary flows.
Furthermore, Shih's quadratic model provides an explicit formula of the mapping between the scalar invariant~$\bm{\theta}$ and the function~$g$, which serves as an ideal benchmark for evaluating the accuracy of the trained model functions.
In Shih's quadratic model, the $g$ function of the scalar invariant~$\bm{\theta}$ is written as
\begin{equation}
\label{eq:gfunc}
\begin{aligned}
  & g^{(1)}\left(\theta_{1}, \theta_{2}\right)=\frac{-2 / 3}{1.25+\sqrt{2 \theta_{1}}+0.9 \sqrt{-2 \theta_{2}}} \\
  & g^{(2)}\left(\theta_{1} , \theta_{2}\right)=\frac{7.5}{1000+\left(\sqrt{2 \theta_{1}}\right)^{3}} \\
  & g^{(3)}\left(\theta_{1} , \theta_{2}\right)=\frac{1.5}{1000+\left(\sqrt{2 \theta_{1}}\right)^{3}} \\
  & g^{(4)}\left(\theta_{1}, \theta_{2}\right)=\frac{-9.5}{1000+\left(\sqrt{2 \theta_{1}}\right)^{3}} \text{.}
\end{aligned}
\end{equation}
Hence we use the velocity results from Shih's quadratic model as the synthetic truth and show that the method is able to reveal the underlying relationship between the scalar invariant and the tensor basis when the model exists in the form of the tensor bases.
Moreover, we aim to compare the adjoint-based and the present ensemble-based methods in terms of the training accuracy and efficiency in this case.

The flow in a square duct is fully developed, and only one cell is used in the stream-wise direction.
Moreover, one-quarter of the domain is used due to the symmetry, and the mesh grid is $50 \times 50$.
As for the architecture of the neural network in this case, two scalar invariants are used as input features, and four $g$ functions~$g^{(1-4)}$ are used in the output layer.
The input features of the synthetic truth are shown in Figure~\ref{fig:hills_feature}.
Since the stream-wise velocity $u_x$ is dominant, the first two scalar invariants are approximately equal in magnitude but with opposite signs.
The slight difference between the scalar invariants~$\theta_1$ and $\theta_2$ is caused by the secondary flow in the plane.
We also provide the plot of $|\theta_1| - |\theta_2|$, which indicates the relative importance of the strain rate and the vorticity.
The stream-wise velocity gradient is relatively small near the center of the duct, leading to the negligible scalar invariant~$\theta_1$.
Moreover, the shear strain rate is dominant near the duct center, while there is a pair of vortexes indicating the strong rotation rate.
Besides, it can be seen that the range of the input features is from $0$ to approximately $7$.
We draw $50$ samples of the neural network weights in this case. 
In the neural network, we use $2$ hidden layers with $5$ neurons per layer.
A sensitivity study of the training algorithm to the neural network architecture and the observation data is provided in Appendix~\ref{sec:sensitivity}.

\subsection{Separated flow over periodic hills}
The flow over periodic hills is a canonical separated flow for the numerical investigation of turbulence models.
There is no ground truth for the model function which is able to capture the flow characteristics accurately.
Here we use the DNS results~\citep{xiao2020flows} as the training data and learn the neural network-based model by using the ensemble-based method.
Further, we validate the generalizability of the learned model in similar configurations with varying slopes~\citep{xiao2020flows}.
Specifically, the hill geometry is parameterized with the slope coefficient~$\alpha$.
The separation extent decreases as the slope~$\alpha$ increases from $0.5$ to $1.5$.
The case with slope parameter $\alpha=1$ is used as the training case, and the cases with other slopes of~$\alpha = 0.5, 0.8, 1.2, 1.5$ are used to test the generalizability of the learned model in the scenarios having different levels of flow separation. 
The mesh is set as $149$ cells in stream-wise direction and $99$ cells in normal direction after grid-independence tests.
We use the $k$--$\varepsilon$ model~\citep{launder1974application} as the baseline model.
The model learned from direct data is also provided for comparison. 
The implementation of the direct learning method is illustrated in Appendix~\ref{sec:direct_learning}.

\begin{figure}
    \centering
    \subfloat[square duct: $\theta_1$]{\includegraphics[width=0.27\textwidth]{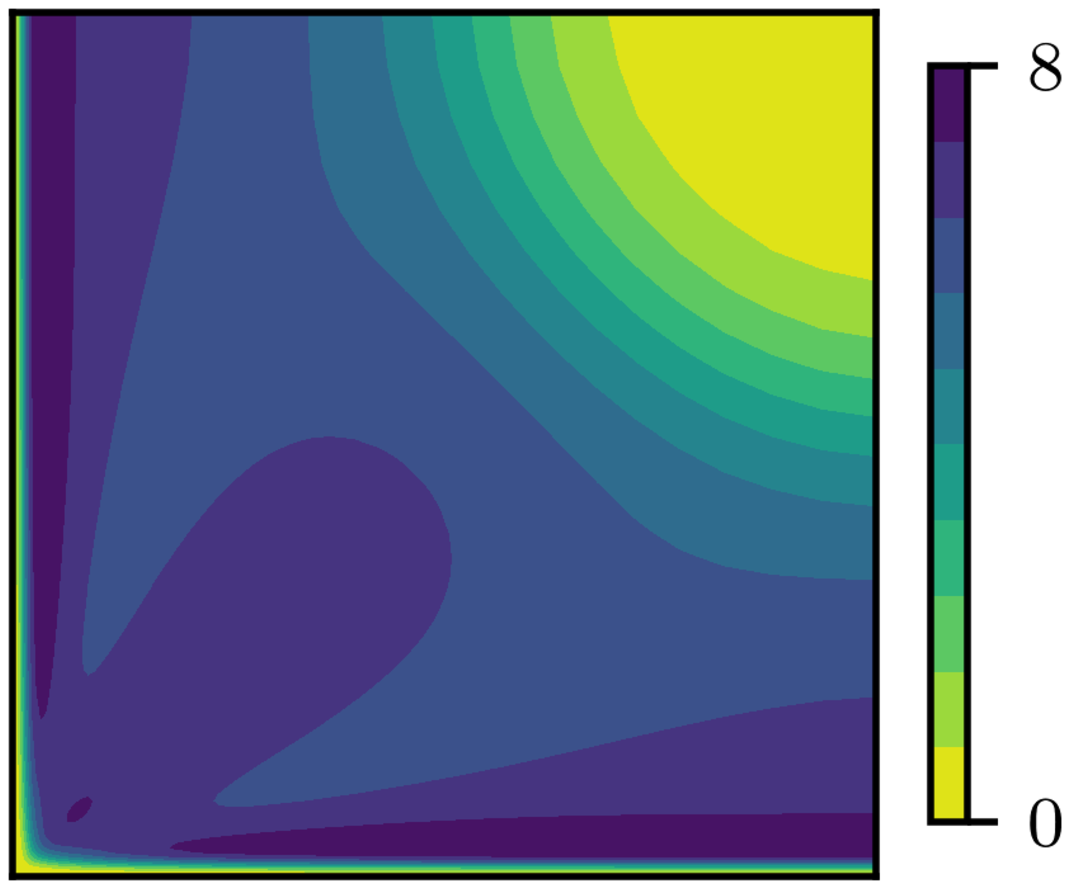}}
    \hspace{15mm}
    \subfloat[square duct: $|\theta_1|-|\theta_2|$]{\includegraphics[width=0.27\textwidth]{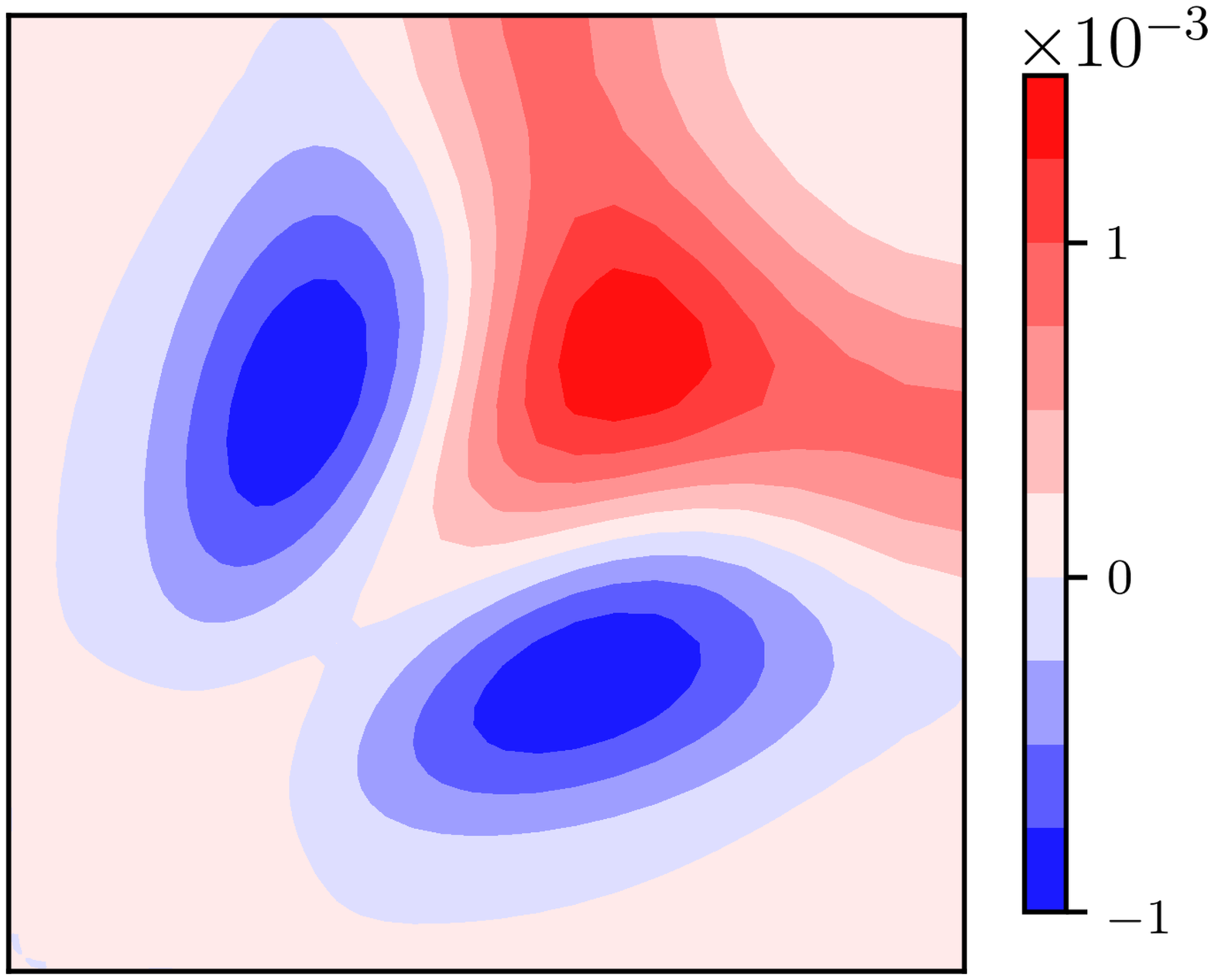}} \\
    \subfloat[periodic hills: $\theta_1$]{\includegraphics[width=0.28\textwidth]{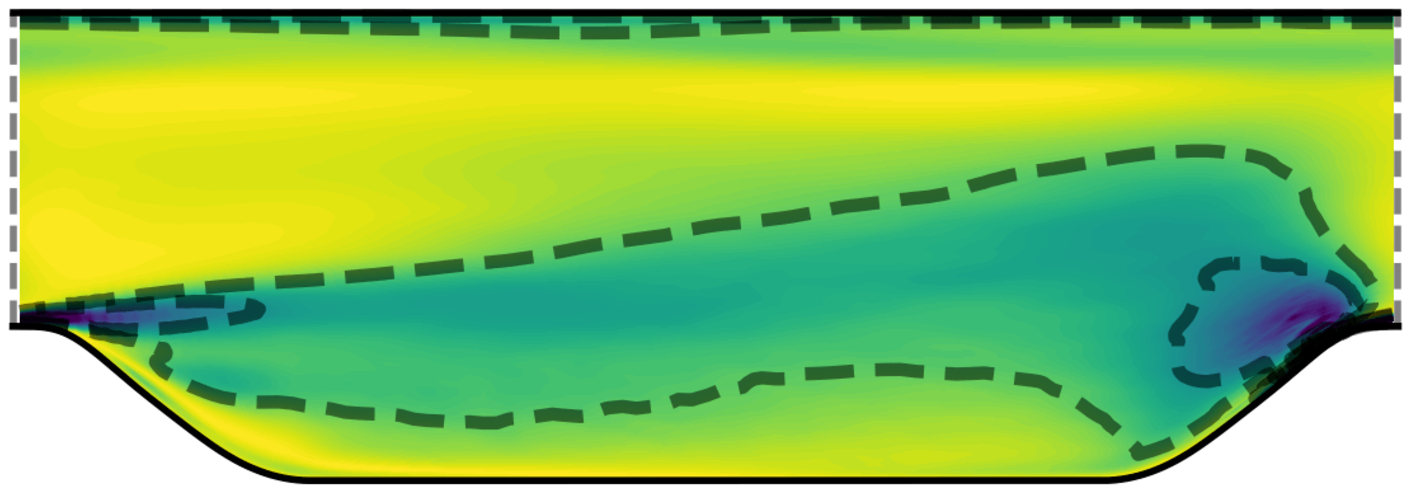}}
    \includegraphics[scale=0.16]{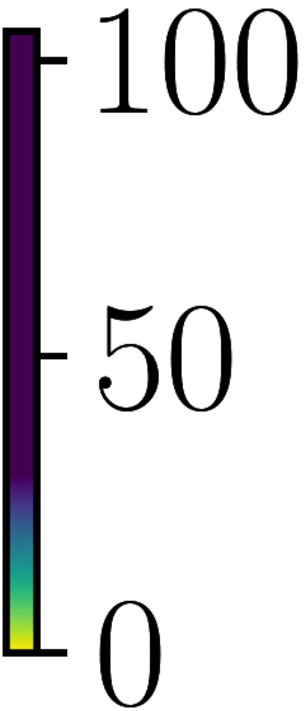}
    \subfloat[periodic hills: $\theta_2$]{\includegraphics[width=0.28\textwidth]{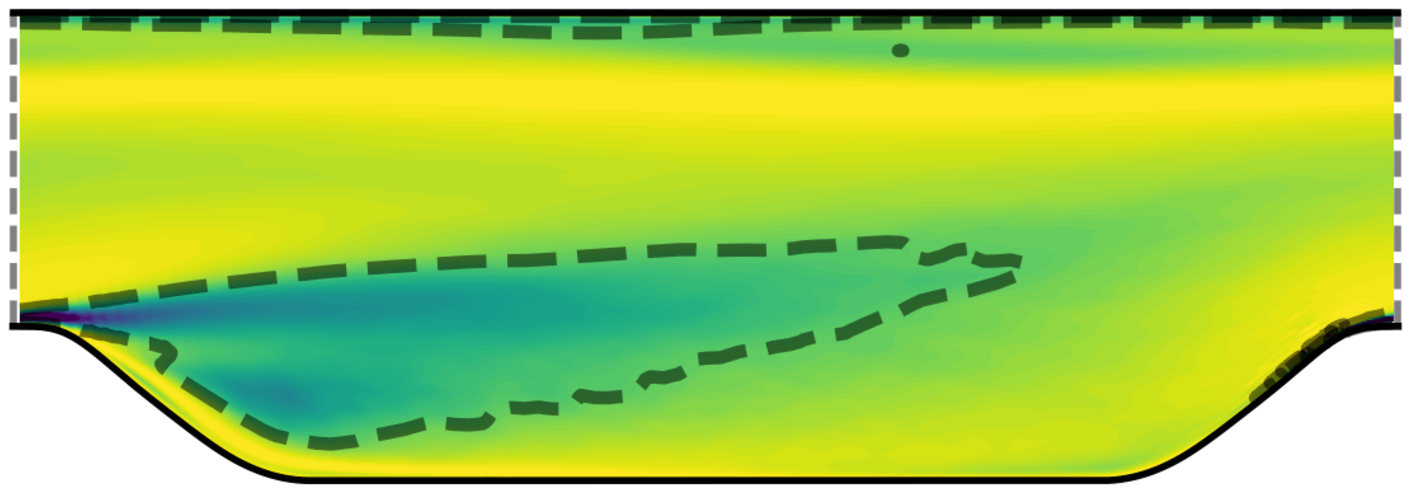}}
    \includegraphics[scale=0.16]{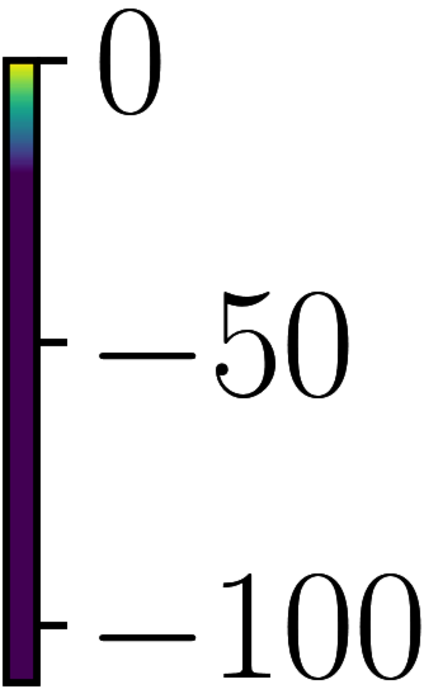}
    \subfloat[periodic hills: 
    $|\theta_1|-|\theta_2|$]{\includegraphics[width=0.28\textwidth]{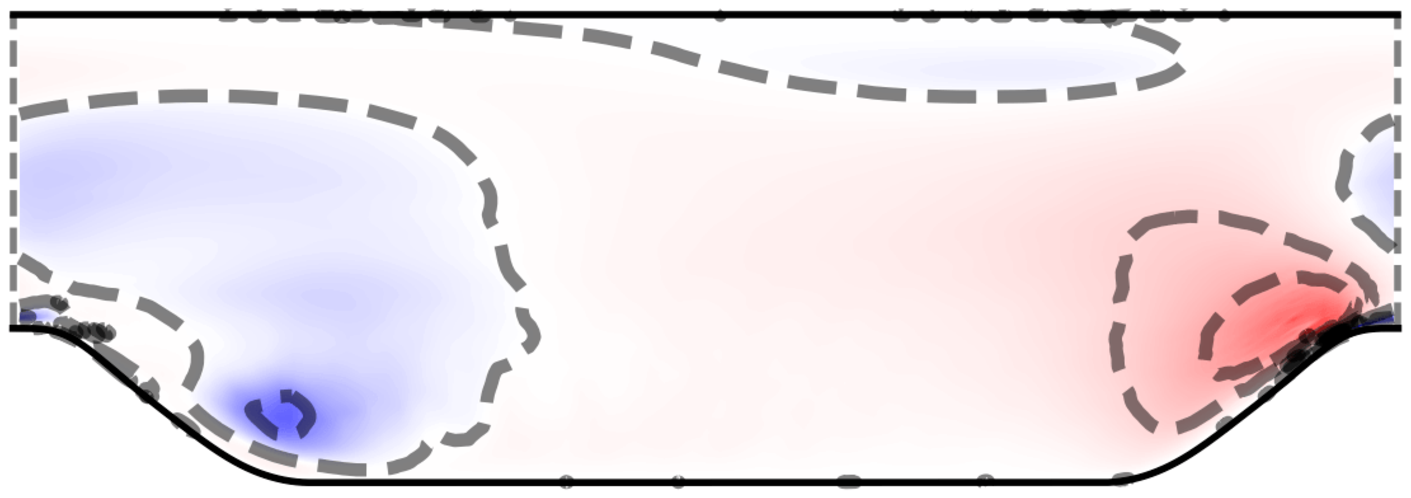}}
    \includegraphics[scale=0.16]{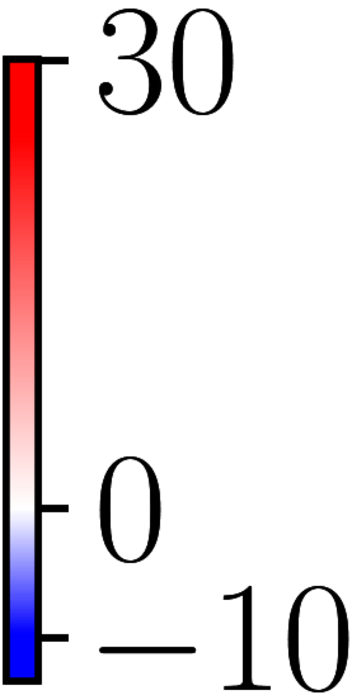}
    \caption{Contour plots of input features of the reference data for the square duct case and periodic hill case}
    \label{fig:hills_feature}
\end{figure}

For the two-dimensional incompressible flow, there are only the first two scalar invariants and independent tensors after merging the third tensor basis into the pressure term in the RANS equation~\citep{strofer2021end}.
The input features of the DNS data are shown in Figure~\ref{fig:hills_feature}, scaled with the RANS predicted time scale.
The plot of the first scalar invariant~$\theta_1$ indicates the large strain rate in the free shear layer and the windward side of the hill. 
The second scalar invariant~$\theta_2$ shows the vorticity mainly in the flow separation region at the leeward side of the hill.
From the plot of $|\theta_1| - |\theta_2|$, it can be seen that the magnitude of the first two scalars is equivalent in most areas.
The strong vorticity in the downhill is caused by the flow separation, while near the uphill region the shear strain rate is dominant due to the channel contraction.
Compared to the square duct case, the separated flow over periodic hills has a wider range in the magnitude of the input features, which is from $0$ to about $100$.
That is because in the square duct case, the magnitude of the scalar invariant is mainly determined by the stream-wise velocity~$u_x$, while in the periodic hill case, both $u_x$ and $u_y$ have considerable effects on the input features.
Moreover, the magnitude of the time scale in the periodic hill is much larger than that in the square duct flow.
Concretely, the maximum value for the periodic hill case is about $490$, while that for the square duct case is about $10$.
Hence, we use a deeper neural network of $10$ hidden layers with $10$ neurons per layer compared to the square duct case based on the sensitivity analysis of the neural network architecture as shown in Appendix~\ref{sec:sensitivity}.
We draw $50$ samples of the neural network weights in this case. 
The training data set is summarized in Table~\ref{tab:dataset}.

\begin{table}
    \begin{center}
    \begin{tabular}{c c c}
         Cases & Flow configuration & Training data \\ \\
         Case 1 & Square duct &  Shih's quadratic model \\ \\
         Case 2 & Periodic hills & DNS~\citep{xiao2020flows} \\ & ($\alpha=0.5, 0.8, 1.0, 1.2, 1.5$) &  \\
    \end{tabular}
    \caption{Summary of the configurations and the training data}
    \label{tab:dataset}
    \end{center}
\end{table}

\section{Results}
\label{sec:results}

\subsection{Flow in a square duct: learning underlying closure functions}
\label{sec:result_squareduct}

\begin{figure}
    \centering
    \begin{tabular}{ccccc}
        & $u$ & $u_x$ & $u_y$ & $\text{error}(\bm{u})$\\
        \rotatebox[origin=c]{90}{ground truth} &
        \raisebox{-.5\height}{\includegraphics[scale=0.23]{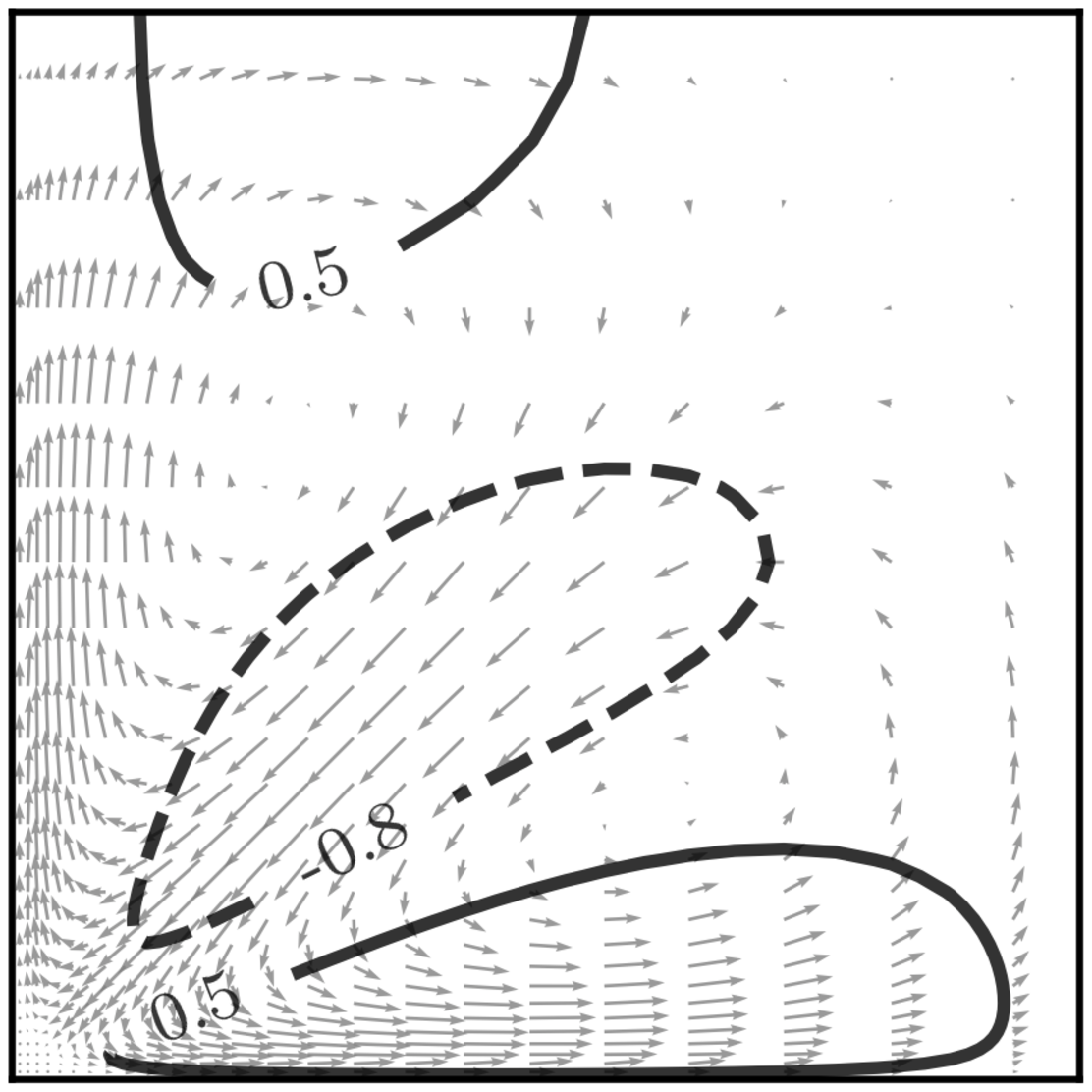}} & 
        \raisebox{-.5\height}{\includegraphics[scale=0.23]{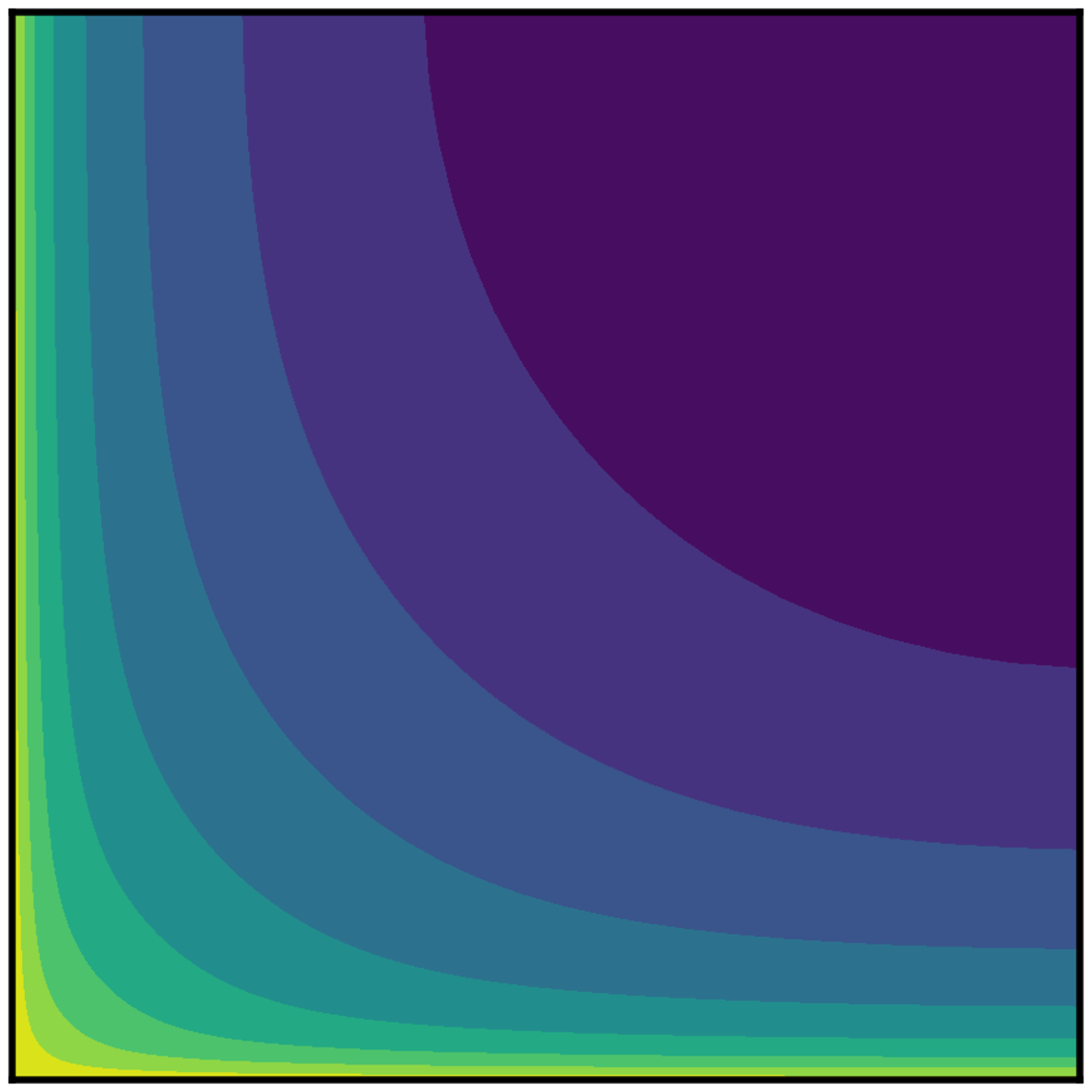}} &
        \raisebox{-.5\height}{\includegraphics[scale=0.23]{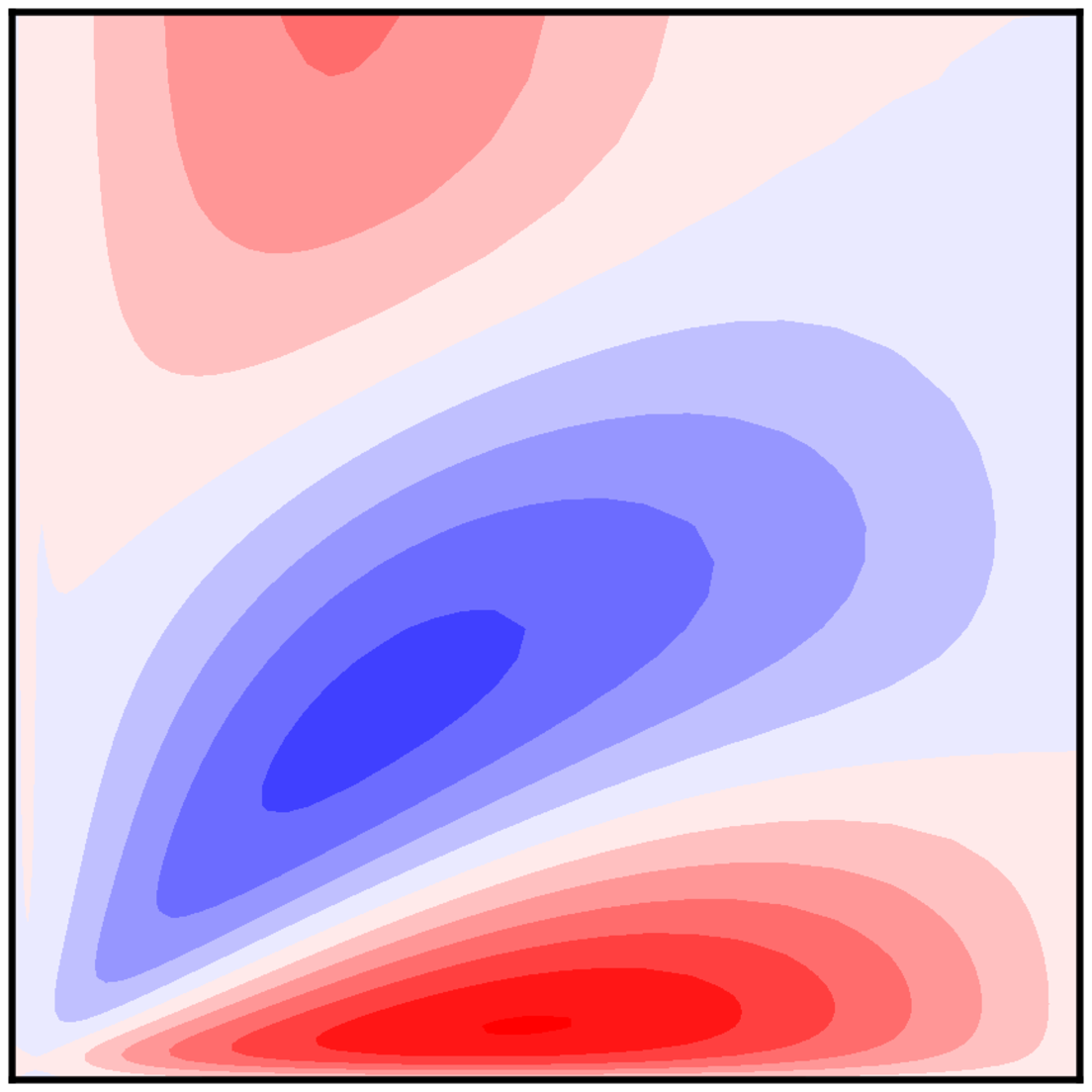}} &
        
        \\
        \rotatebox[origin=c]{90}{adjoint-based} & 
        \raisebox{-.5\height}{\includegraphics[scale=0.23]{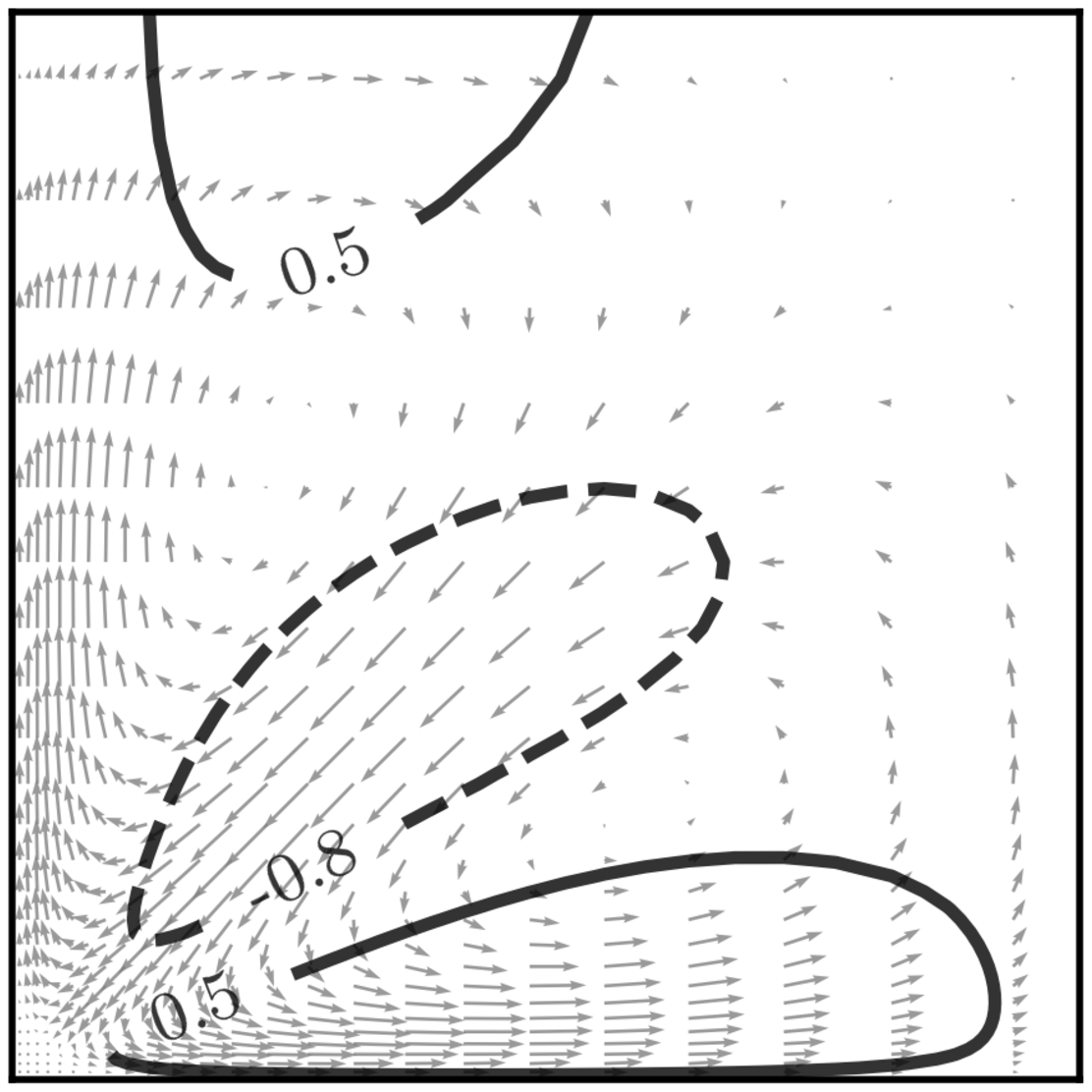}} &
        \raisebox{-.5\height}{\includegraphics[scale=0.23]{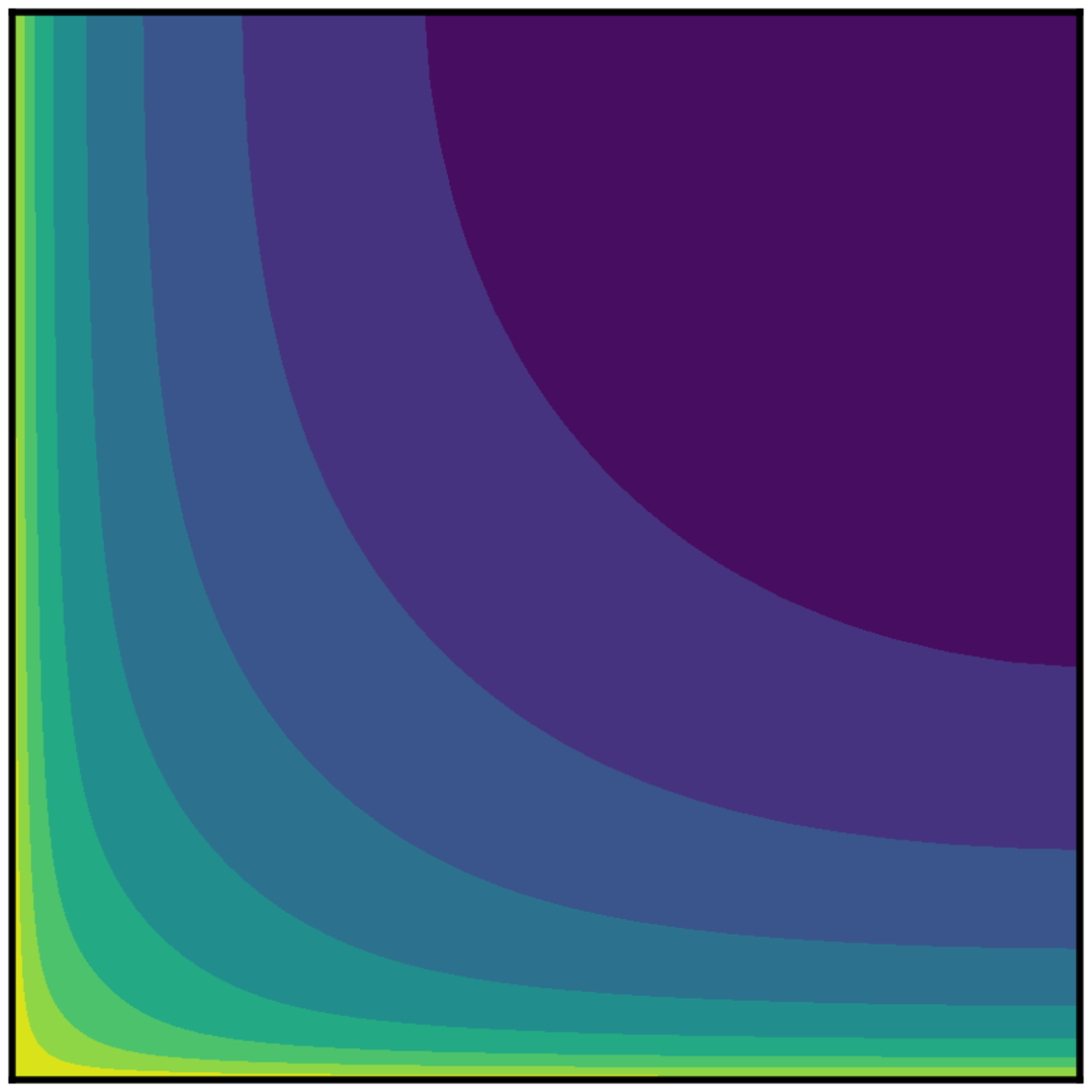}} &
        \raisebox{-.5\height}{\includegraphics[scale=0.23]{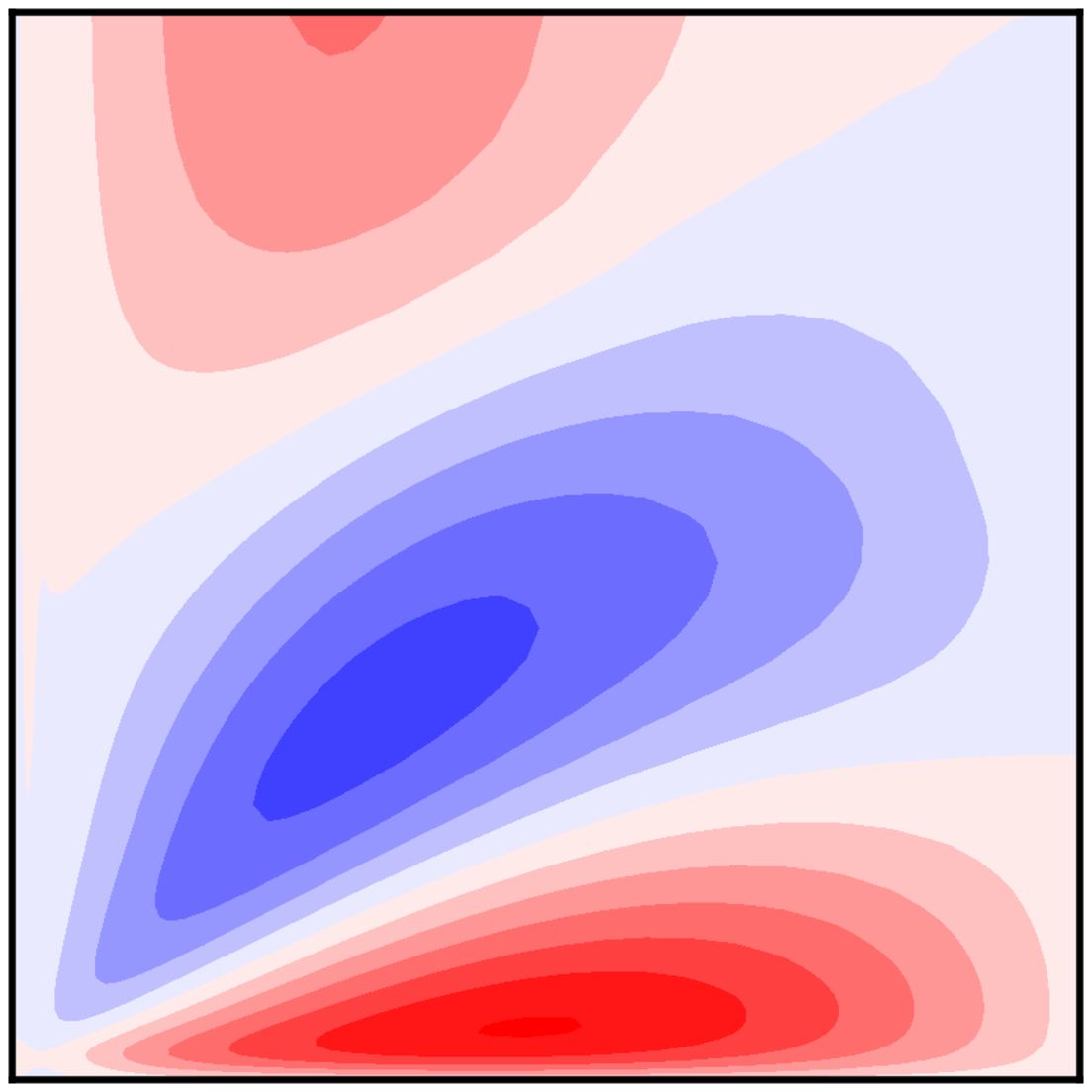}} &
        \raisebox{-.5\height}{\includegraphics[scale=0.23]{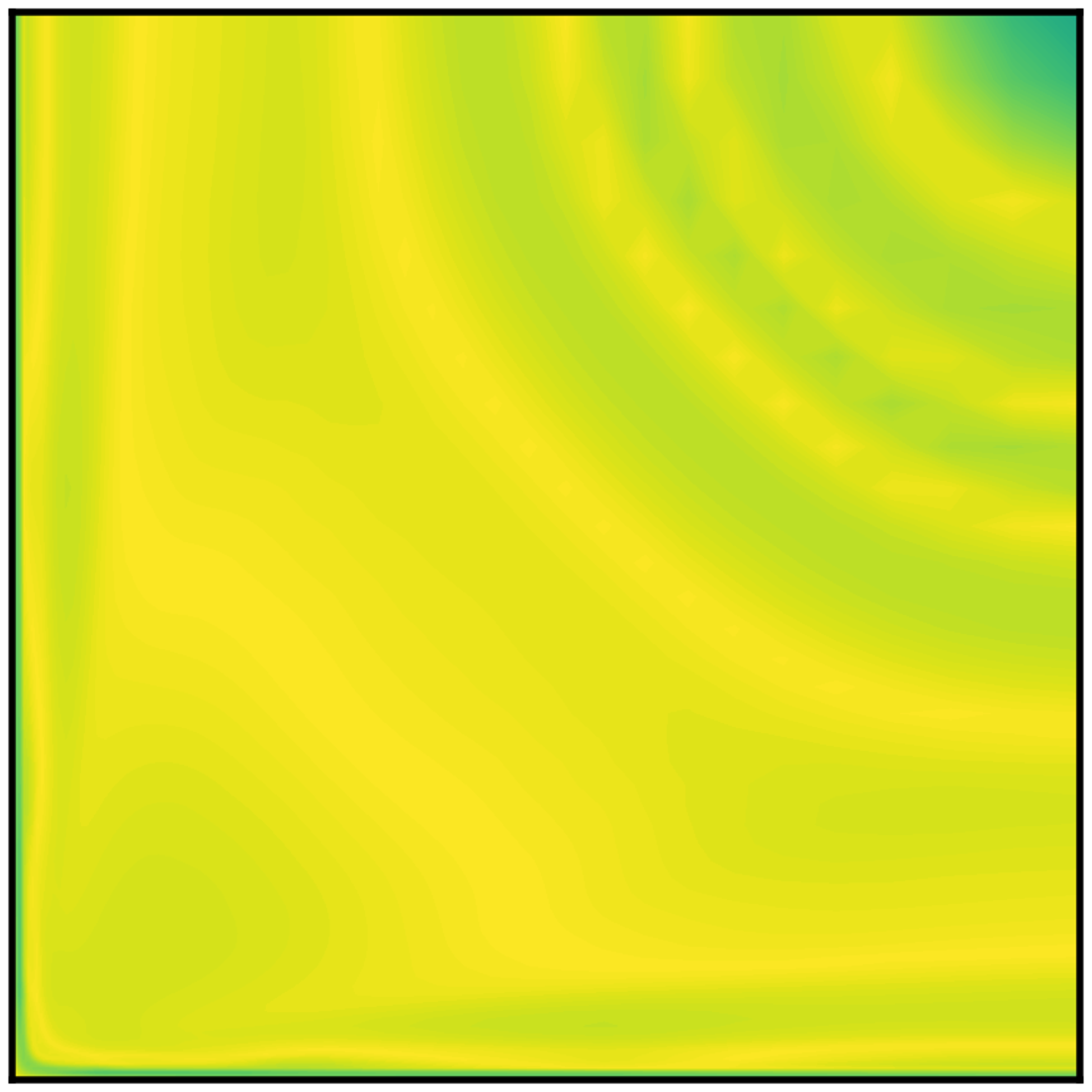}}
        \\
        \rotatebox[origin=c]{90}{ensemble-based} &
        \raisebox{-.5\height}{\includegraphics[scale=0.23]{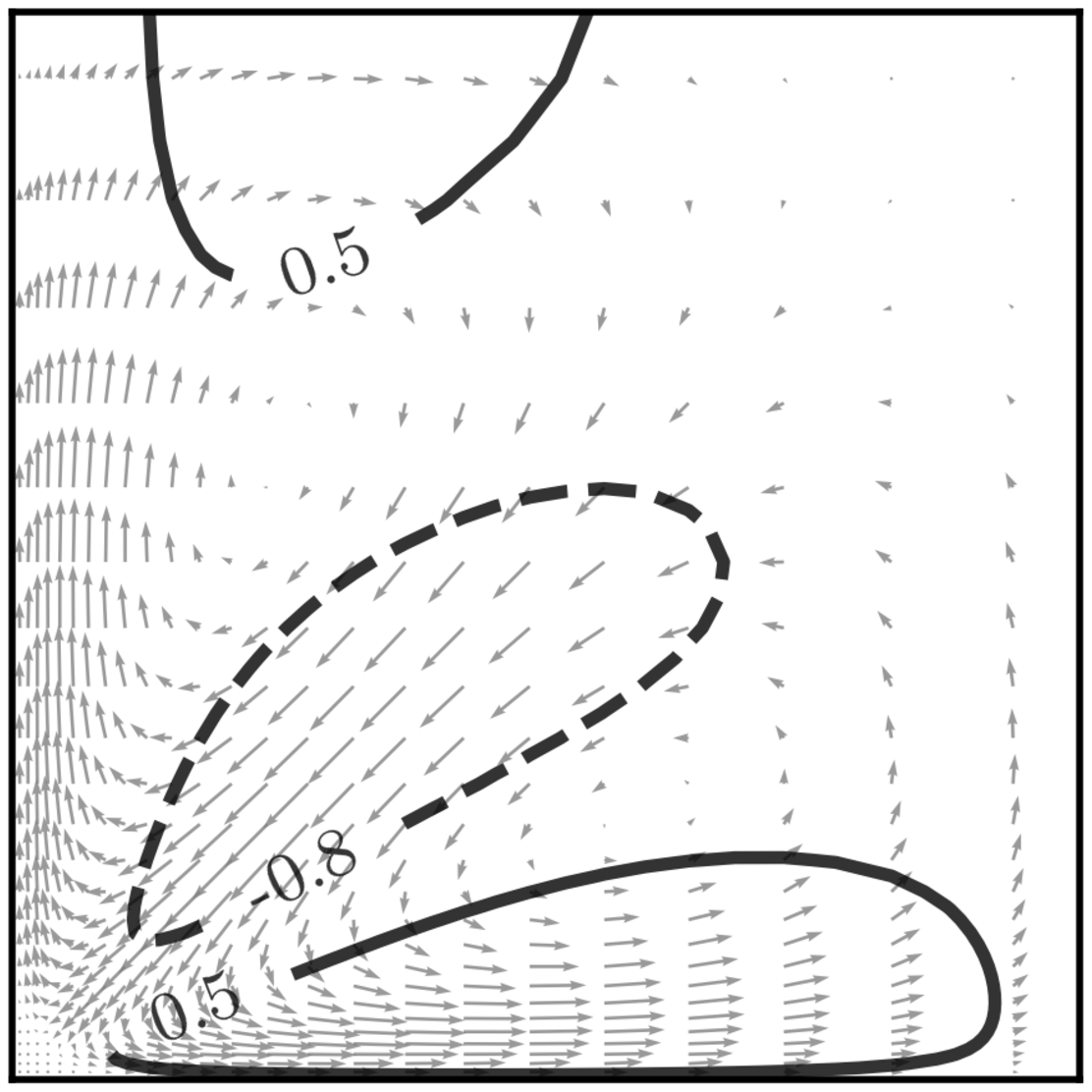}} &
        \raisebox{-.5\height}{\includegraphics[scale=0.23]{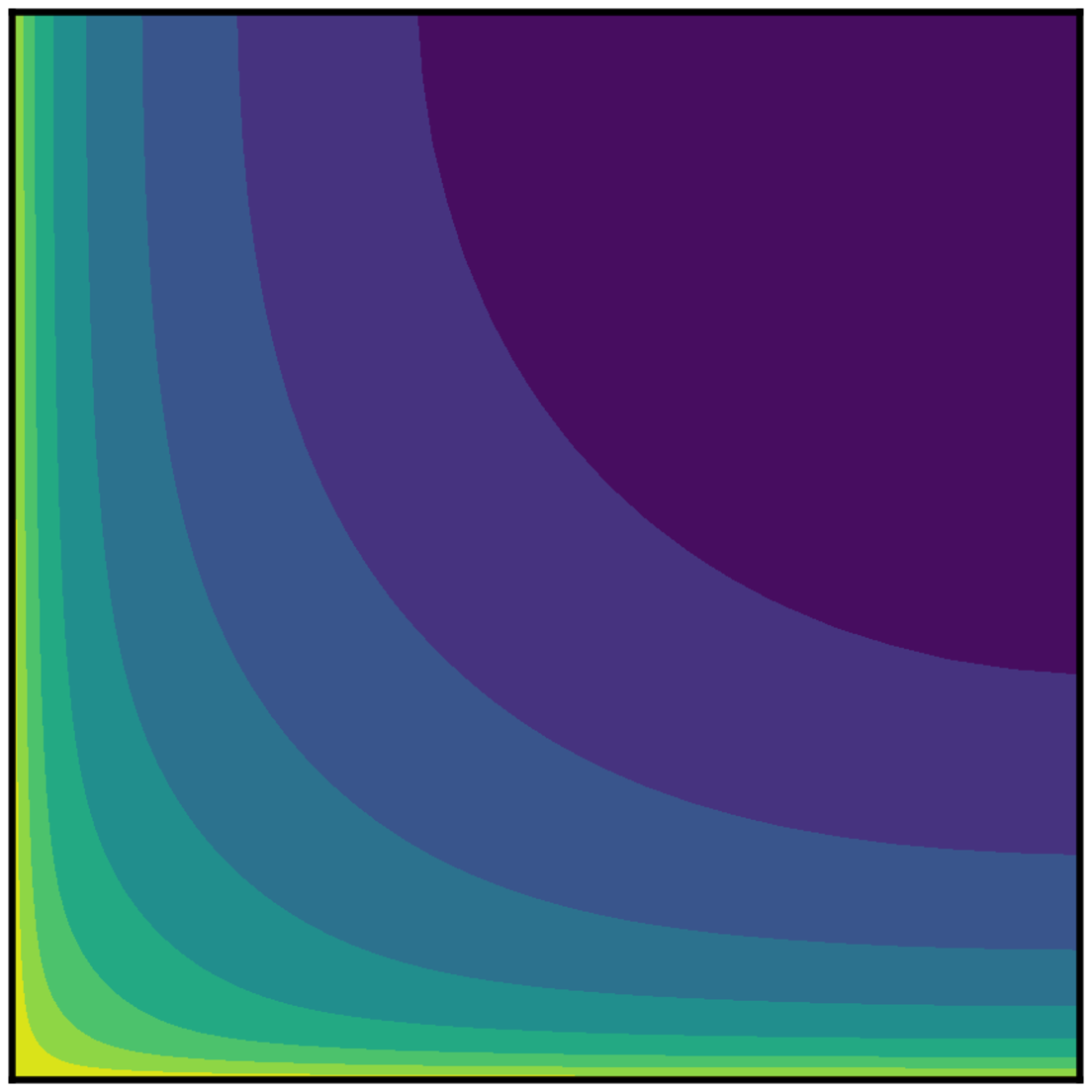}} &
        \raisebox{-.5\height}{\includegraphics[scale=0.23]{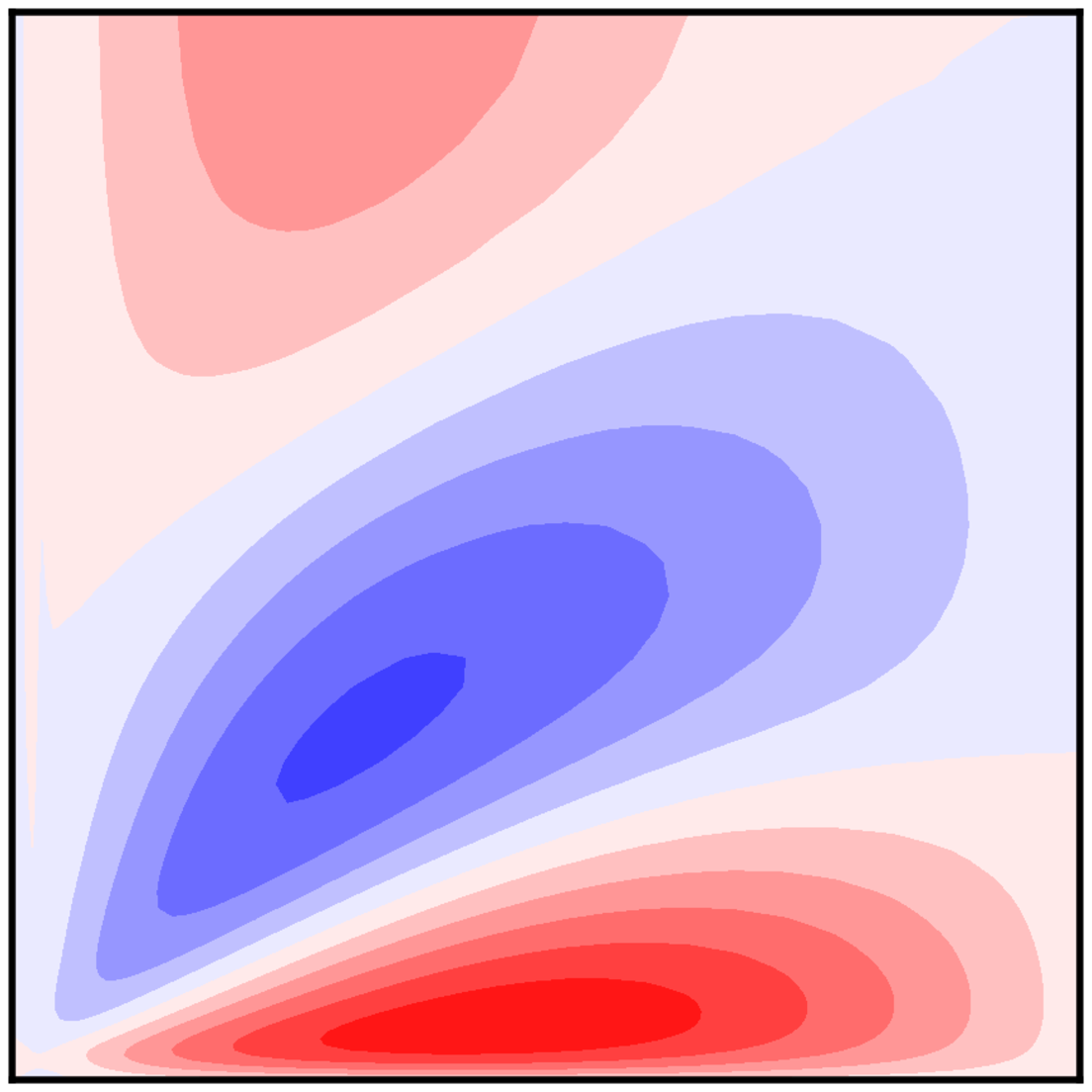}} &
        \raisebox{-.5\height}{\includegraphics[scale=0.23]{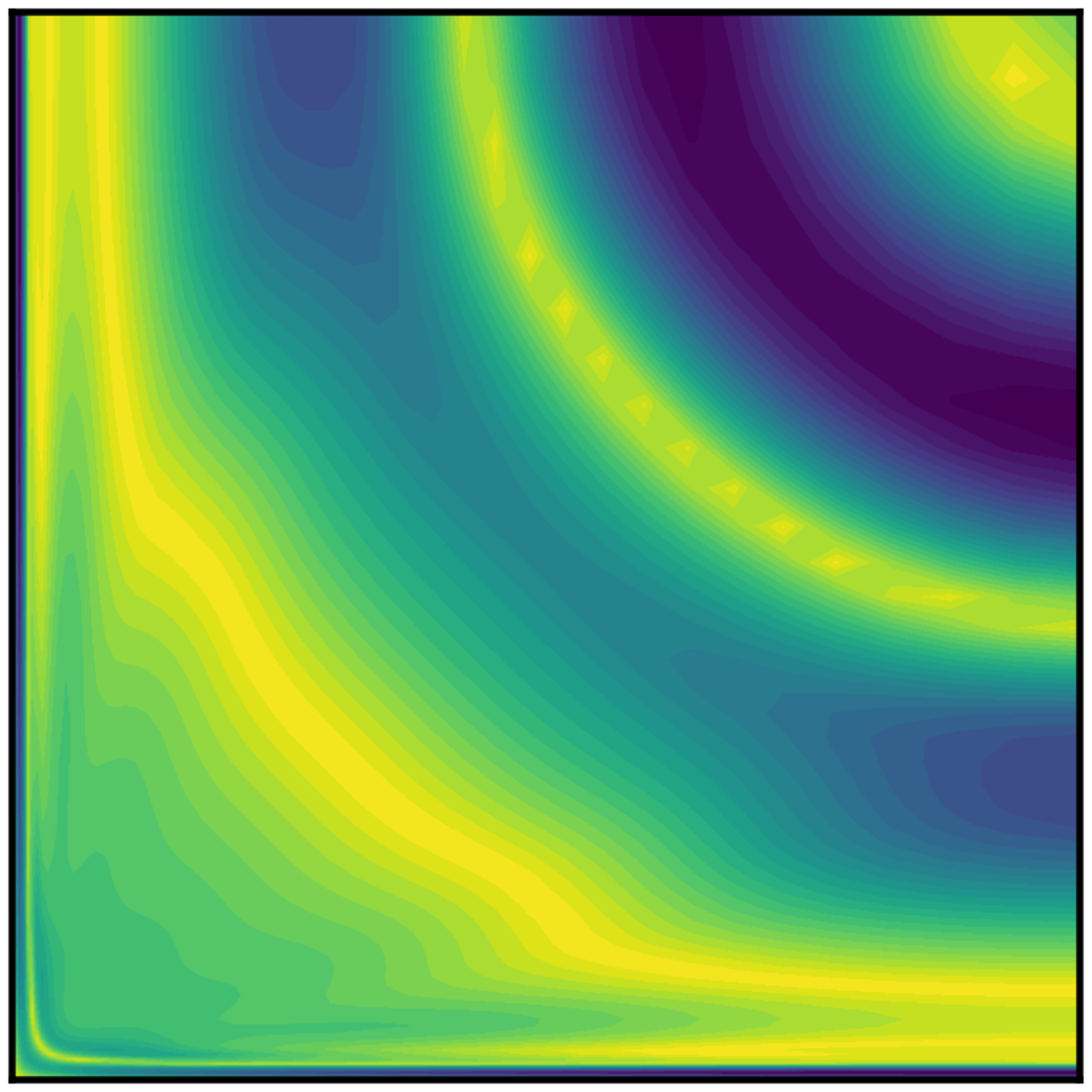}} 
        \\
        & 
        & \raisebox{-.5\height}{\includegraphics[scale=0.33]{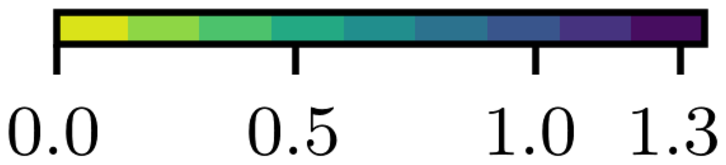}} 
        & \raisebox{-.5\height}{\includegraphics[scale=0.33]{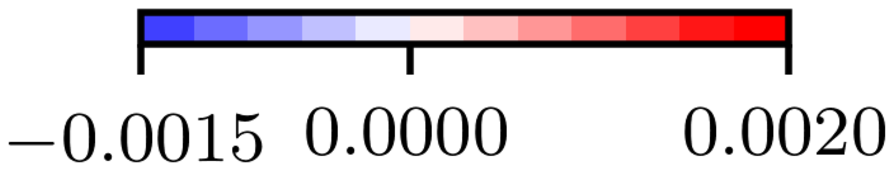}} 
        & \raisebox{-.5\height}{\includegraphics[scale=0.26]{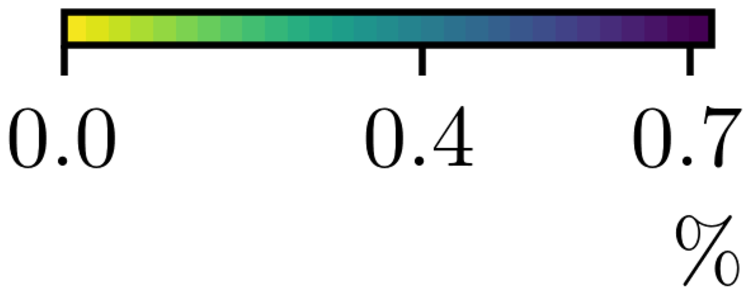}} 
    \end{tabular}
    \caption{ 
     Plots of the velocity vector and components $u_{x}$ and $u_y$ in the square duct predicted from the models learned by the adjoint (center row) and ensemble method (bottom row), compared against the ground truth (top row). 
     The velocity vectors are plotted along with contours of the in-plane velocity $u_y$ scaled by a factor of 1000.
     The error contour is plotted based on $\|\bm{u} - \bm{u}^\text{truth}\|$ normalized by the maximum magnitude of $\bm{u}^\text{truth}$.}
    \label{fig:duct_u}
\end{figure}

\begin{figure}
    \centering
    \begin{tabular}{cccccc}
        & $\tau_{xy}$ & $\tau_{yz}$ & $\tau_{yy}$ & $\tau_{yy}-\tau_{zz}$ & $\text{error}(\bm{\tau})$ \\
        \rotatebox[origin=c]{90}{ground truth} & 
        \raisebox{-.5\height}{\includegraphics[scale=0.2]{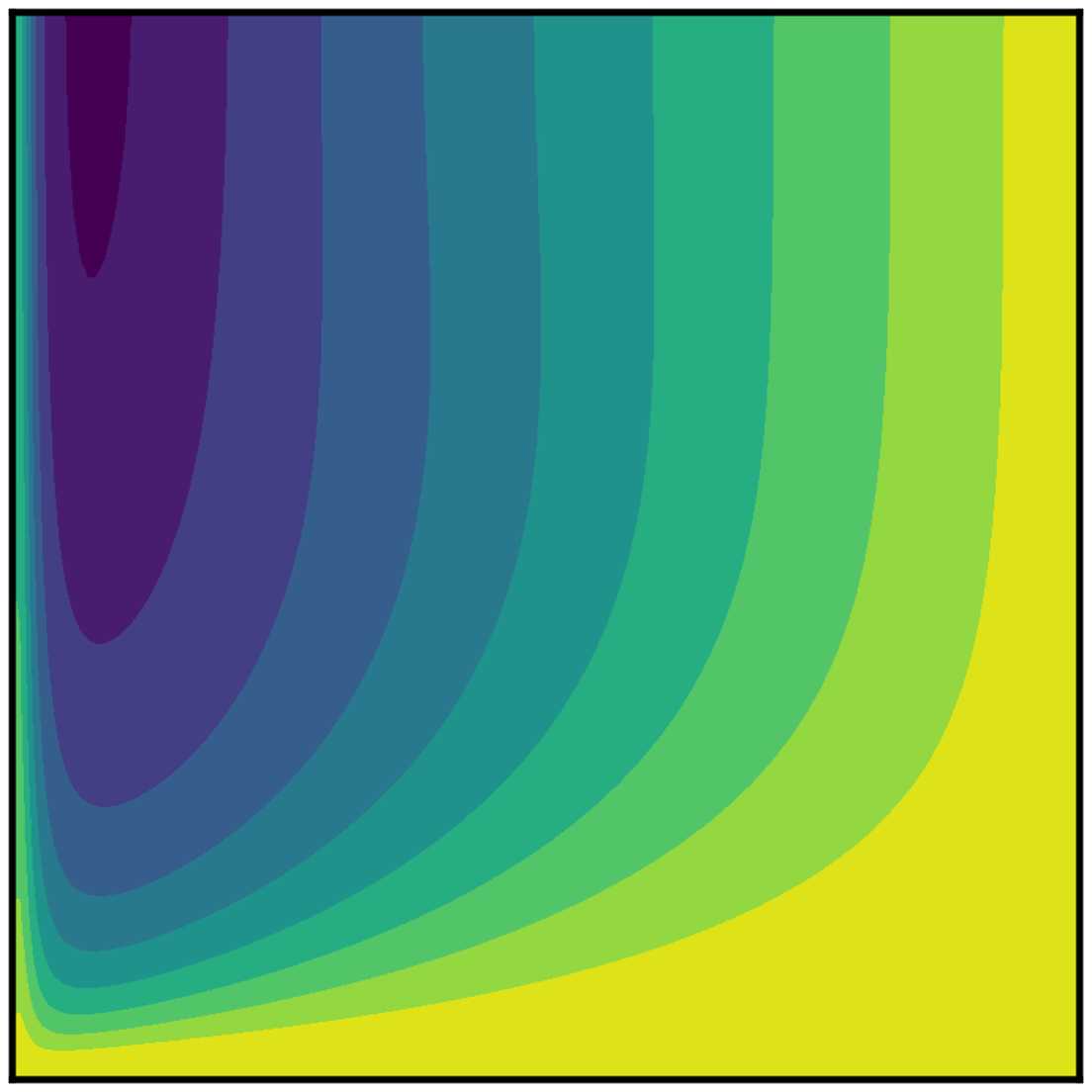}} &
        \raisebox{-.5\height}{\includegraphics[scale=0.2]{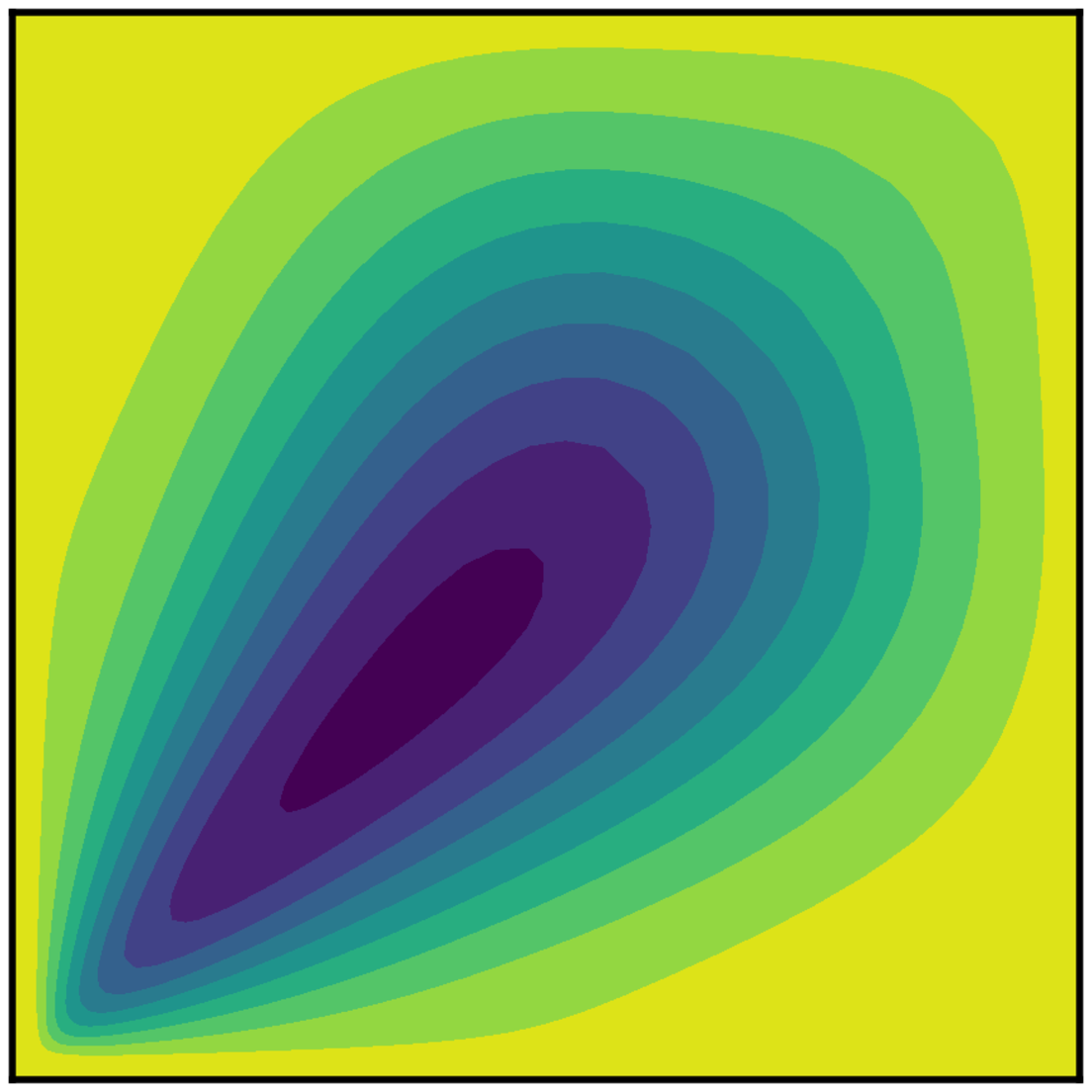}} &
        \raisebox{-.5\height}{\includegraphics[scale=0.2]{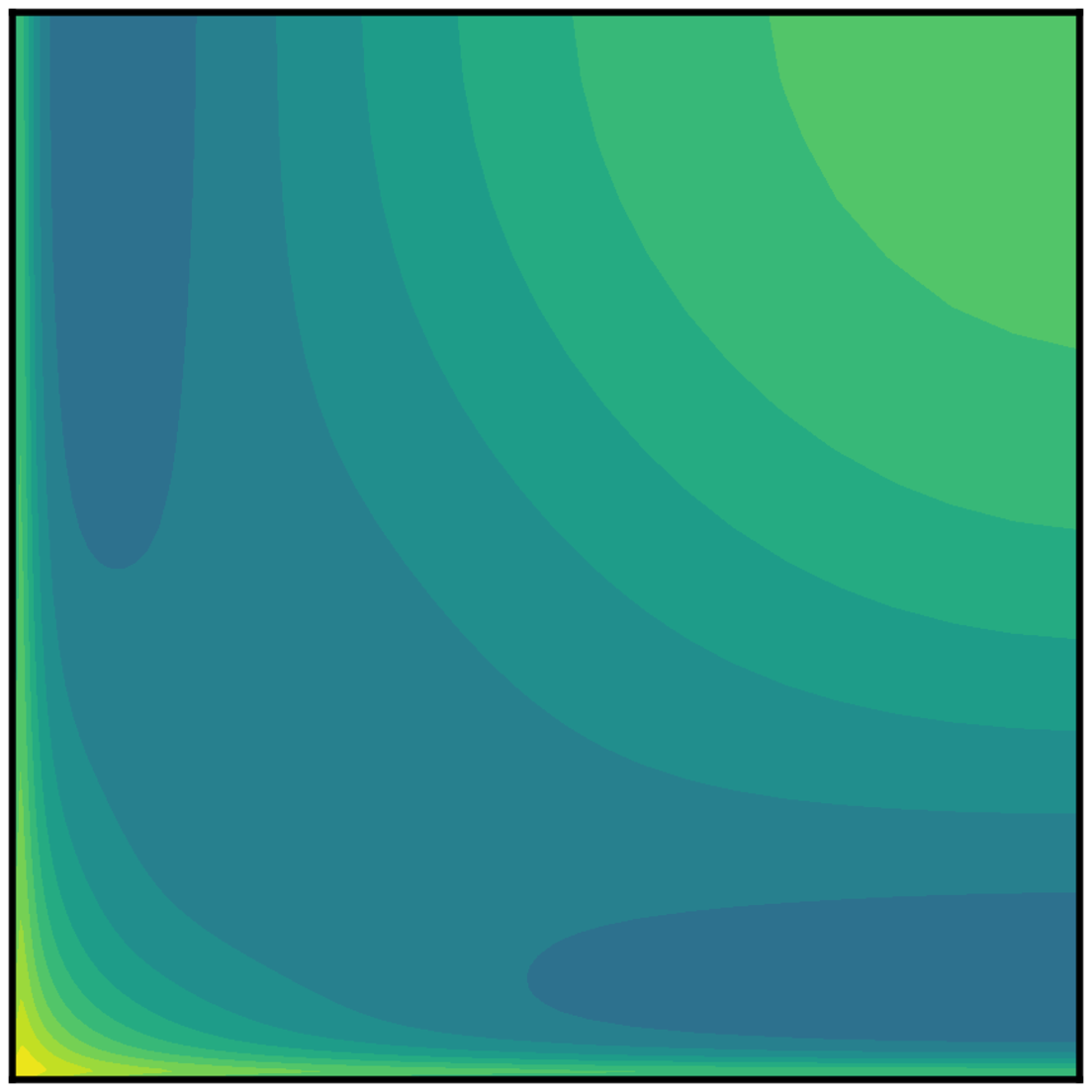}} &
        \raisebox{-.5\height}{\includegraphics[scale=0.2]{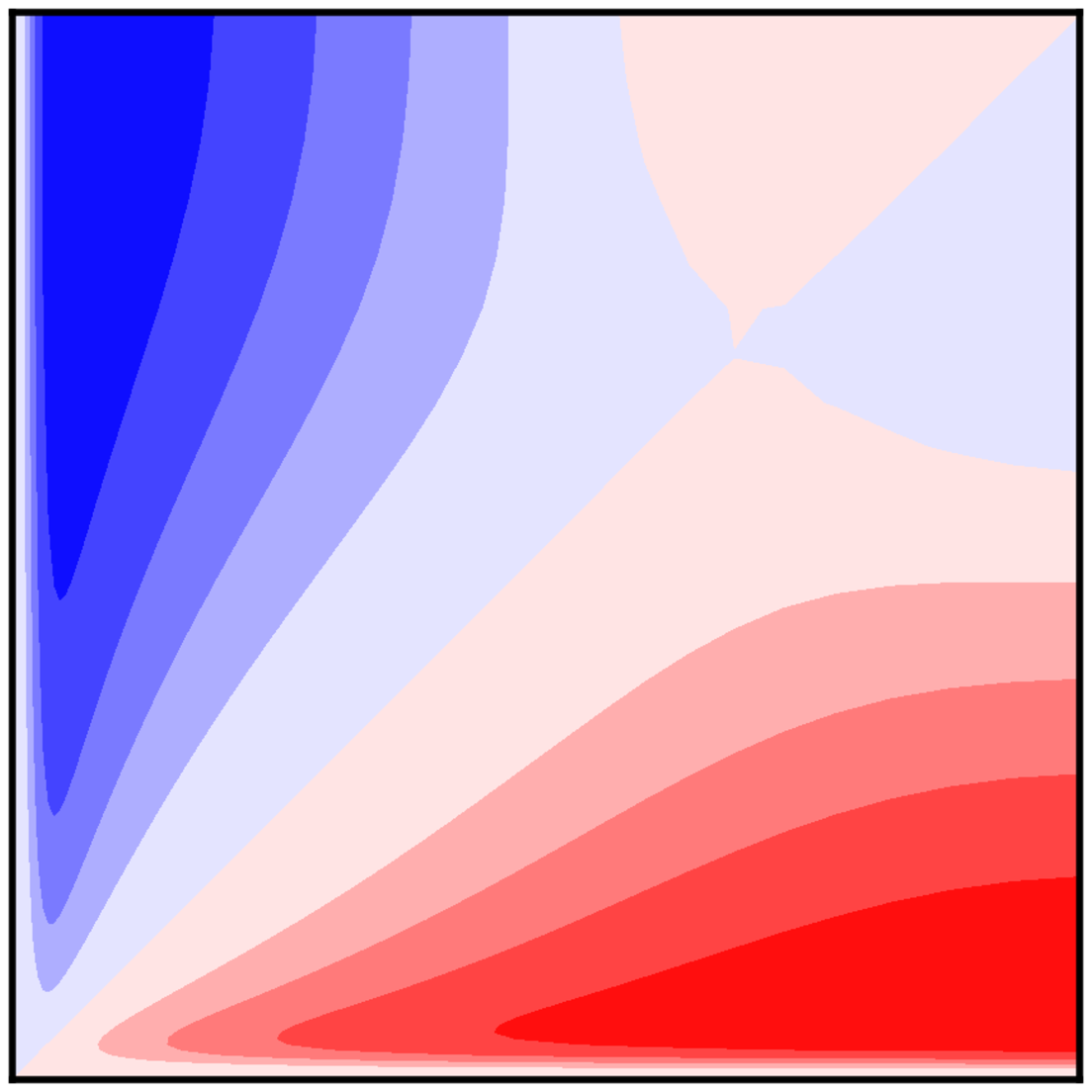}} &
        \\
        \rotatebox[origin=c]{90}{adjoint-based} & 
        \raisebox{-.5\height}{\includegraphics[scale=0.2]{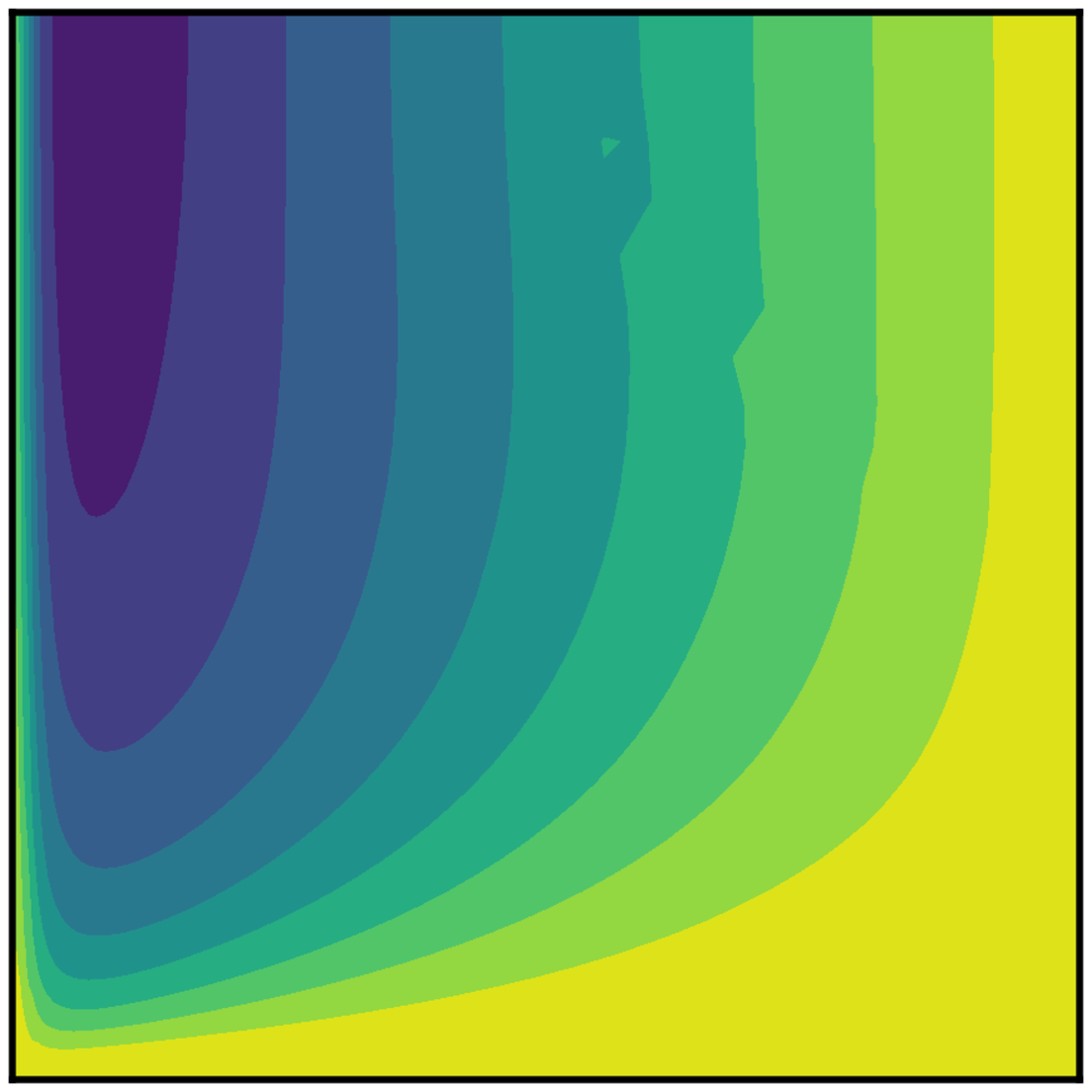}} &
        \raisebox{-.5\height}{\includegraphics[scale=0.2]{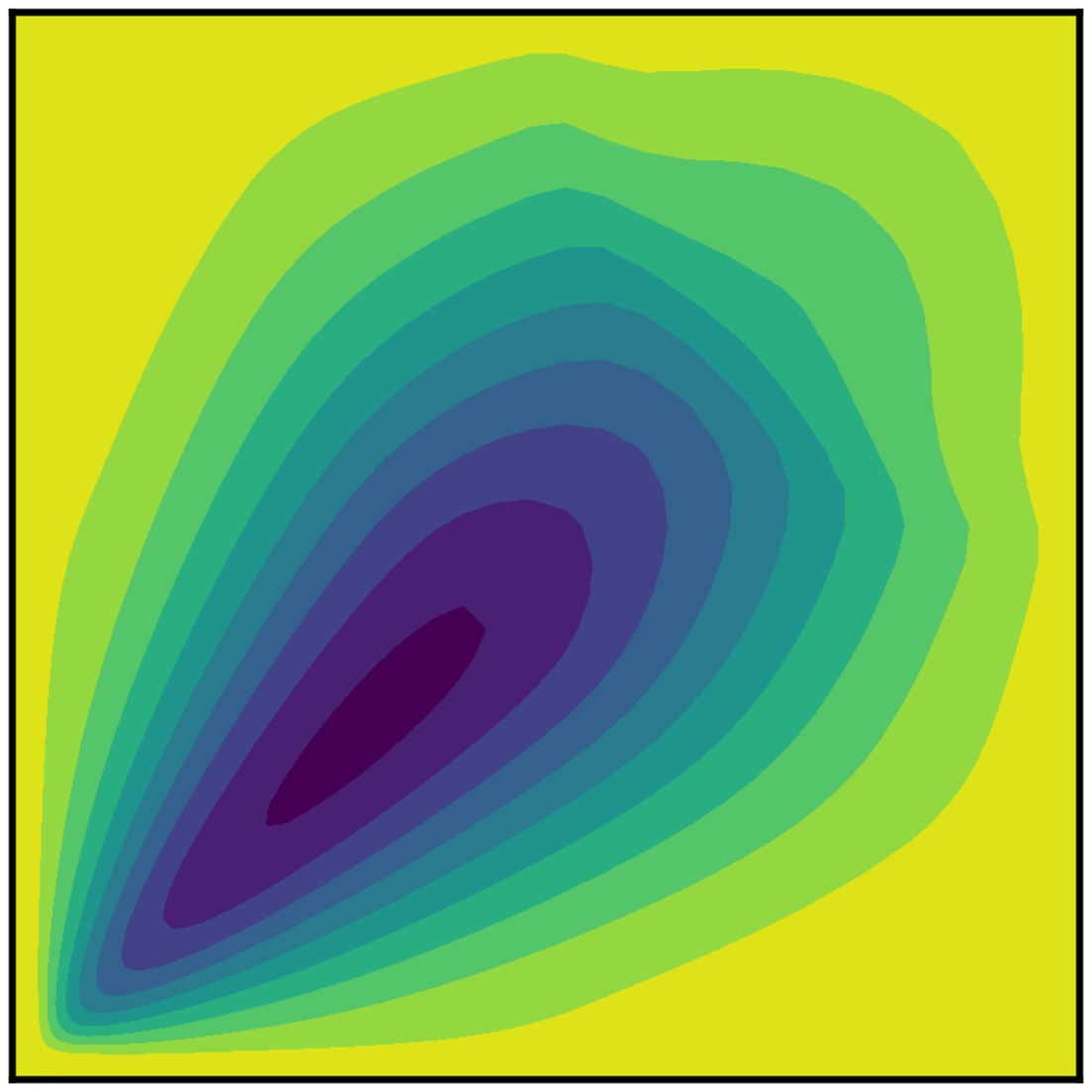}} &
        \raisebox{-.5\height}{\includegraphics[scale=0.2]{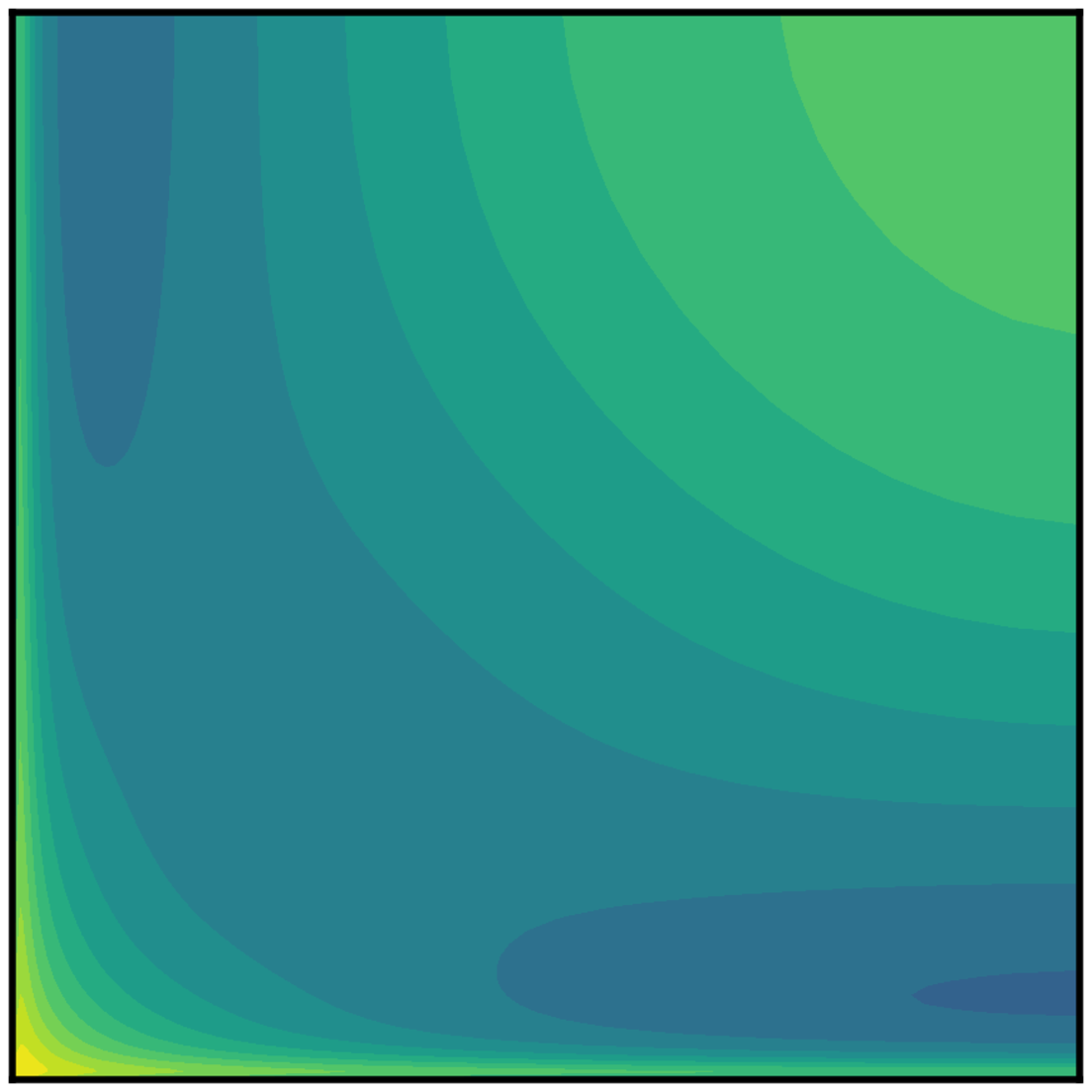}} &
        \raisebox{-.5\height}{\includegraphics[scale=0.2]{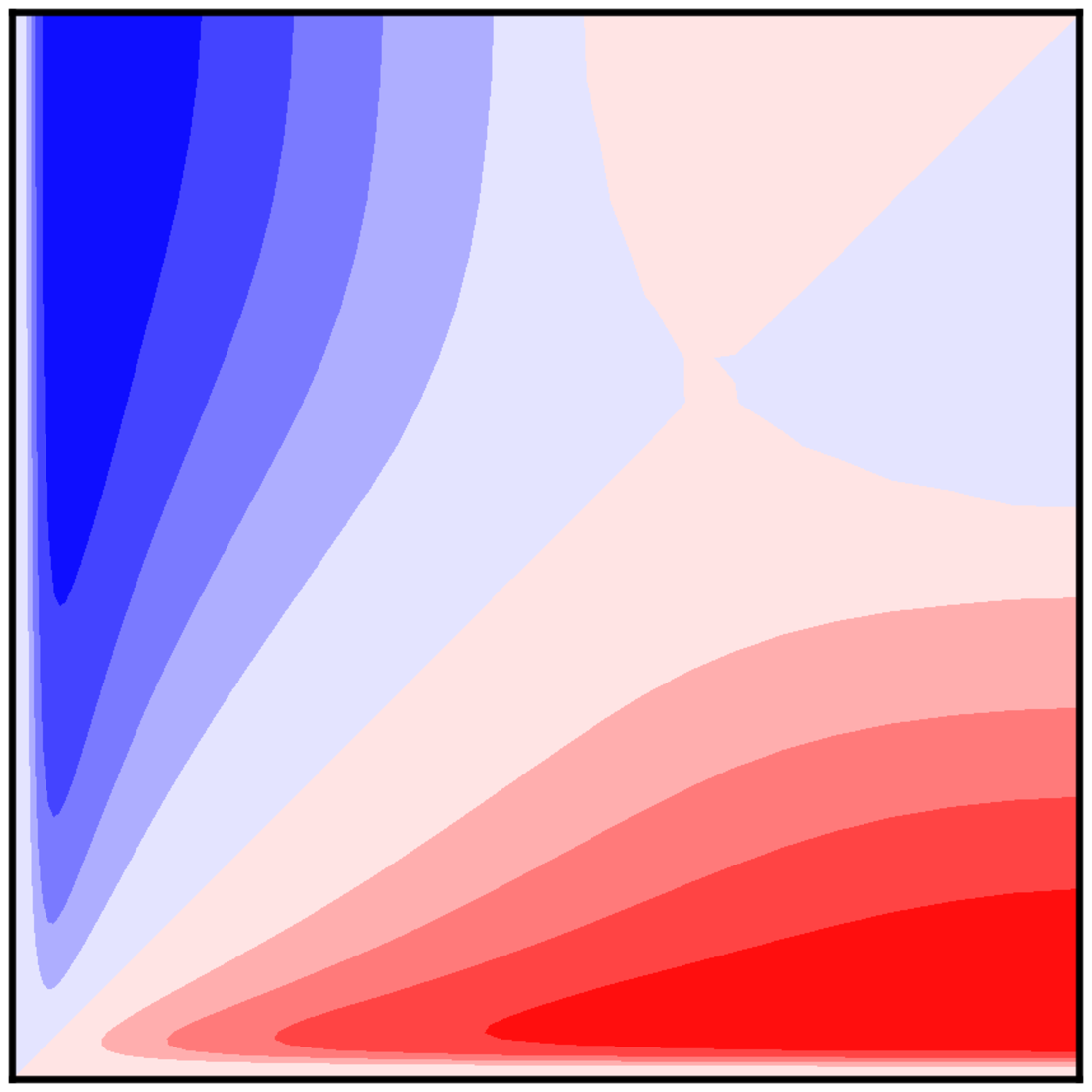}} &
        \raisebox{-.5\height}{\includegraphics[scale=0.2]{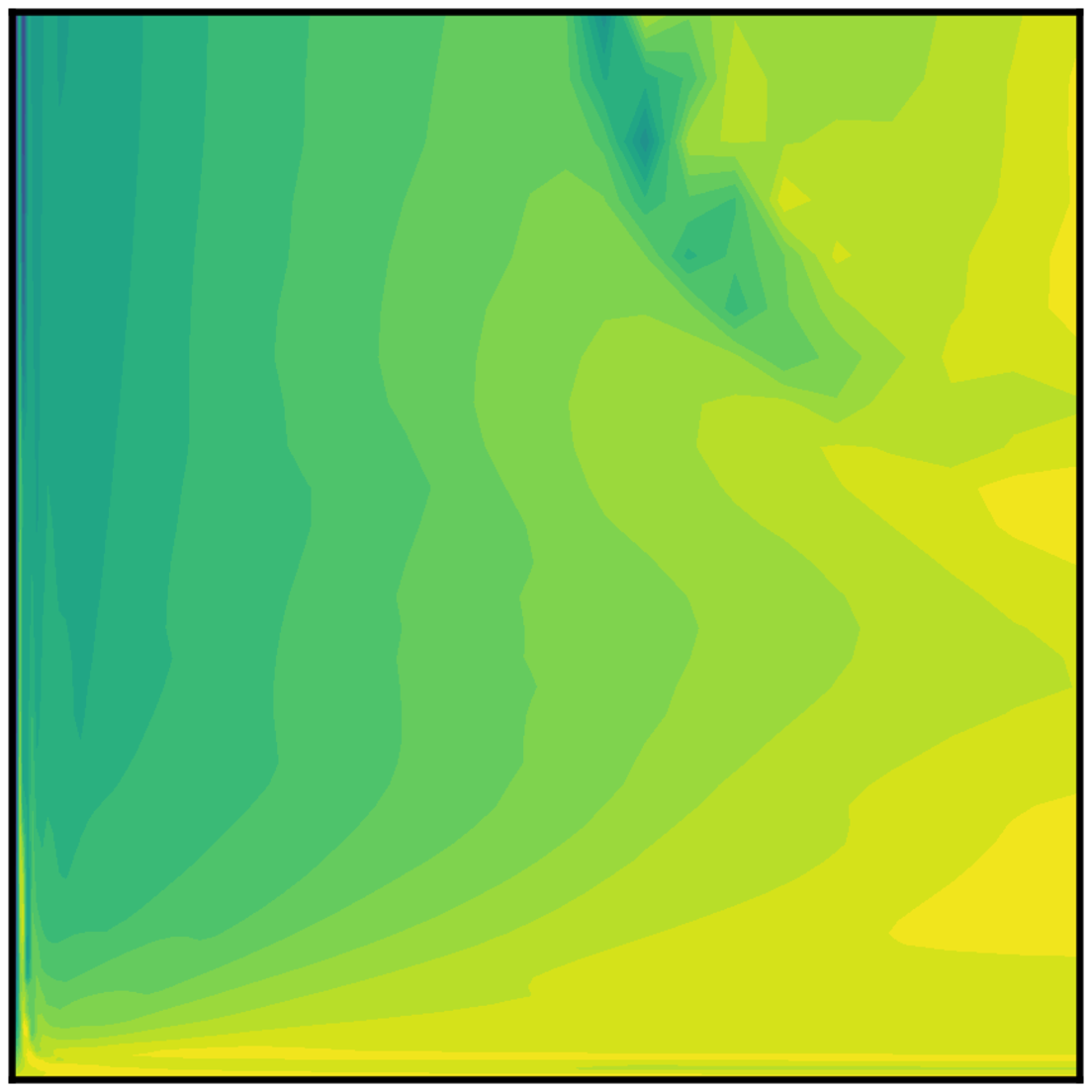}}
        \\
        \rotatebox[origin=c]{90}{ensemble-based} &
        \raisebox{-.5\height}{\includegraphics[scale=0.2]{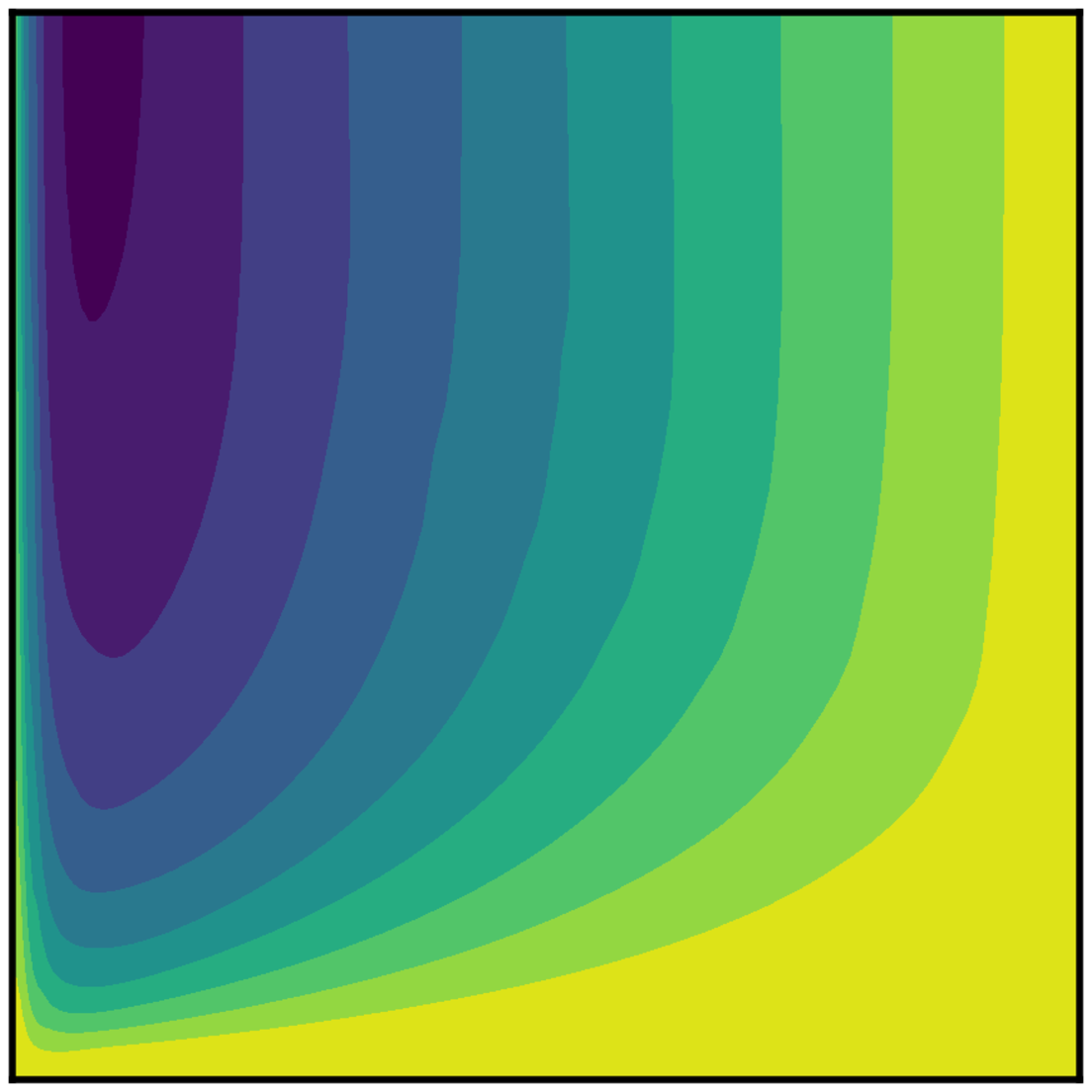}} &
        \raisebox{-.5\height}{\includegraphics[scale=0.2]{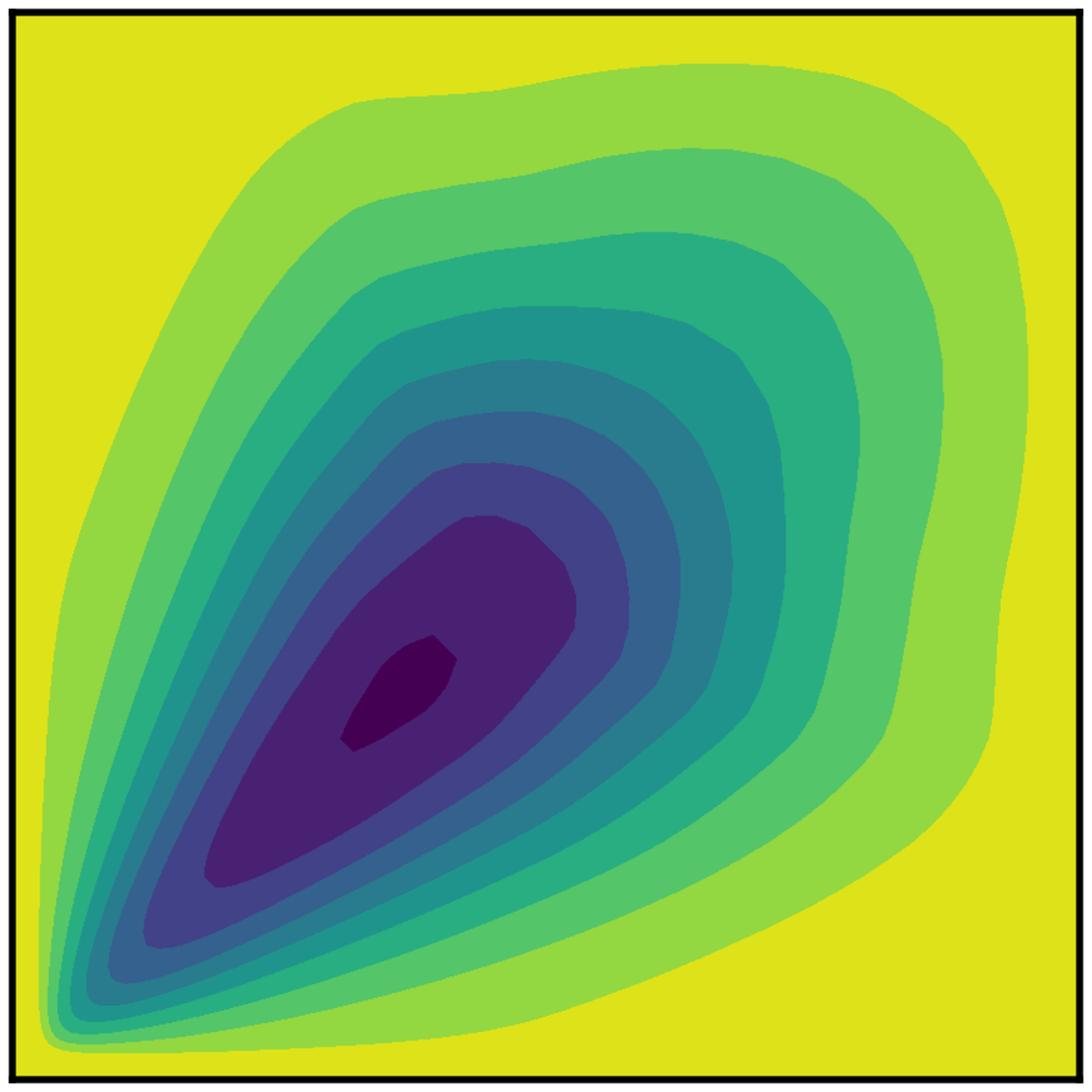}} &
        \raisebox{-.5\height}{\includegraphics[scale=0.2]{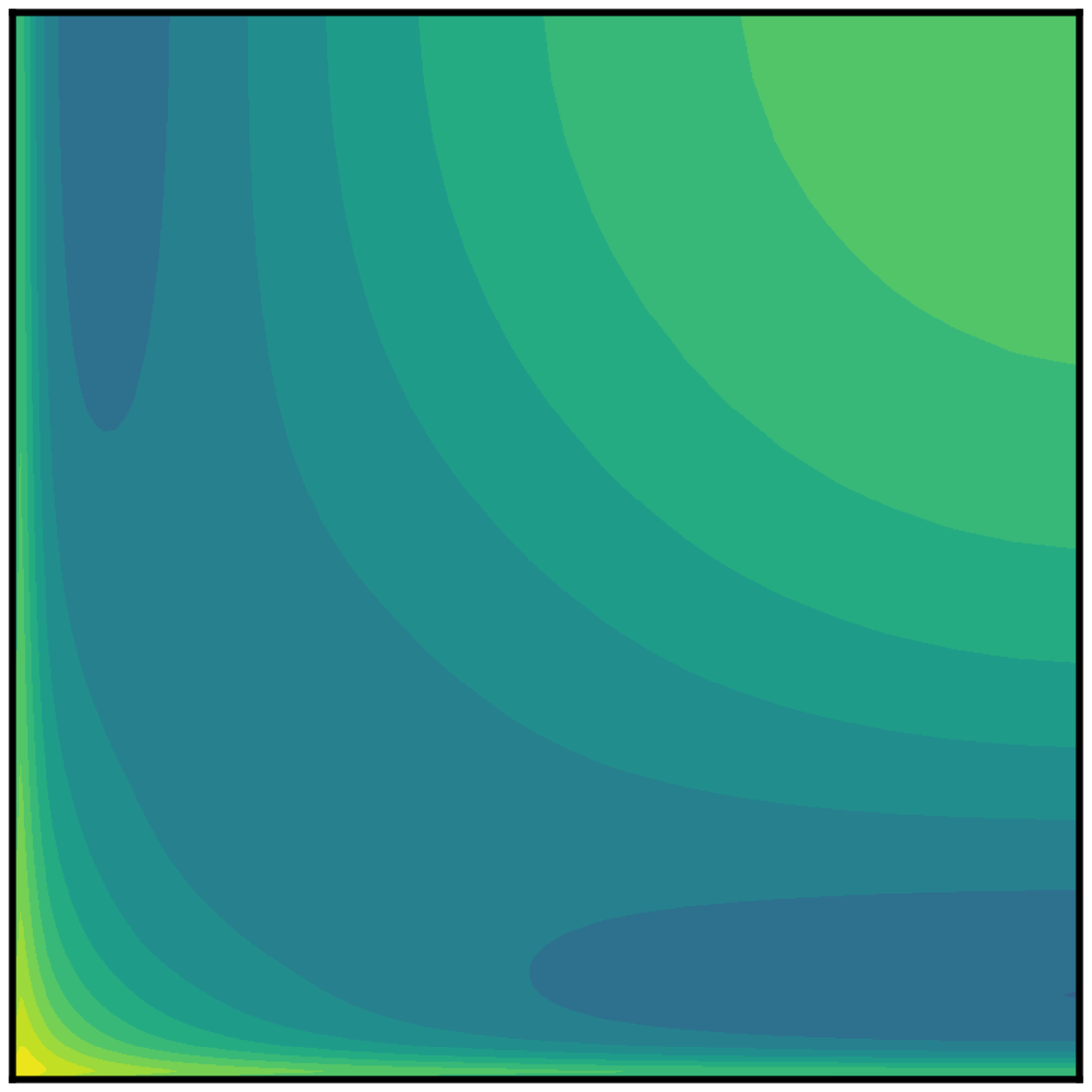}} &
        \raisebox{-.5\height}{\includegraphics[scale=0.2]{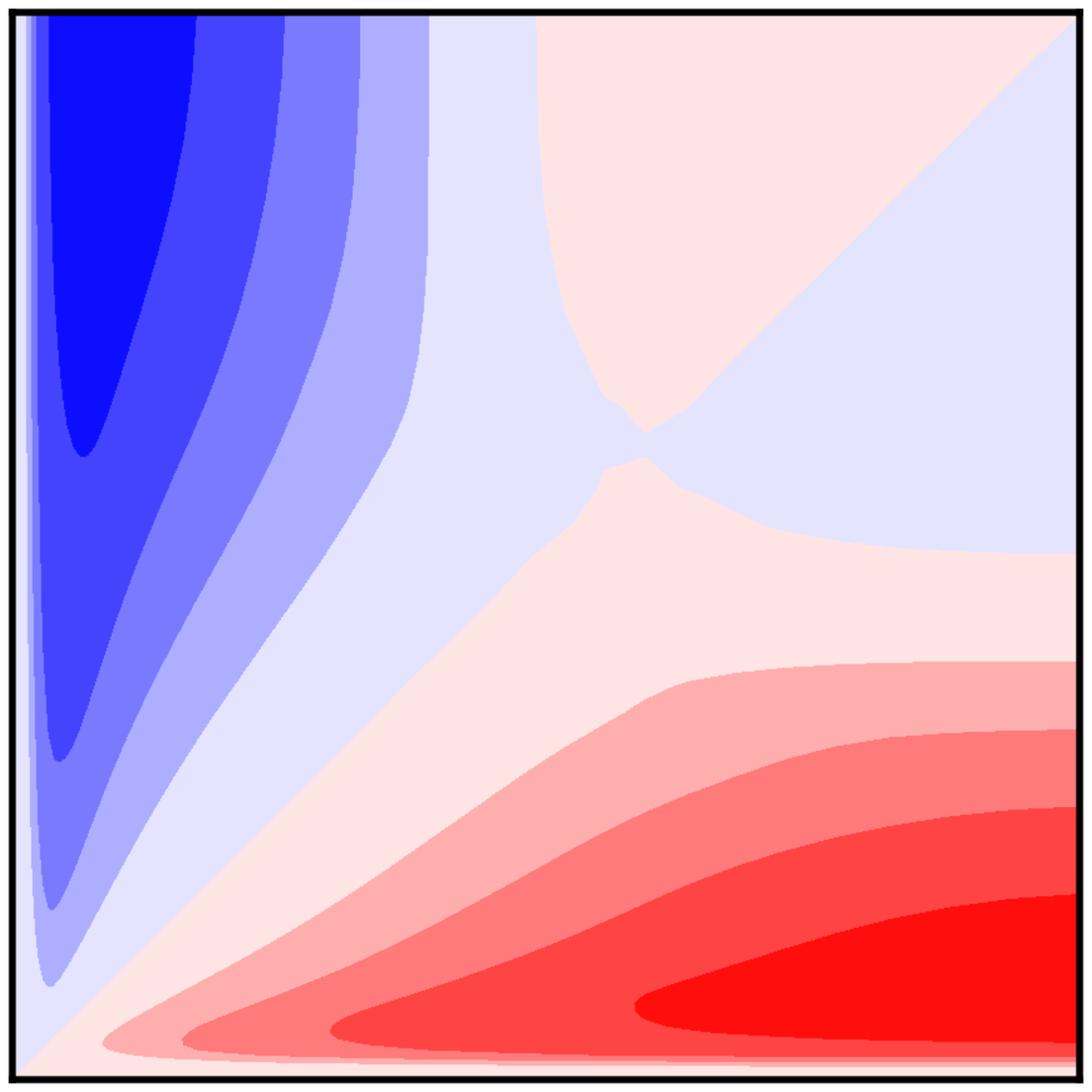}} &
        \raisebox{-.5\height}{\includegraphics[scale=0.2]{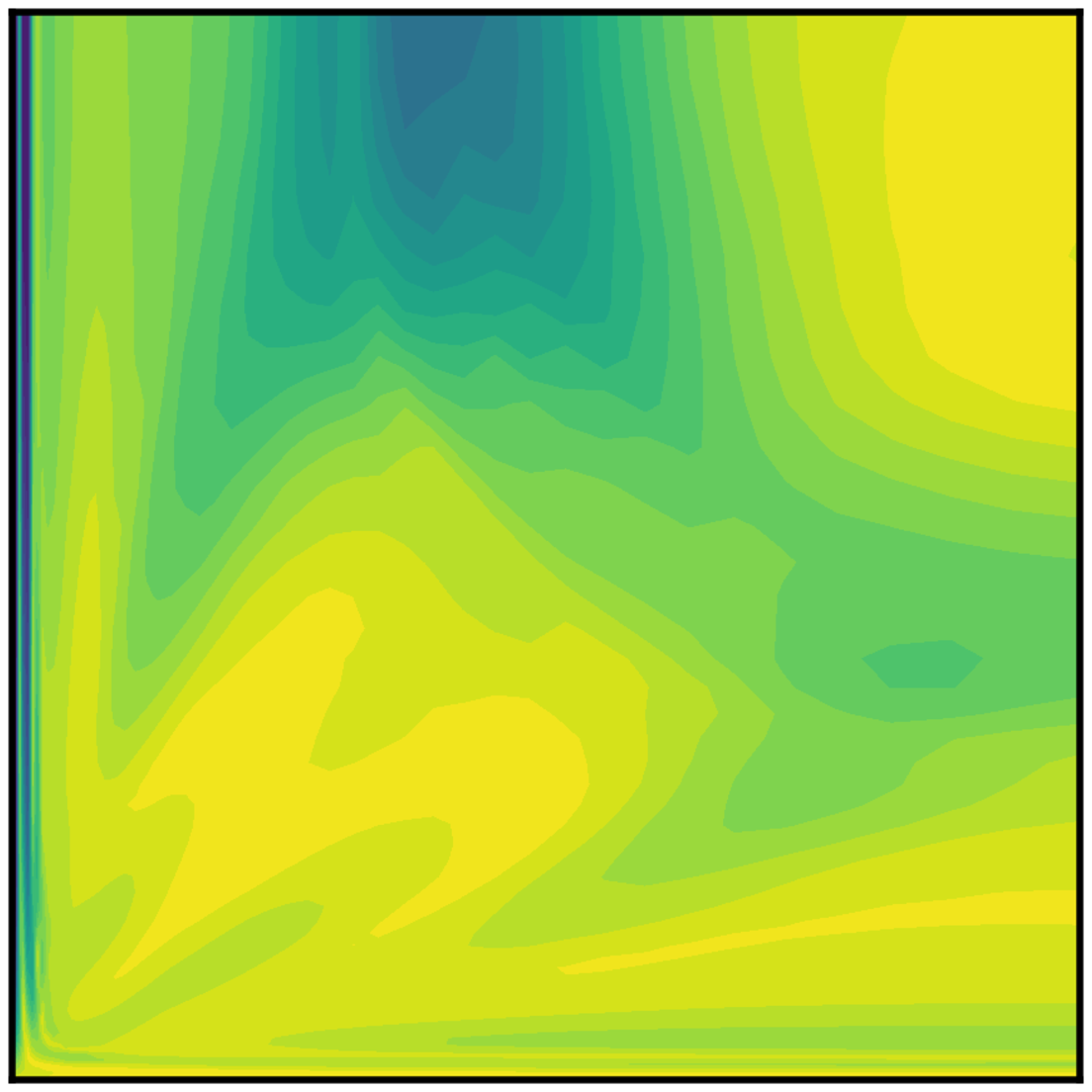}}
        \\
        &\raisebox{-.5\height}{\includegraphics[scale=0.25]{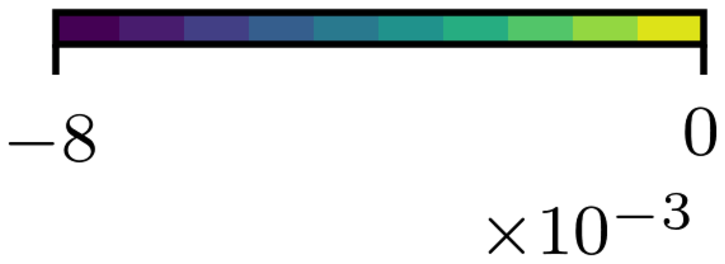}} 
        &\raisebox{-.5\height}{\includegraphics[scale=0.25]{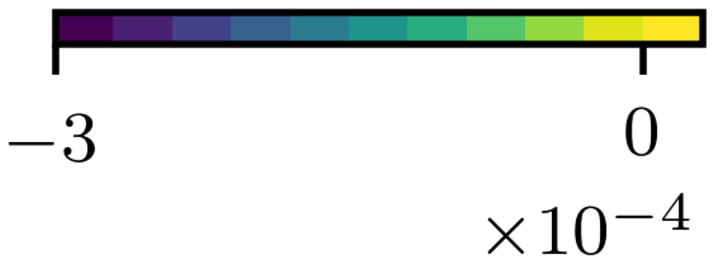}}
        &\raisebox{-.5\height}{\includegraphics[scale=0.22]{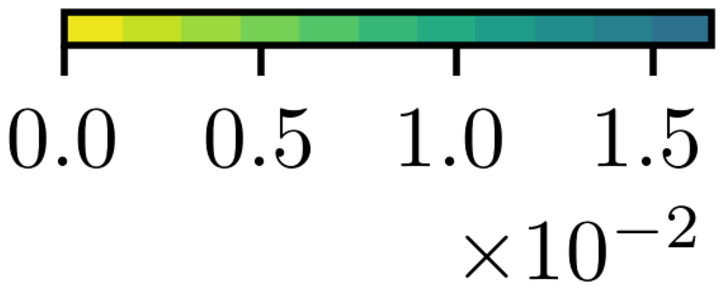}}
        &\raisebox{-.5\height}{\includegraphics[scale=0.25]{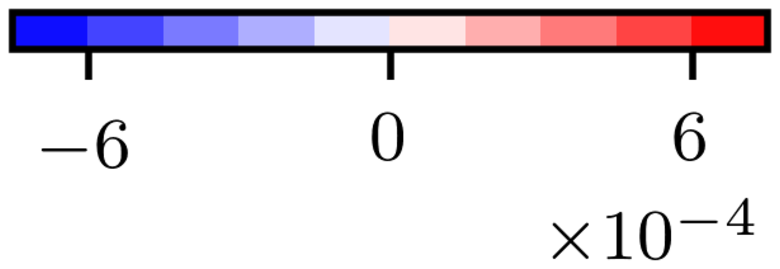}}
        &\raisebox{-.5\height}{\includegraphics[scale=0.22]{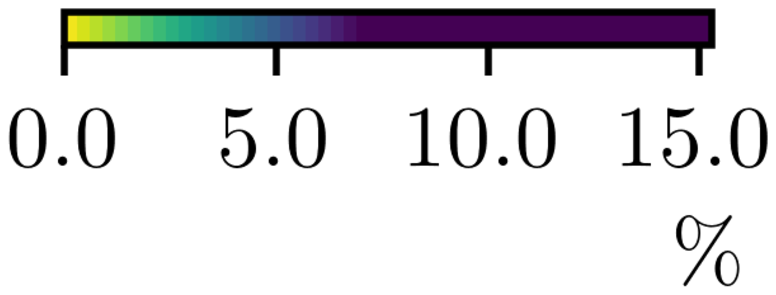}}
    \end{tabular}
    \caption{ 
     Plots of Reynolds shear stresses $\tau_{xy}$ and $\tau_{yz}$, normal stress $\tau_{yy}$, and normal stresses imbalance $\tau_{yy}-\tau_{zz}$ in the square duct predicted from the models learned by the adjoint (center row) and ensemble method (bottom row), compared against the ground truth (top row). The error contour is plotted based on $\|\bm{\tau} - \bm{\tau}^\text{truth}\|$ normalized by the maximum magnitude of $\bm{\tau}^\text{truth}$.
    }
    \label{fig:duct_rs}
\end{figure}

We first use the proposed ensemble-based method to train the turbulence model for flows in a square duct, and the results show that the predicted Reynolds stress has a good agreement with the synthetic ground truth (Equation~\ref{eq:gfunc}). 
The plots of the velocity and the Reynolds stress are presented in Figures~\ref{fig:duct_u} and~\ref{fig:duct_rs} with comparison to the adjoint-based method and the ground truth.
The contour lines for $u_y$ are indicated in the velocity vector plot to clearly show similar patterns among the ground truth, the adjoint method, and the ensemble-based method.
The contour plots of the Reynolds stress in $\tau_{xy}$, $\tau_{yz}$, and $\tau_{yy}$ are used to demonstrate the ability of the ensemble method in discovering the underlying Reynolds stress model given velocity data.
The in-plane velocity is driven by Reynolds normal stresses imbalance $\tau_{yy}-\tau_{zz}$, which is evident from the vorticity transport equation~\citep{launder2002closure}.
As such, the imbalance $\tau_{yy}-\tau_{zz}$ is also presented in Figure~\ref{fig:duct_rs}, demonstrating that the Reynolds stress field is accurately learned from the in-plane velocities.
The learned model with the proposed method achieves similar results in both the velocity and Reynolds stress to those of the adjoint-based method.
The error contours are provided to show the error distribution of the adjoint-based and ensemble-based methods in the estimation of velocity and Reynolds stress. 
It is noticeable that the adjoint-based method can achieve better velocity estimation than the ensemble-based method.
As for the Reynolds stress, the adjoint-based and ensemble-based methods lead to similar results.
It is noted that in this case the entire field is used as the training data.
By using fewer observations, e.g., only velocity data on the anti-diagonal line (upper-right corner to lower-left corner), the full velocity field can be also recovered and the Reynolds stresses are correctly learned,  but the errors are larger, especially in velocity. This is presented in Appendix~\ref{sec:sensitivity}.
The results demonstrate that the proposed method is able to learn the underlying turbulence model, which in turn provides good estimations of velocities and Reynolds stresses.

To clearly show the performance of the trained model, we provide the error in the estimation of velocity and Reynolds stress.
The error over the computational domain is defined as
\begin{equation}
    \mathcal{E}(\boldsymbol{q}) = \frac{\| \boldsymbol{q}^\text{predict} - \boldsymbol{q}^\text{truth} \|}{\| \boldsymbol{q}^\text{truth} \|} \text{.}
\end{equation}
The comparison between the adjoint and ensemble-based methods in the error of velocity and Reynolds stress as well as the training efficiency is provided in Table~\ref{tab:compare_adjoint_ensemble}.
The results confirm that both adjoint and ensemble-based methods are able to achieve satisfactory agreements in the velocities and to predict the Reynolds stresses well.
By contrast, the adjoint-based method provides slightly better estimation than the ensemble method.
Specifically, the errors in velocity and Reynolds stress with the adjoint-based method are $0.1\%$ and $4.5\%$, respectively, while those for the ensemble method are $0.47\%$ and $5.8\%$, respectively.

As for the training efficiency, the adjoint-based method is more time-consuming compared to the ensemble-based method as shown in Table~\ref{tab:compare_adjoint_ensemble}.
Specifically, the adjoint-based method requires approximately $1000$ iterations which significantly increase the wall time to about $133$ hours in this case.
In contrast, the ensemble-based method is efficient to obtain comparable results within $3.6$ hours.
To achieve the error reduction of $\mathcal{E}(\boldsymbol{u}) < 0.005$, the adjoint method requires $238$ steps and a wall time of $32$ hours, while the ensemble-based method can reach the same error within only $0.6$ hours.
That is likely due to the use of Hessian information and the covariance inflation factor $\gamma$, which dynamically adjusts the relative weight of the cost function to accelerate the convergence~\citep{nocedal2006numerical}.
Here we emphasize that the adjoint-based learning for comparison is based on our particular implementation of the continuous adjoint method~\citep{othmer2008continuous}.
The used adjoint solver is publicly available in the Github repository~\citep{zhang2022dafi}.
Recent developments of the adjoint method, such as the online adjoint method~\citep{sirignano2021online}, would have significant potential to improve the efficiency of the adjoint-based learning method.
However, the present work aims to introduce the ensemble Kalman method for learning turbulence models, and comprehensive comparisons with the state-of-the-art adjoint method are out of the scope of this paper.
It is also noted that this work mainly focuses on the steady-state RANS problem where the data size is small.
In scenarios with large data sets such as unsteady three-dimension flow fields, the present algorithm would be computationally expensive compared to the adjoint method, since the update scheme requires the inversion of a matrix with the rank as the dimensionality of observation data.
In view of this limitation, dimension reduction techniques such as the truncated singular value decomposition are usually incorporated into the ensemble method~\citep{evensen2009data}, enabling it to handle large data sets.
This strategy has been widely applied in large-scale reservoir applications~\citep{chen2013levenberg, luo2018correlation}.

\begin{table}
    \begin{center}
    \begin{tabular}{l cccccc}
         Method & $\mathcal{E}(\boldsymbol{u})$ & $\mathcal{E}(\boldsymbol{\tau})$ & Total steps & Wall time & \shortstack{Steps \\ ($\mathcal{E}(\boldsymbol{u}) < 0.005$)}& \shortstack{Wall time \\ ($\mathcal{E}(\boldsymbol{u}) < 0.005$)} \\ \\
         Adjoint-based & $0.1\%$ & $4.5\%$ & $1000$ & $133$ hours & $238$ & $32$ hours \\ \\
         Ensemble-based & $0.47\%$ & $5.8\%$ & $50$ & $3.6$ hours & $8$ & $0.6$ hours \\
    \end{tabular}
    \caption{Comparison of the estimation error and time cost between adjoint-based and ensemble-based learning}
    \label{tab:compare_adjoint_ensemble}
    \end{center}
\end{table}

\begin{figure}
    \centering
    \subfloat[contours of invariants ]{\begin{tabular}{ccccl} 
        & adjoint & ensemble & truth & \\
        \rotatebox[origin=c]{90}{$\theta_1$} & 
        \raisebox{-.5\height}{\includegraphics[scale=0.15]{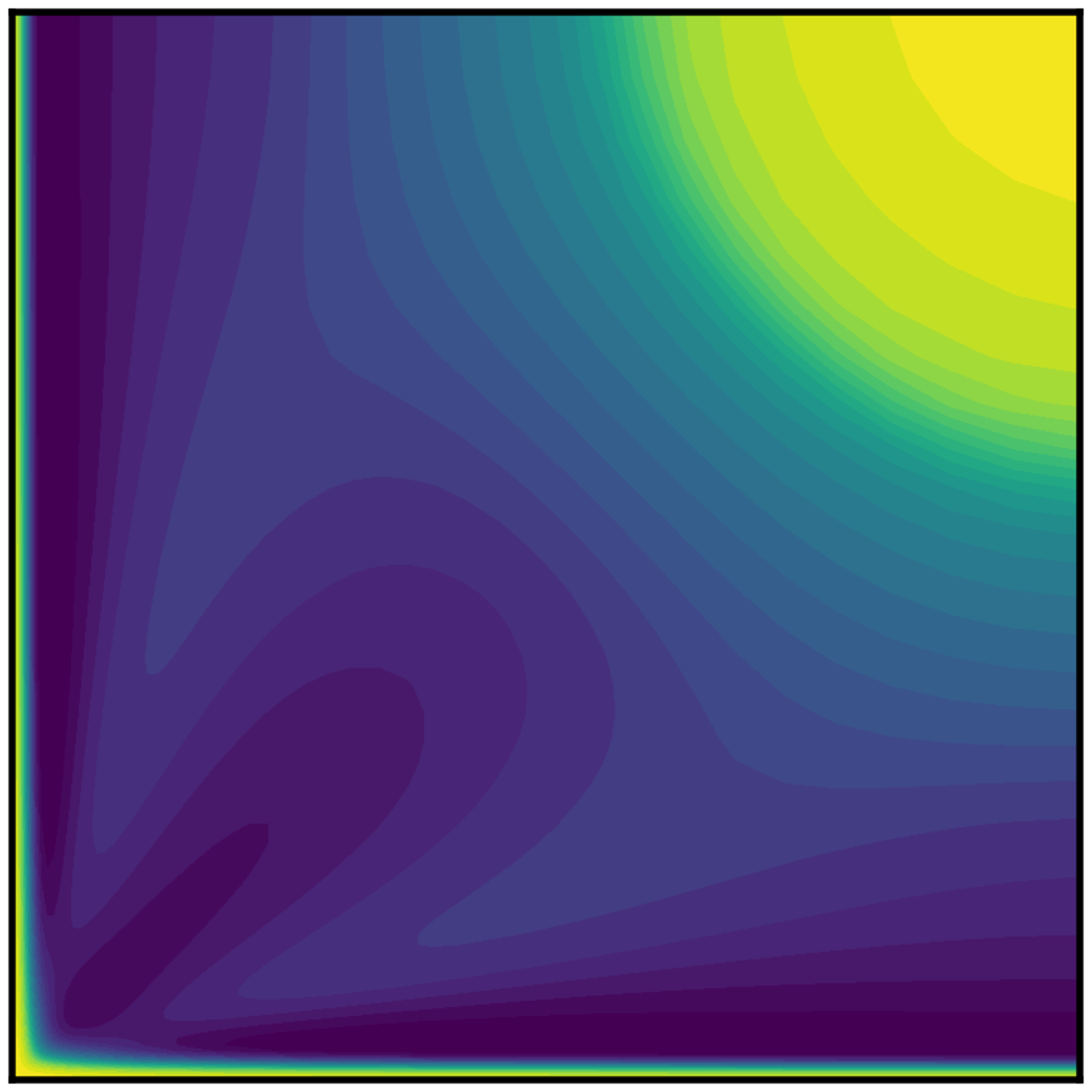}} &
        \raisebox{-.5\height}{\includegraphics[scale=0.15]{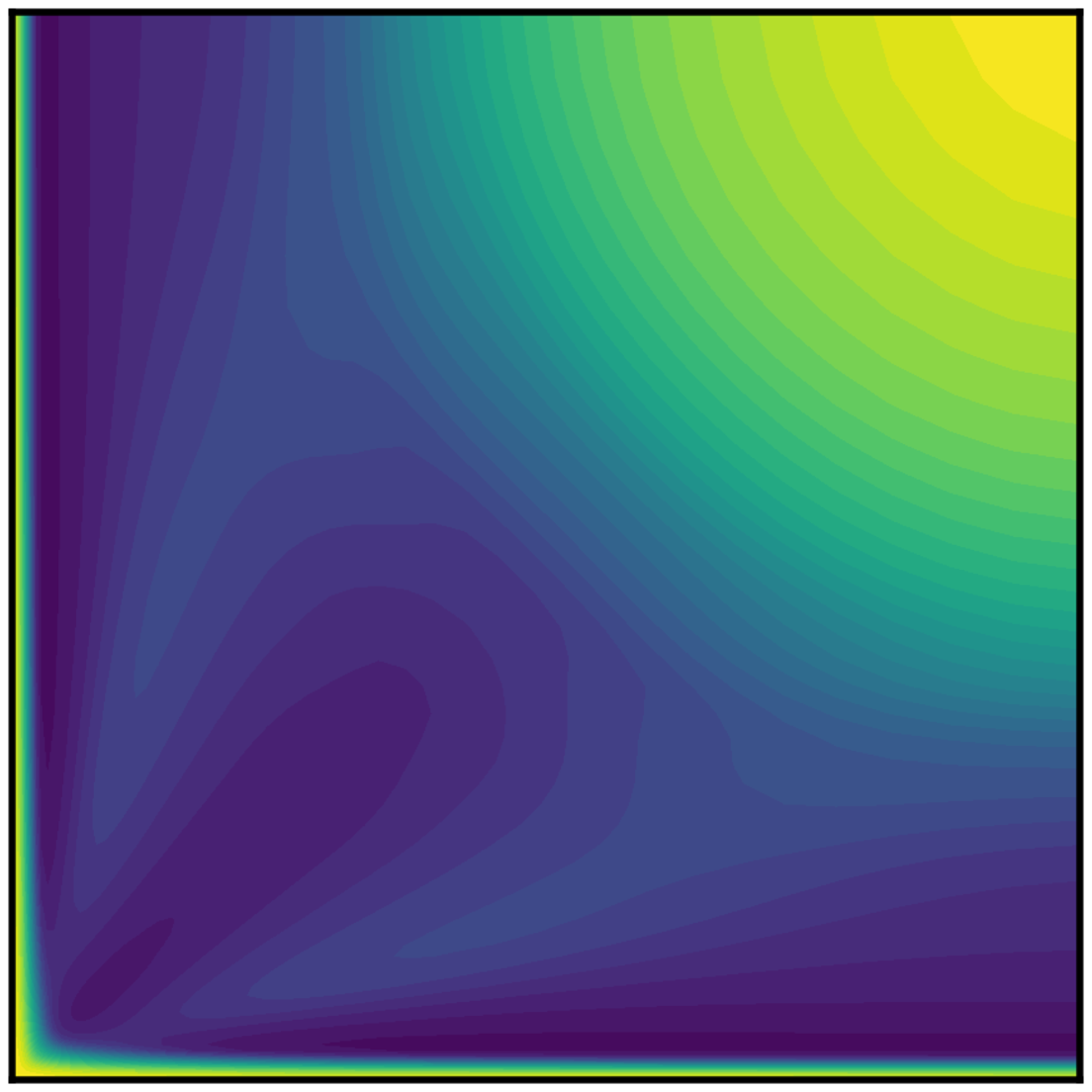}} &
        \raisebox{-.5\height}{\includegraphics[scale=0.15]{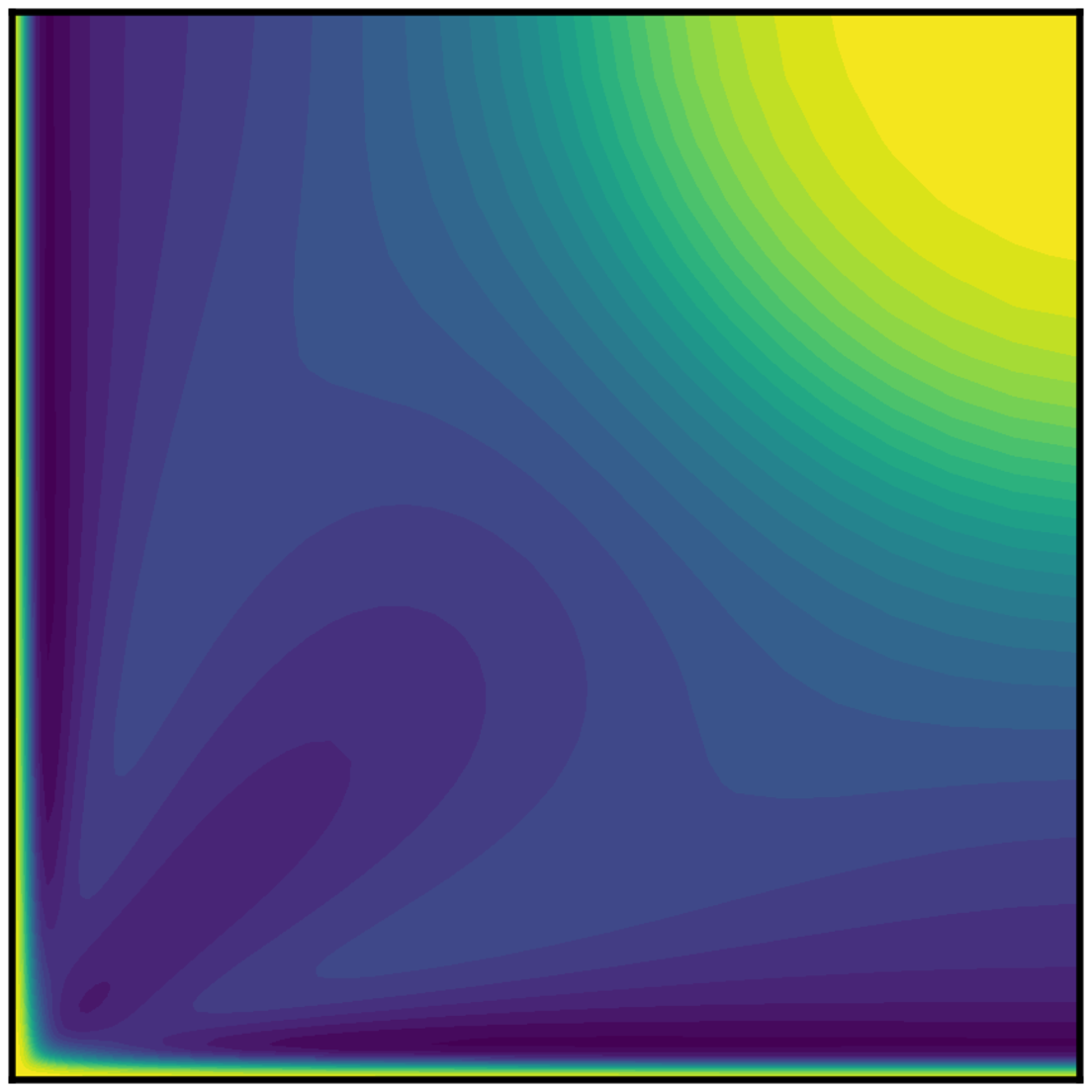}} & 
        \raisebox{-.5\height}{\includegraphics[scale=0.2]{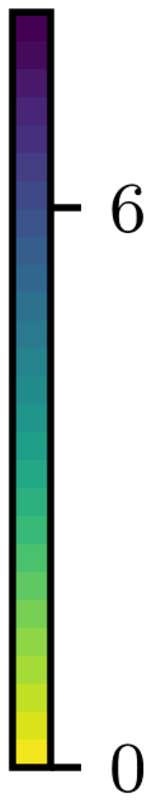}}
        \\
        \rotatebox[origin=c]{90}{$|\theta_1|-|\theta_2|$} & 
        \raisebox{-.5\height}{\includegraphics[scale=0.15]{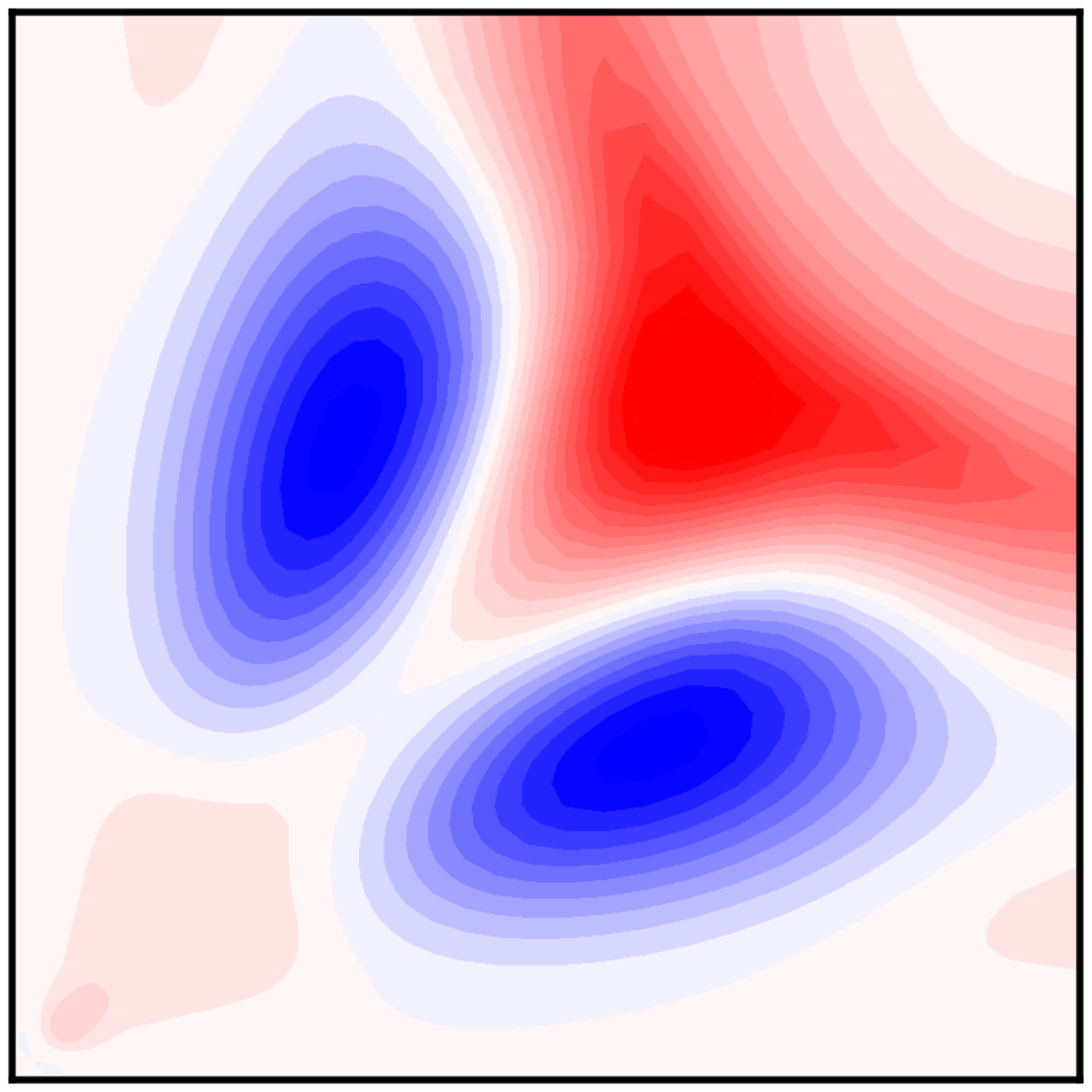}} &
        \raisebox{-.5\height}{\includegraphics[scale=0.15]{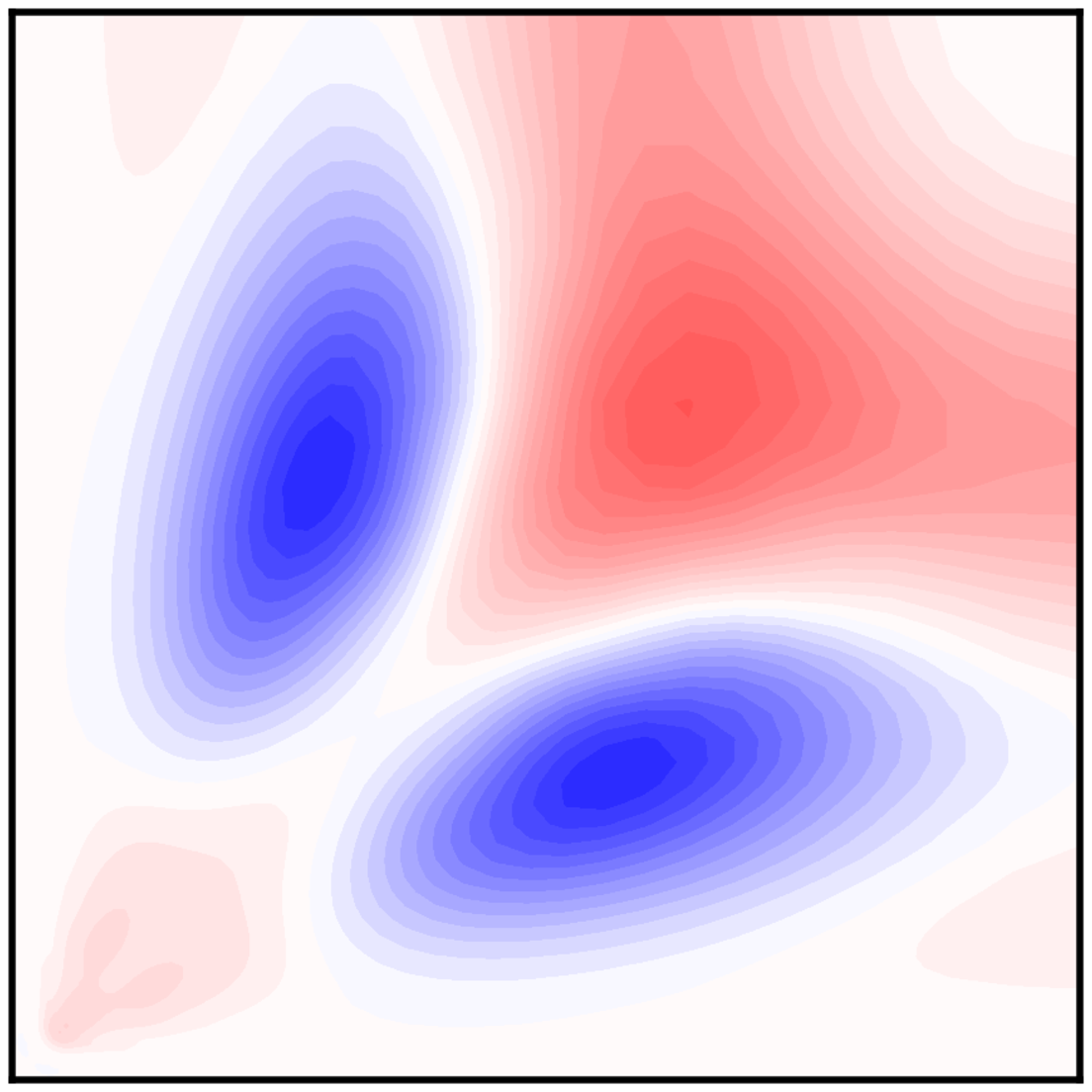}} &
        \raisebox{-.5\height}{\includegraphics[scale=0.15]{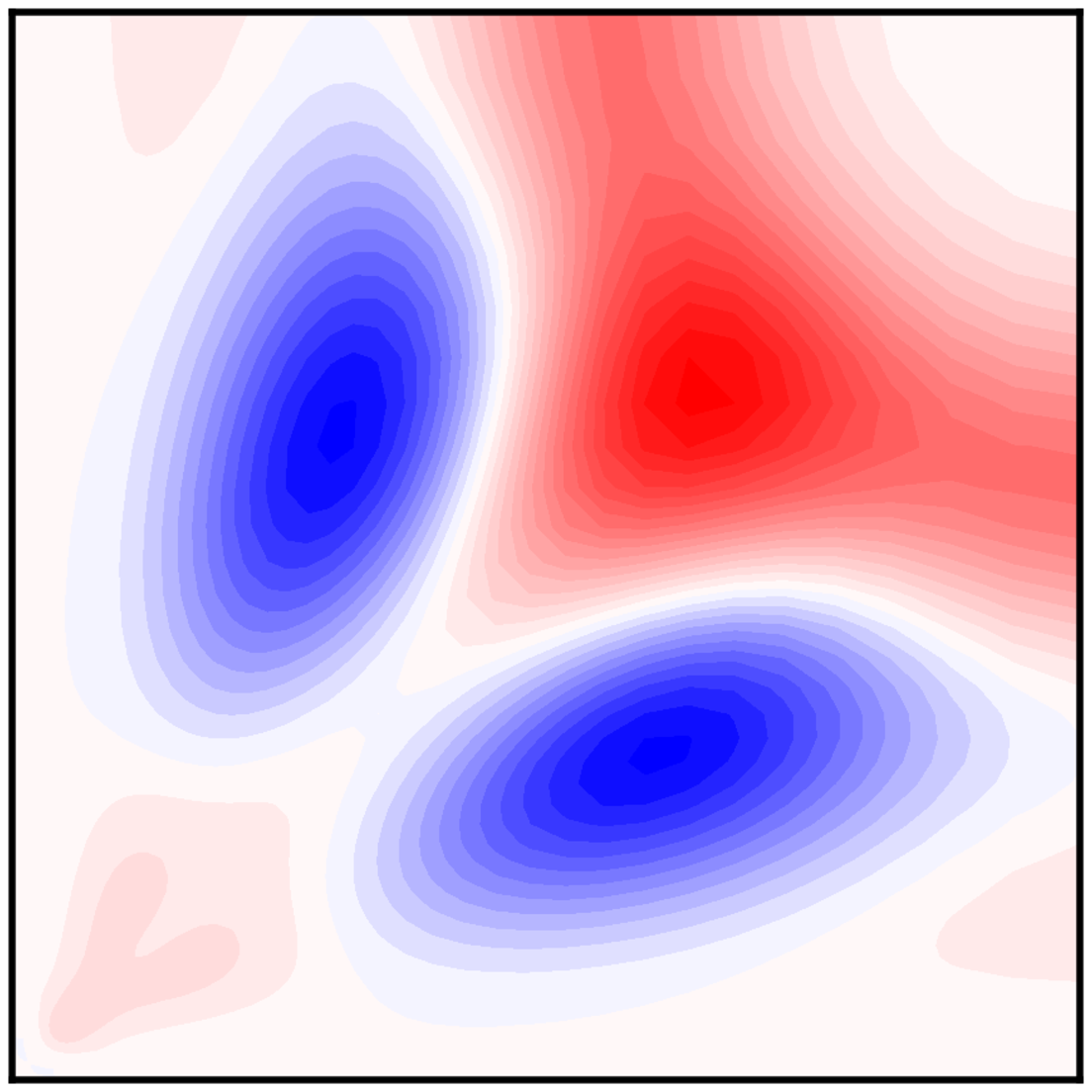}} & 
        \raisebox{-.5\height}{\includegraphics[scale=0.2]{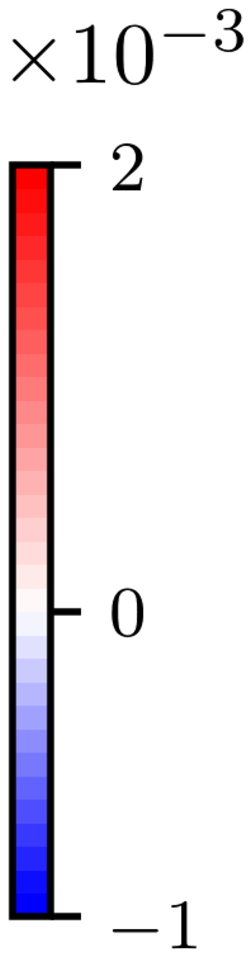}}
    \end{tabular}
    }
    \subfloat[kernel density of $\theta_1$]{\includegraphics[width=0.35\textwidth]{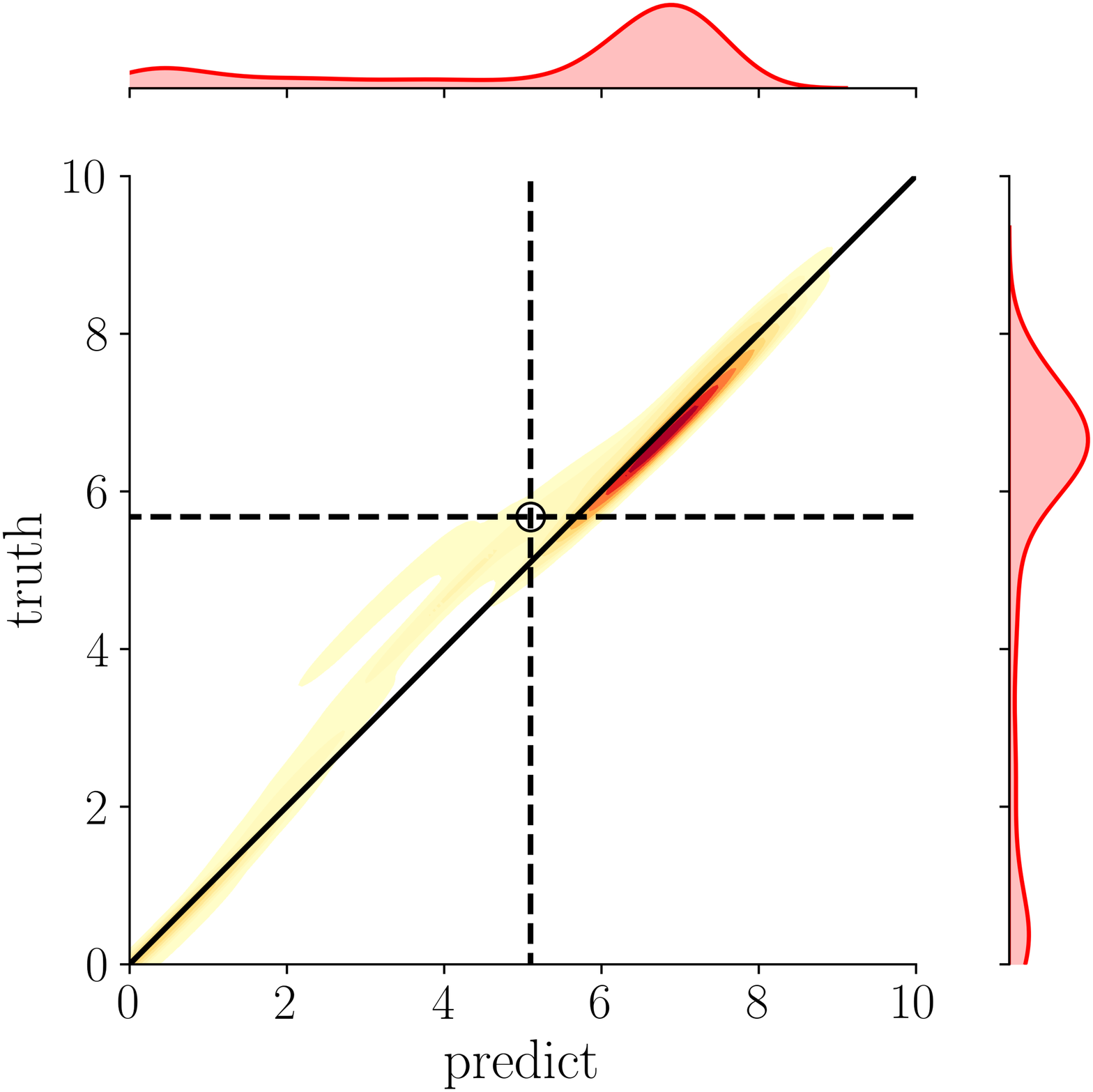}}
    \caption{(a) Comparison of scalar invariant $\theta_1$ and $|\theta_1|-|\theta_2|$ among the adjoint-based learned model, the ensemble-based learned model, and the truth; (b) Kernel density plot of $\theta_1$ from the truth and the estimation with the ensemble-based learned model. The circle indicates the $30\%$ quantile (i.e., $30\%$ of the cells have $\theta_1$ smaller than this value). The probability densities of the truth and the estimation are plotted on the margins.}
    \label{fig:duct_theta_pdf}
\end{figure}

We further show the good reconstruction in the scalar invariant~$\theta_1$ and $|\theta_1| - |\theta_2|$ with the ensemble-based method compared to the ground truth.
The contour plots of the scalar invariant are presented in Figure~\ref{fig:duct_theta_pdf}.
The predicted scalar invariant with the learned model agrees well with the ground truth.
The difference between the initial and the truth is mainly due to the in-plane secondary flow that cannot be captured by the linear eddy viscosity model.
With the learned models, the flow field in the $y$-$z$ plane is well predicted, which further improves the estimate of the scalar invariant.
It is observed that slight differences exist near the duct center. 
In that region, there are mainly small values of the scalar invariant~$\bm{\theta}$, due to the negligible stream-wise velocity gradient.
Additionally, we provide the estimated scalar invariant compared to the ground truth, which clearly shows the good agreements between the estimation and the truth.
The probability density function (PDF) of the scalar invariant~$\bm{\theta}$ is also plotted in Figure~\ref{fig:duct_theta_pdf}, showing the significantly small probability for $\bm{\theta}$ less than about $5$.
The $30\%$ quantile is located approximately at $5.1$, indicating that only $30\%$ of the cells in the domain have $\theta_1$ smaller than this value.

\begin{figure}
        \centering
        \subfloat[$g^{(1)}$]{
        \includegraphics[width=0.33\textwidth]{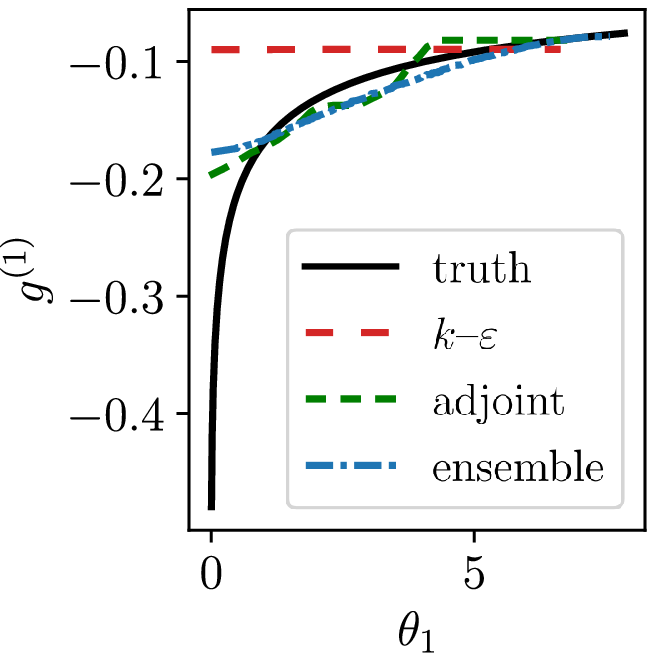}}
    	\hspace{1cm}
        \subfloat[$g^{(2)} - 0.5 g^{(3)} + 0.5 g^{(4)}$]{
        \includegraphics[width=0.33\textwidth]{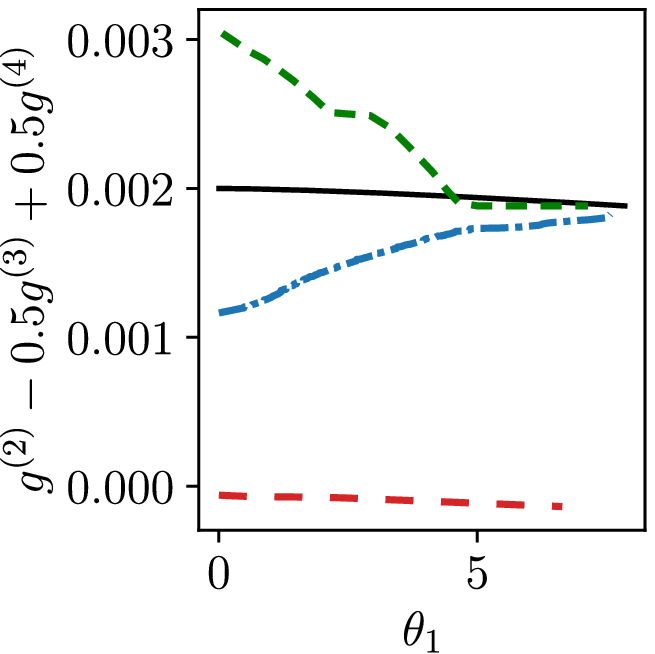}}\\
        \subfloat[$g^{(2)}$]{
        \includegraphics[width=0.33\textwidth]{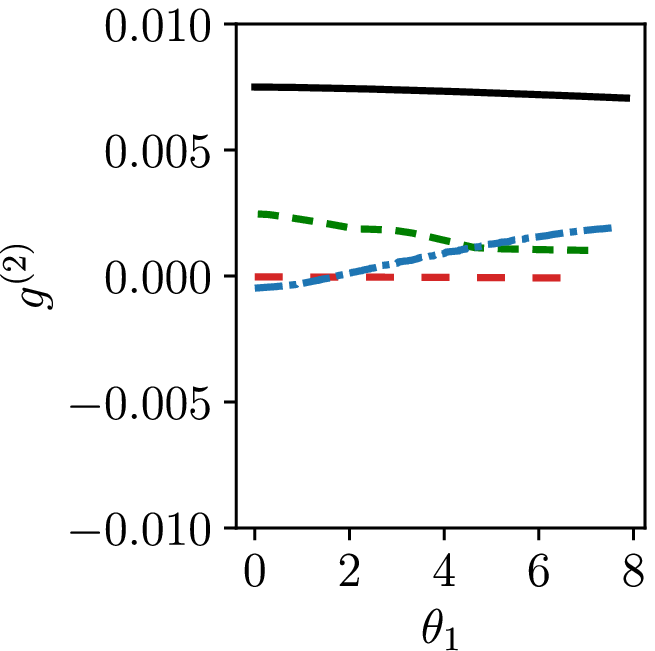}}
        \subfloat[$g^{(3)}$]{
        \includegraphics[width=0.33\textwidth]{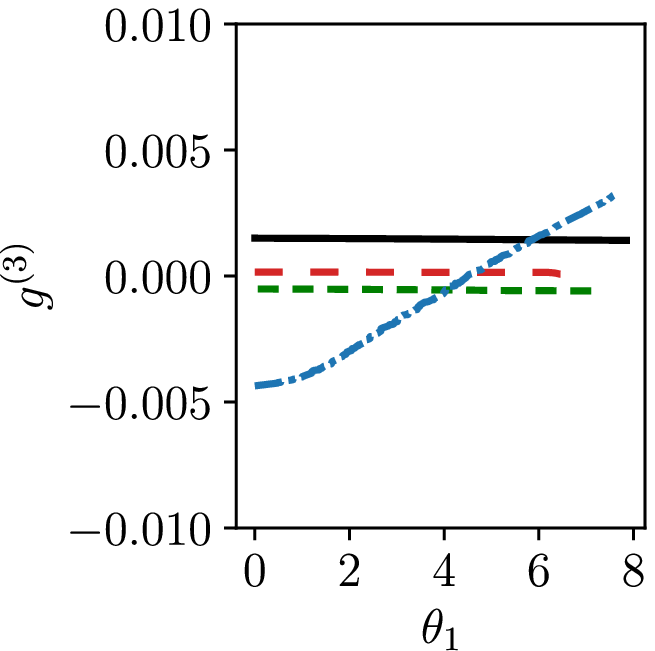}}
        \subfloat[$g^{(4)}$]{
        \includegraphics[width=0.33\textwidth]{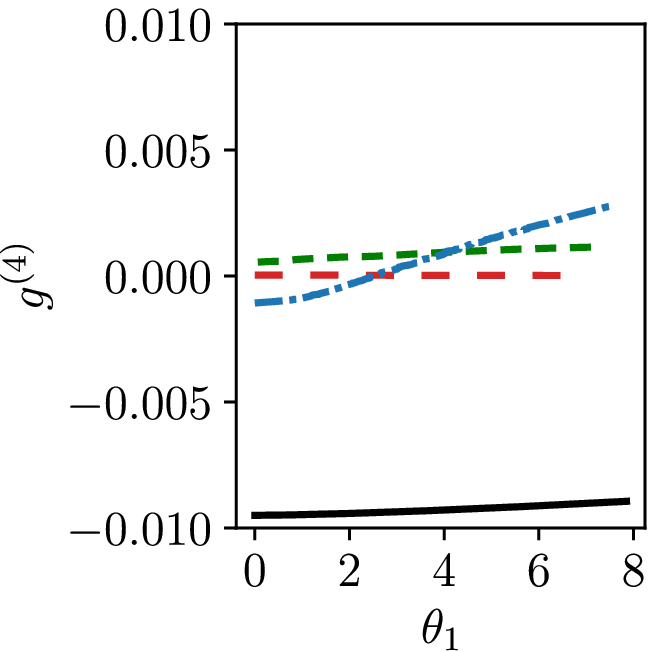}}
        \caption{Comparison plots of the functional mapping between the scalar invariant and the tensor coefficient $g$ among the truth, the baseline $k$--$\varepsilon$ model, and the models learned with adjoint and ensemble methods. 
        }
        \label{fig:duct_g}
    \end{figure}

The learned functional mapping between the scalar invariant~$\bm{\theta}$ and the tensor basis coefficient~$\bm{g}$ also have a good agreement with the ground truth.
This is illustrated in Figure~\ref{fig:duct_g}.
Since the two invariants are linearly correlated ($\theta_1 \approx - \theta_2$), we only show the plot of the mapping from the scalar invariant $\theta_1$ to the coefficients $\bm{g}$.
It can be seen that the learned mapping can have a good agreement with the ground truth (the $\bm{g}(\bm{\theta})$ in Equation~\ref{eq:gfunc}) implied by Shih's quadratic model.
Although the combination~$g^{(2)}-0.5g^{(3)}+0.5g^{(4)}$ is close to the ground truth, the components~$g^{(2)-(4)}$ have significant discrepancies.
That is because, in the duct flow, the in-plane velocity is affected by the linear combination $g^{(2)}-0.5g^{(3)}+0.5g^{(4)}$ of the $g$ functions.
We note that relatively large differences exist in the region with small values of~$\theta_1$, particularly for the combination $g^{(2)}-0.5g^{(3)}+0.5g^{(4)}$.
That is because the velocity is affected by the product of the $g$ function and the tensor bases $\mathbf{T}$.
In the region with small $\theta_1$ (near the center of the duct), the magnitudes of the tensor
bases $\mathbf{T}^{(1)}$ and $\mathbf{T}^{(2)}$ (even after normalization with $k/\varepsilon$) are small, and thus the velocities are no longer sensitive to the $\bm{g}$ functions.
Moreover, small values of $\theta_1$ are represented by only a small number of cells in the domain, which is evident from Figure~\ref{fig:duct_theta_pdf}(b). 
Only $30\%$ of the cells in the domain have $\theta_1$ smaller than around $5$, which is likely response for the large discrepancies of the $\bm{g}$ functions in the range of $\theta_1 < 5$.
This lack of representation makes it difficult to learn the underlying mapping in the region with small $\theta_1$.
However, we note that the ensemble method achieves qualitatively similar results (albeit with errors of opposite signs) to the adjoint-based method in the functional mapping. This suggests that the bottleneck for learning the complete mapping lies in the intrinsic ill-conditioning of the problem (insensitivity to small $\theta_1$ magnitudes) rather than the lack of analytic gradient. 
Meanwhile, the ill-conditioning problem may be remedied by learning from several flows with a wider range of $\bm{\theta}$.

\subsection{Flow over periodic hills: generalizability test to varying slopes}
\label{sec:result_pehills}

\begin{figure}
    \centering
    \begin{tabular}{ccccl}
        & direct & ensemble & DNS & \\
        \rotatebox[origin=c]{90}{$u_x$} & 
        \raisebox{-.5\height}{\includegraphics[scale=0.25, trim=0 -10 0 -10, clip]{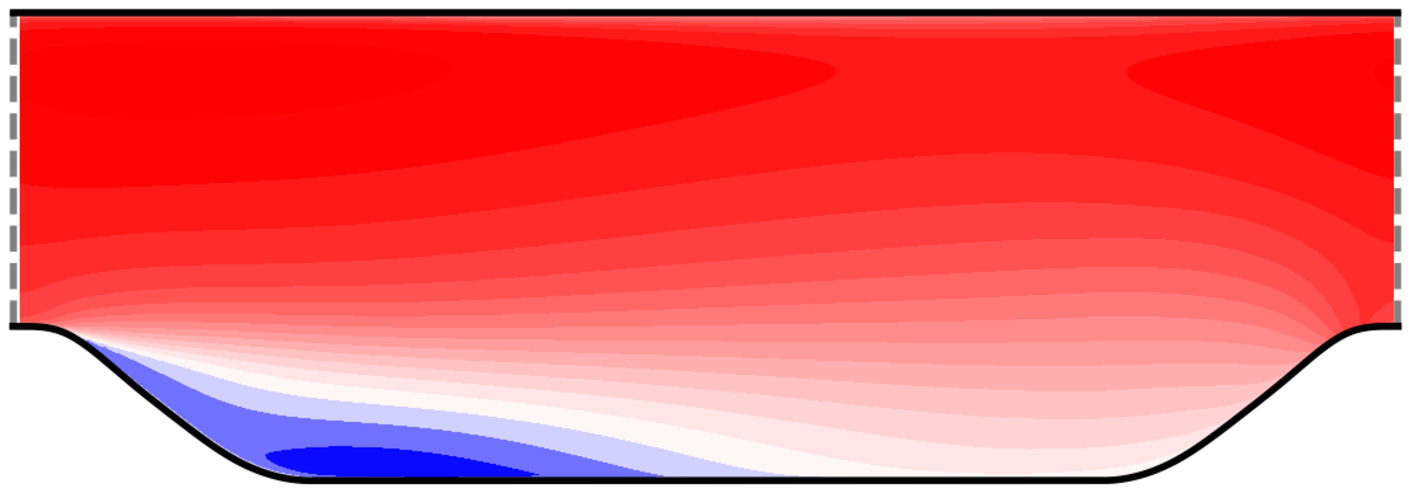}} &
        \raisebox{-.5\height}{\includegraphics[scale=0.25, trim=0 -10 0 -10, clip]{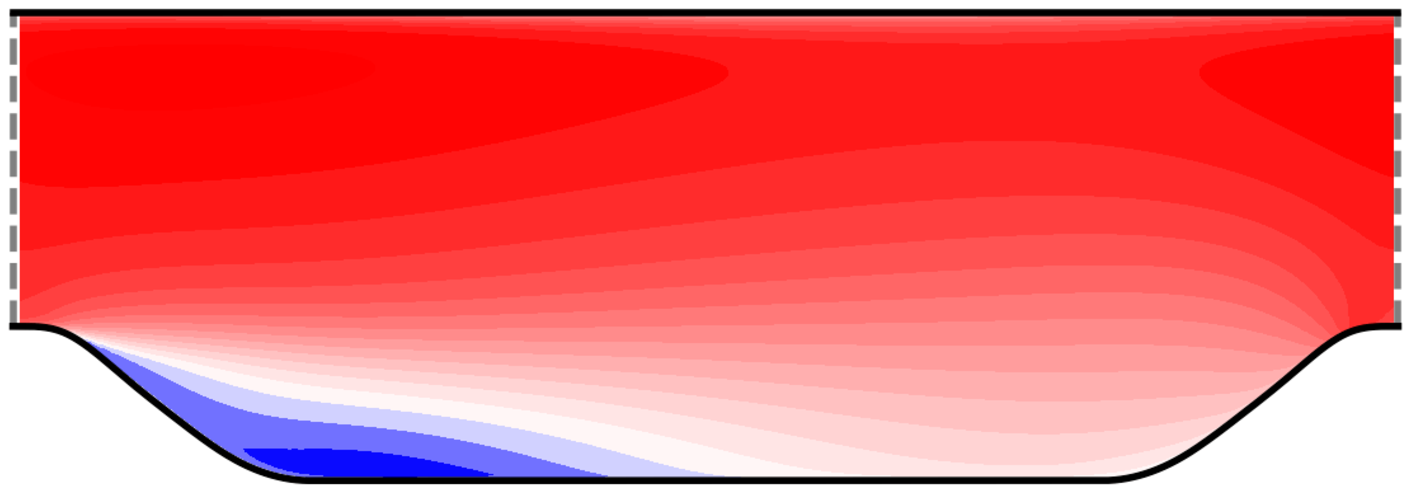}} &
        \raisebox{-.5\height}{\includegraphics[scale=0.25, trim=0 -10 0 -10, clip]{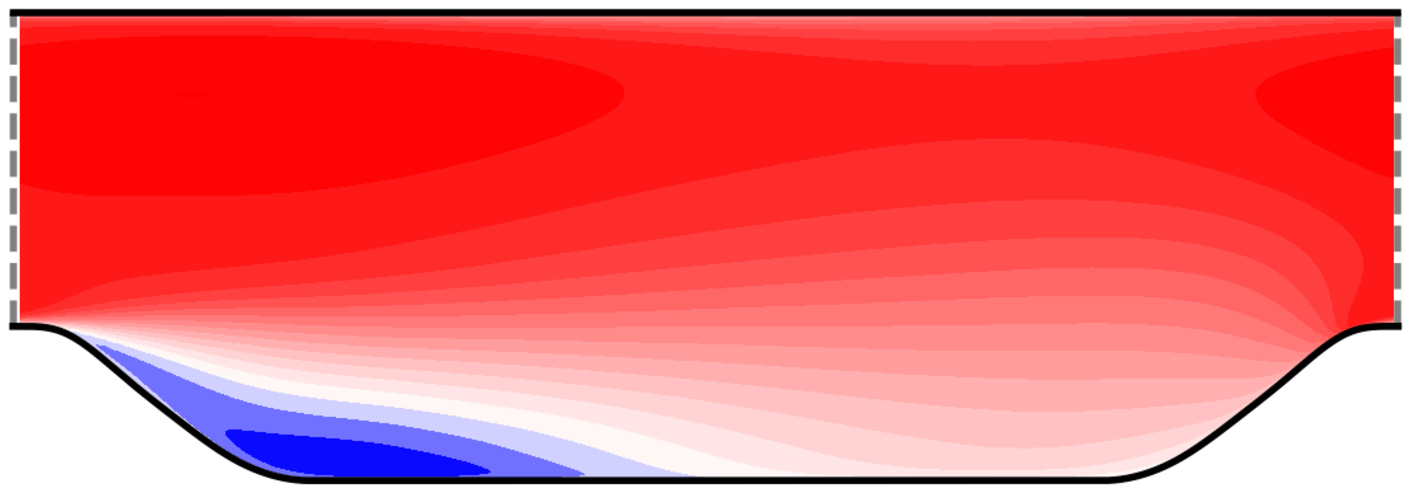}} &
        \raisebox{-.5\height}{\includegraphics[scale=0.17, clip]{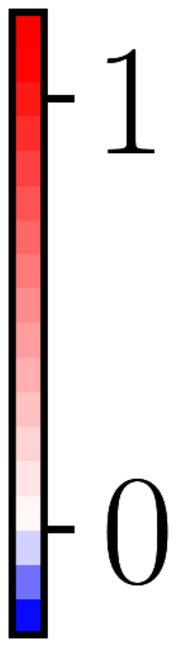}}
        \\
        \rotatebox[origin=c]{90}{$u_y$} & 
        \raisebox{-.5\height}{\includegraphics[scale=0.25, trim=0 -10 0 -10, clip]{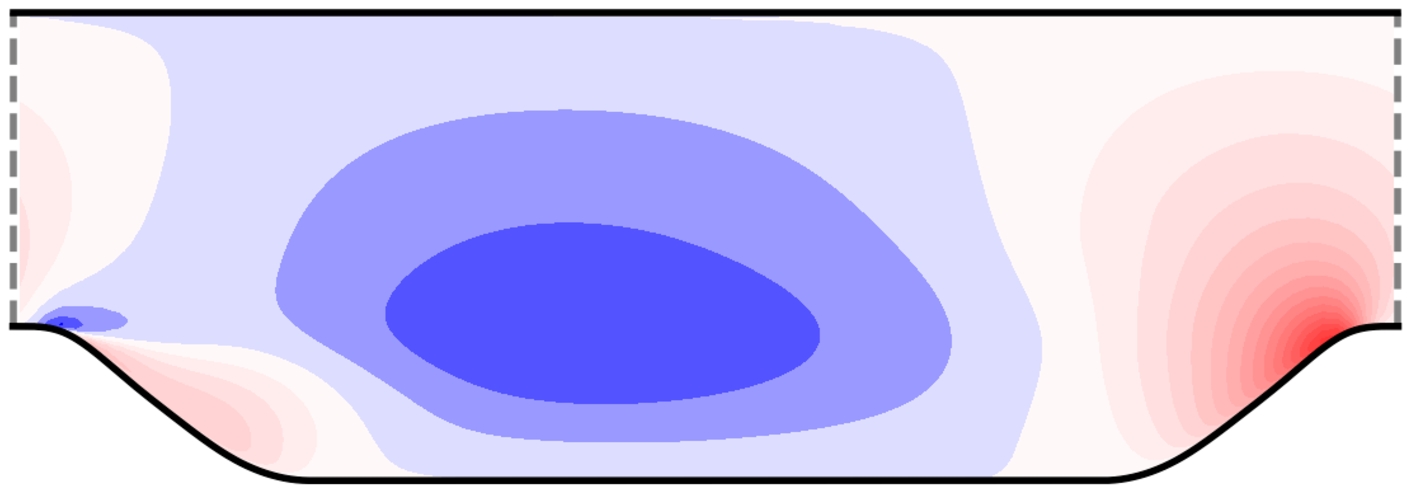}} &
        \raisebox{-.5\height}{\includegraphics[scale=0.25, trim=0 -10 0 -10, clip]{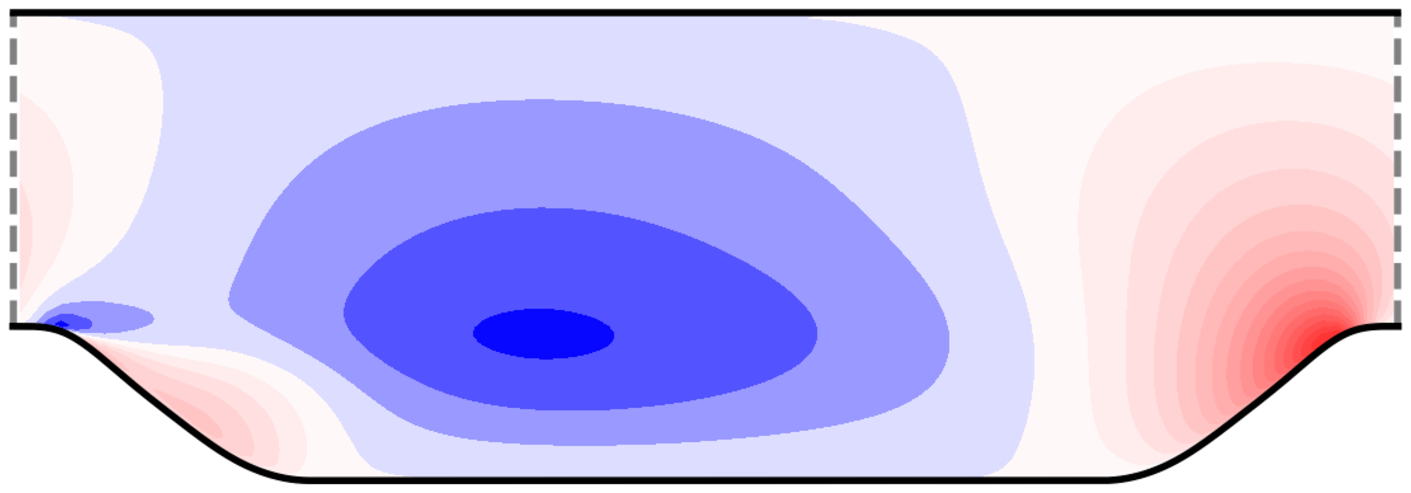}} &
        \raisebox{-.5\height}{\includegraphics[scale=0.25, trim=0 -10 0 -10, clip]{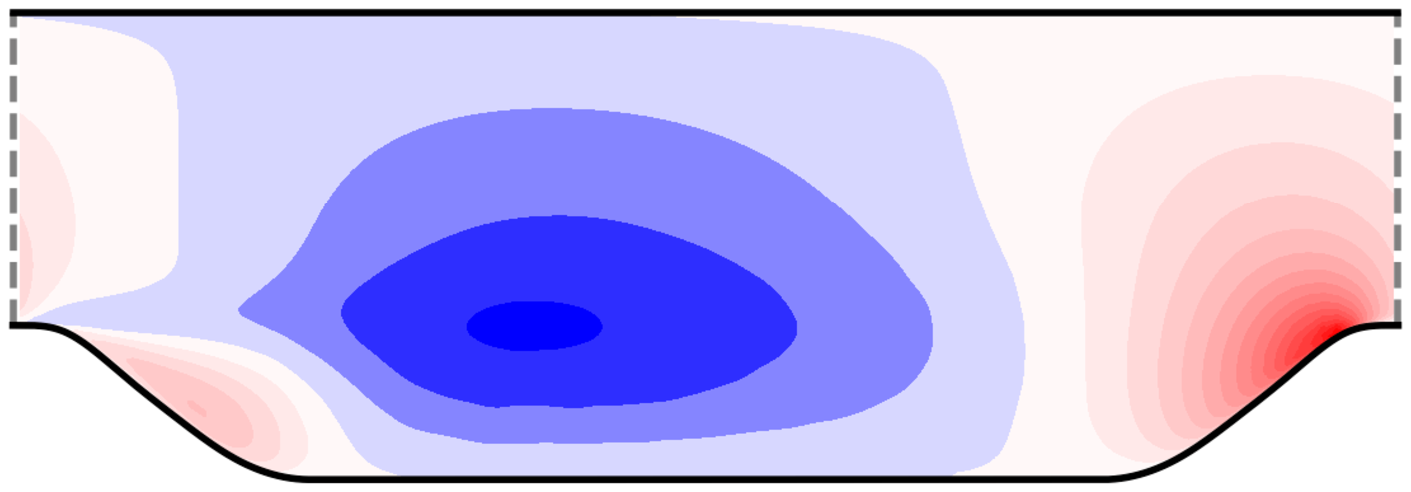}} &
        \raisebox{-.5\height}{\includegraphics[scale=0.17, clip]{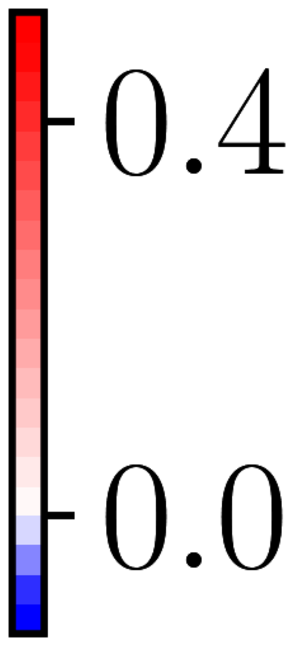}}
        \\
        \rotatebox[origin=c]{90}{$\| \boldsymbol{u} \|$} & 
        \raisebox{-.5\height}{\includegraphics[scale=0.31, trim=0 -10 0 -10, clip]{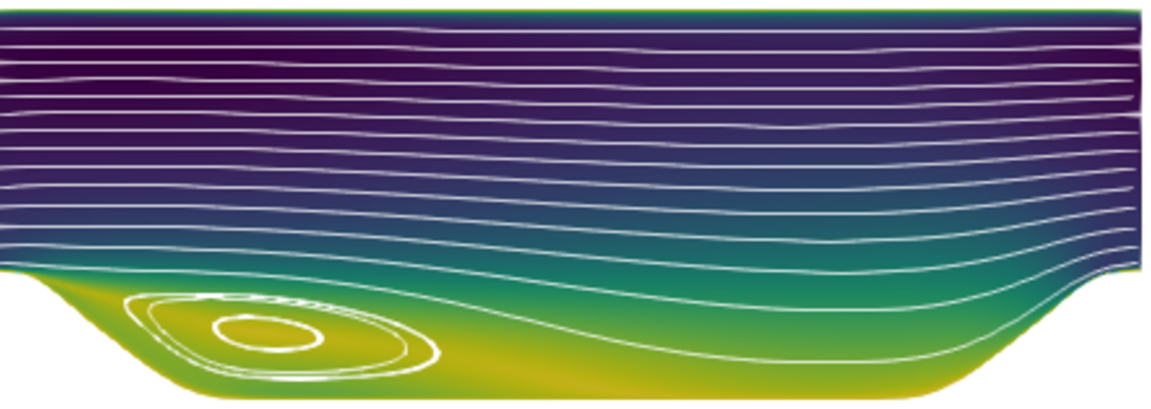}} &
        \raisebox{-.5\height}{\includegraphics[scale=0.31, trim=0 -10 0 -10, clip]{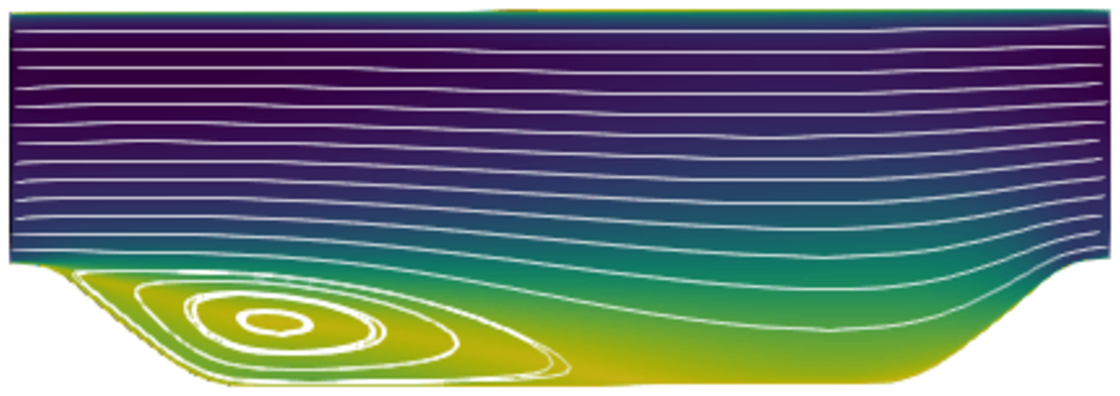}} &
        \raisebox{-.5\height}{\includegraphics[scale=0.31, trim=0 -10 0 -10, clip]{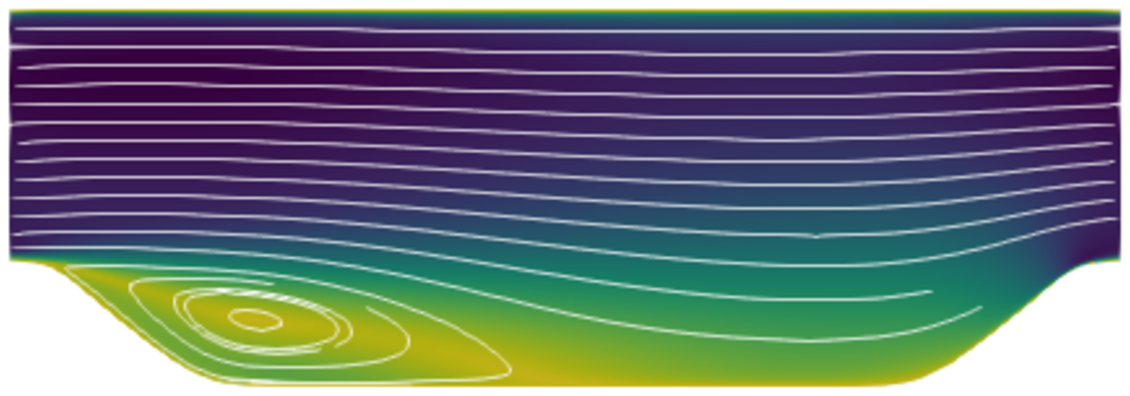}} &
        \raisebox{-.5\height}{\includegraphics[scale=0.2, clip]{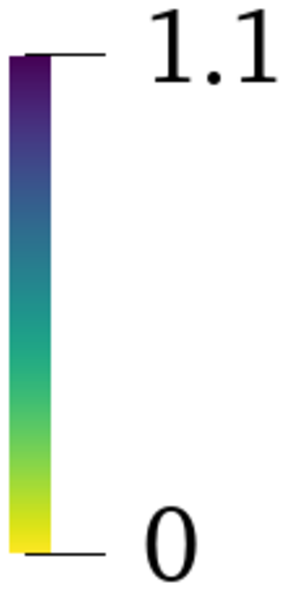}}
        \\
    \end{tabular}
    \caption{Contour plots of the velocity with the direct learned model, the ensemble-based learned model, and DNS for the periodic hill case. Note that the plots are from in-sample tests.}
    \label{fig:pehill_contour}
\end{figure}

The proposed method is further used to train the neural network-based model for the flows over periodic hills.
The flow with the slope of~$\alpha=1$ is used to train the model.
The ensemble-based method is capable of reconstructing the flow field accurately in this case.
This is shown in Figure~\ref{fig:pehill_contour} where the velocity contour is provided with comparison to the results of the direct learning method and the DNS.
It can be seen that the flow characteristics are well captured by minimizing the discrepancies between the velocity estimation and the given data.
It is noted that only four velocity profiles at $x/H=1, 3, 5$, and $7$ are used to achieve the improved reconstruction of the entire field.
The separation bubbles with the direct learned model, the ensemble-based learned model, and the truth are also provided in Figure~\ref{fig:pehill_contour}.
The learned models with the direct learning method and the ensemble method both can well capture the bubble structure.

\begin{figure}
    \centering
    \includegraphics[width=0.8\textwidth]{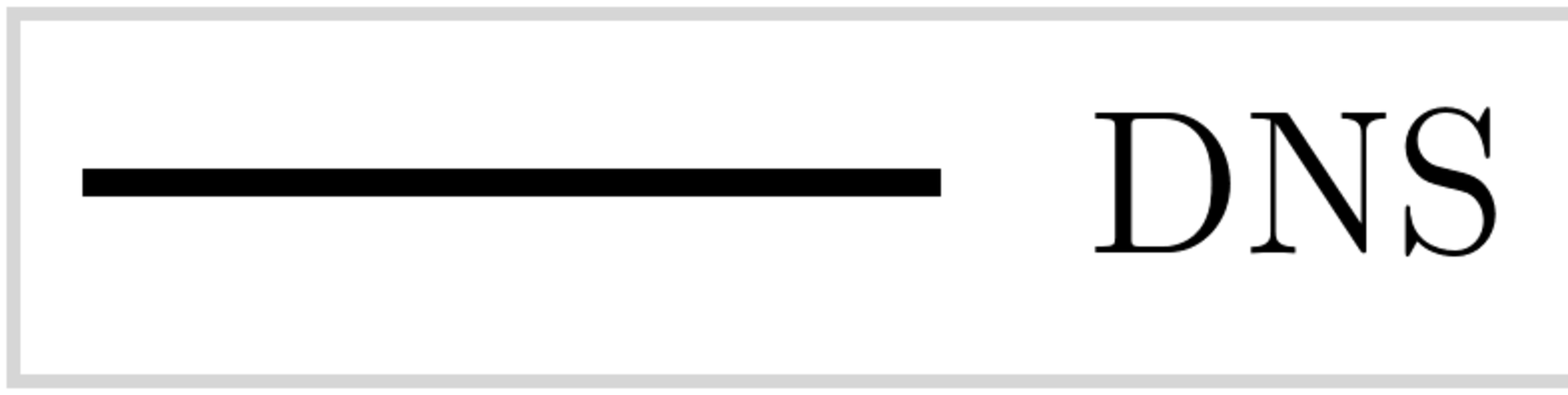} \\
    \subfloat[horizontal velocity $u_x$]{\includegraphics[width=0.5\textwidth, trim=0 0 0 0, clip]{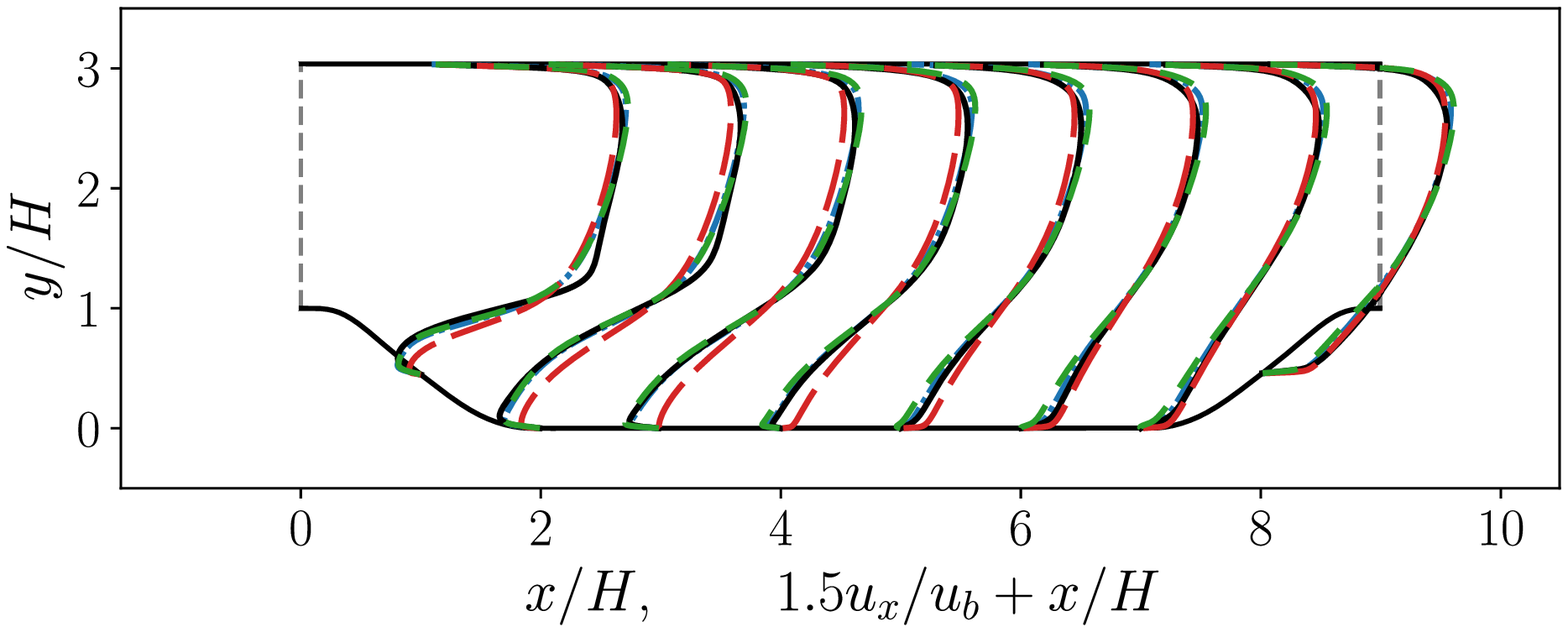}}
    \subfloat[vertical velocity $u_y$]{\includegraphics[width=0.5\textwidth, trim=0 0 0 0, clip]{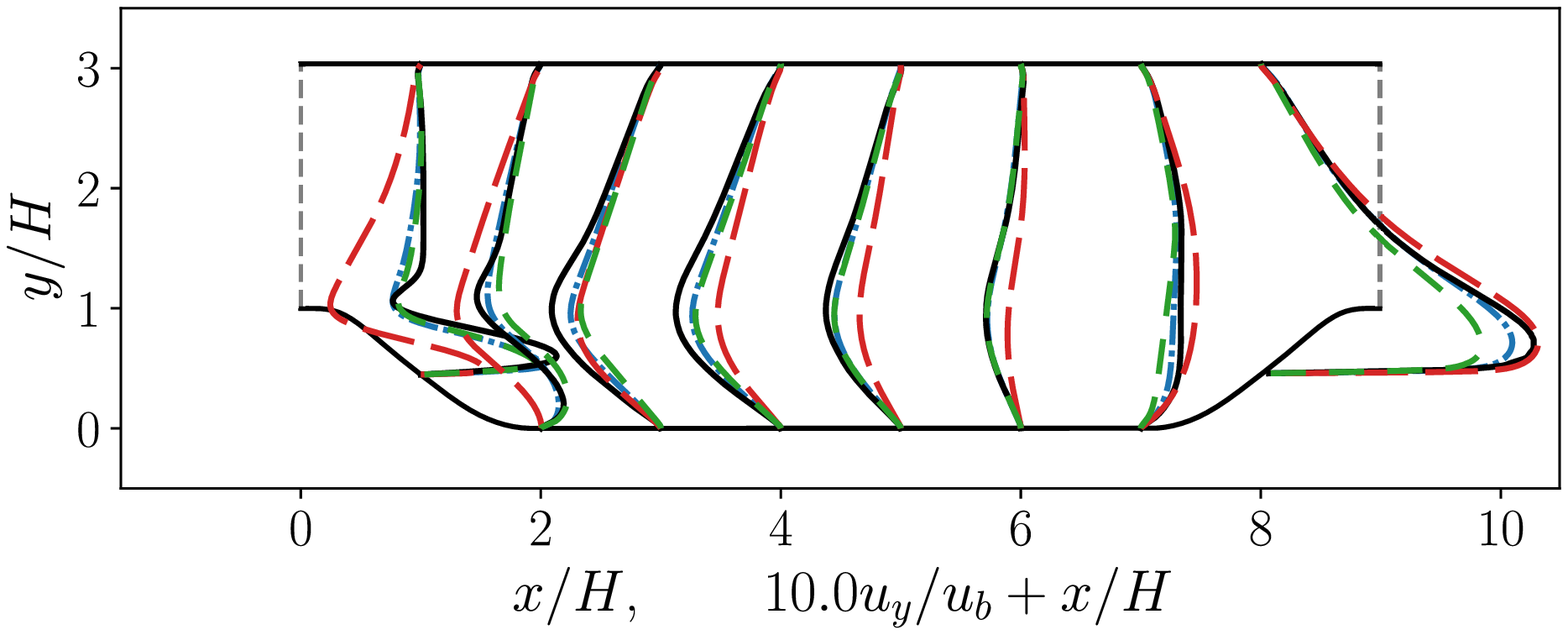}} \\
    \subfloat[friction coefficient $c_f$]{\includegraphics[width=0.5\textwidth]{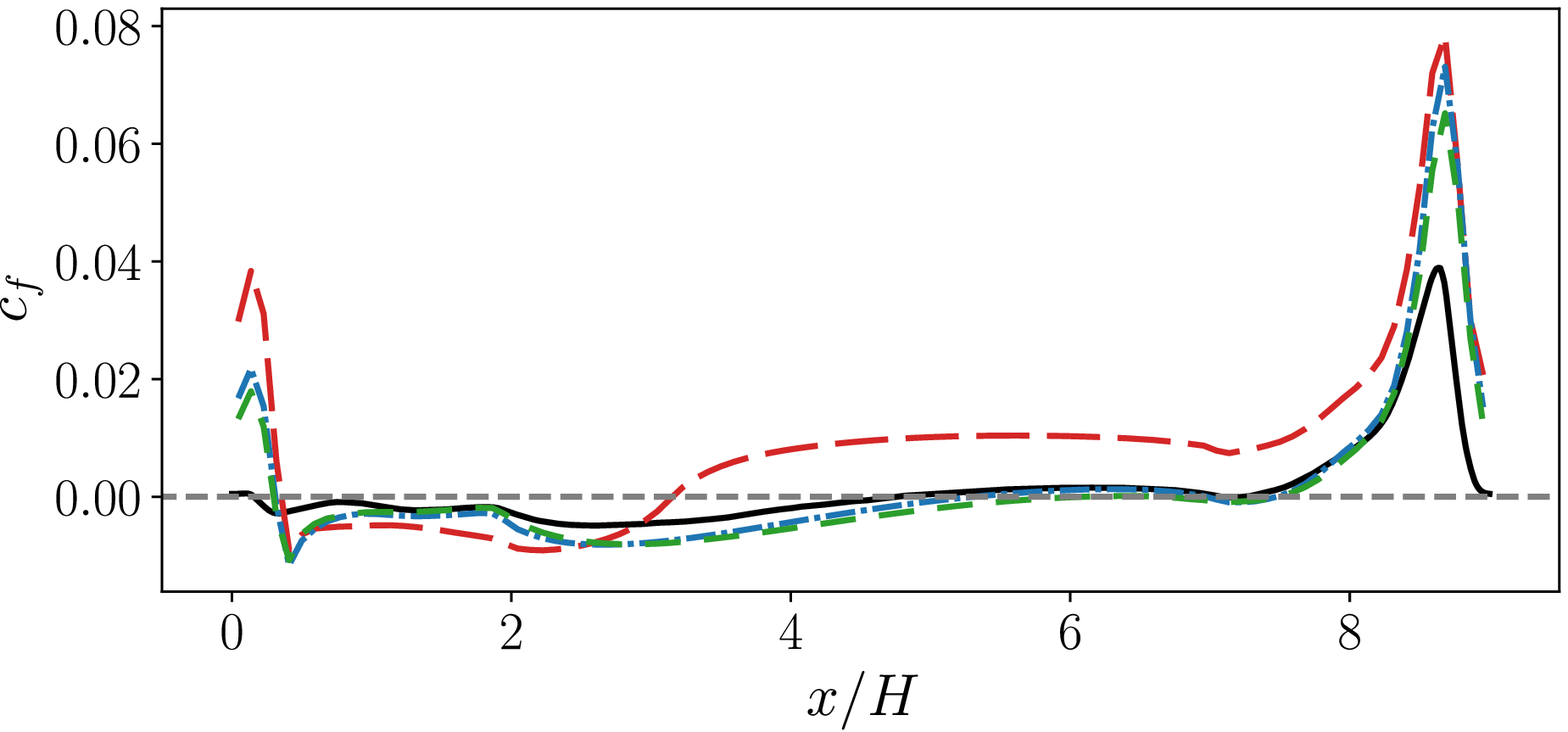}}
    \caption{Comparison of velocity and the friction coefficient~$c_f$ at the bottom wall along profiles among the $k$--$\varepsilon$ model, the direct learned model, the ensemble-based learned model, and the DNS for the periodic hill case.
    Note that the plots are from in-sample tests.
    }
    \label{fig:pehills_profile}
\end{figure}

To clearly show the improvement in the velocity estimation, we present the comparison results along profiles in Figure~\ref{fig:pehills_profile} (a) and (b).
The velocity profiles are improved significantly compared to the $k$--$\varepsilon$ model predictions.
Particularly, in the separation region, both the velocity~$u_x$ and $u_y$ are well predicted in good agreement with the DNS results.
The comparison in the friction coefficient along the bottom wall is plotted in Figure~\ref{fig:pehills_profile}(c).
The position of the reattachment point with the $k$--$\varepsilon$ model deviates substantially from the DNS.
In contrast, the ensemble-based learned model can significantly improve the friction coefficient estimation, and especially the reattachment point is very close to the truth.
The results with the direct learned model are provided for comparison, and the propagated velocity and friction coefficient is also improved noticeably compared to the $k$--$\varepsilon$ model.
The estimation error and training efficiency of the direct learning and ensemble-based methods are shown in Table~\ref{tab:compare_direct_ensemble}.
The two methods achieve similar velocity estimation, while the Reynolds stress with the direct learning method is slightly better than the ensemble method, due to the use of direct data.
As for the training efficiency, the cost of the direct learning method is around 0.2 hours, which is significantly lower than the present method (4.9 hours), mainly because the CFD solver is not involved in the learning process.

\begin{table}
    \begin{center}
    \begin{tabular}{l cccc}
         Method & $\mathcal{E}(\boldsymbol{u})$ & $\mathcal{E}(\boldsymbol{\tau})$ & Total steps & Wall time \\ \\
         Direct learning & $6.6\%$ & $40.0\%$ & $10000$ & $0.2$ hours \\ \\
         Ensemble-based & $6.4\%$ & $44.0\%$ & $50$ & $4.9$ hours  \\ \\
    \end{tabular}
    \caption{Comparison of the estimation error and time cost between the direct learning and ensemble-based learning methods for the periodic hill case.}
    \label{tab:compare_direct_ensemble}
    \end{center}
\end{table}

\begin{figure}
    \centering
    \subfloat[contour of scalar invariants]{\begin{tabular}{ccccl} 
        & direct & ensemble & DNS & \\
        \rotatebox[origin=c]{90}{$\theta_1$} & 
        \raisebox{-.5\height}{\includegraphics[scale=0.25]{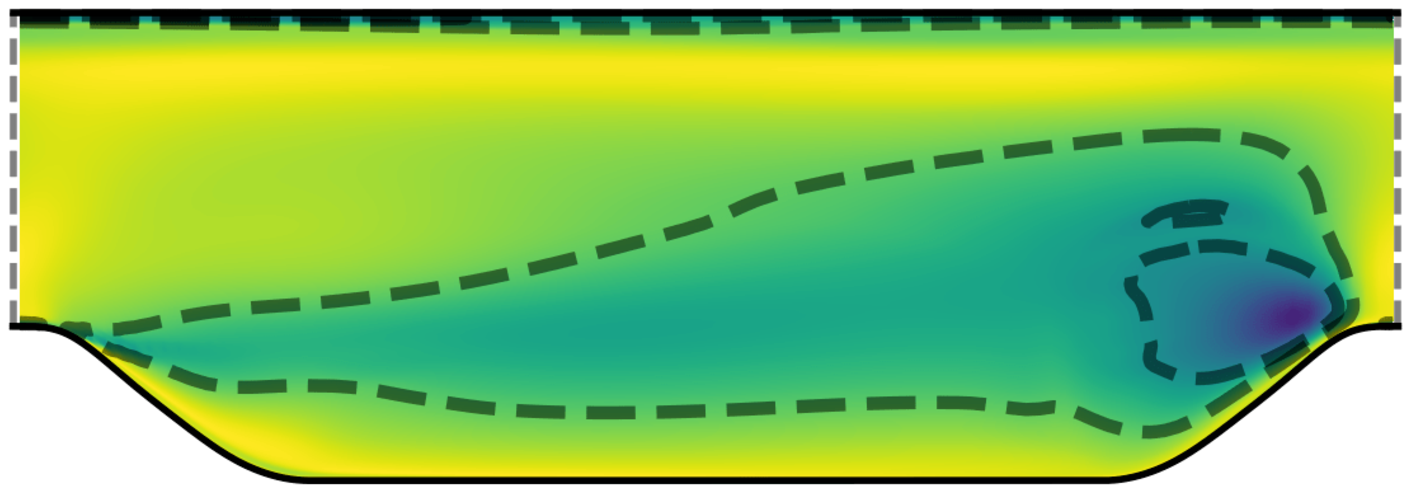}} &
        \raisebox{-.5\height}{\includegraphics[scale=0.25]{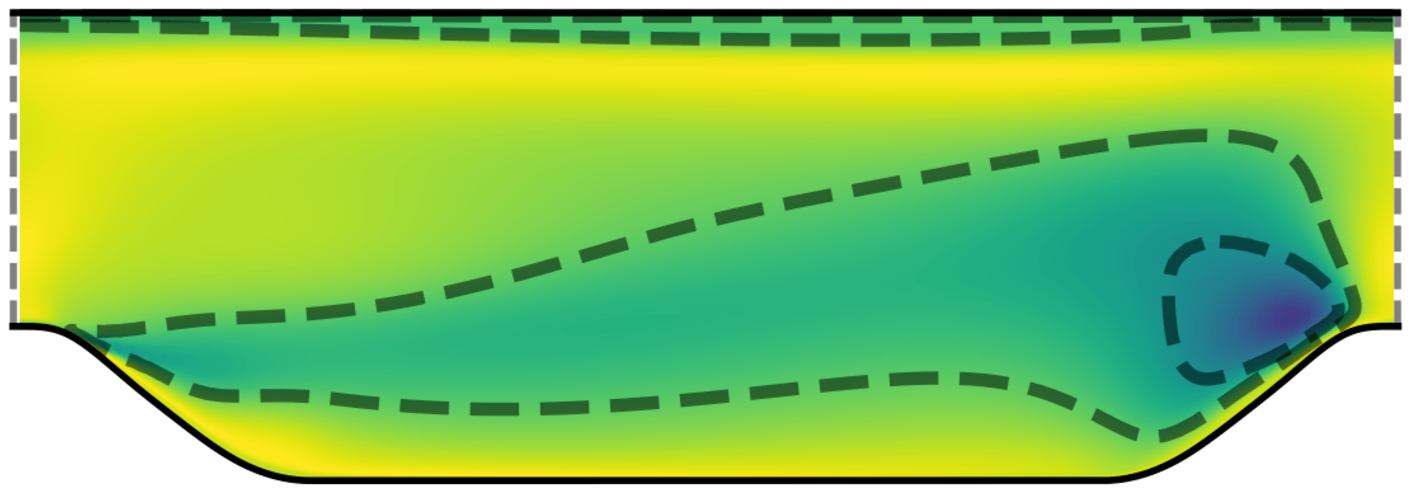}} &
        \raisebox{-.5\height}{\includegraphics[scale=0.25]{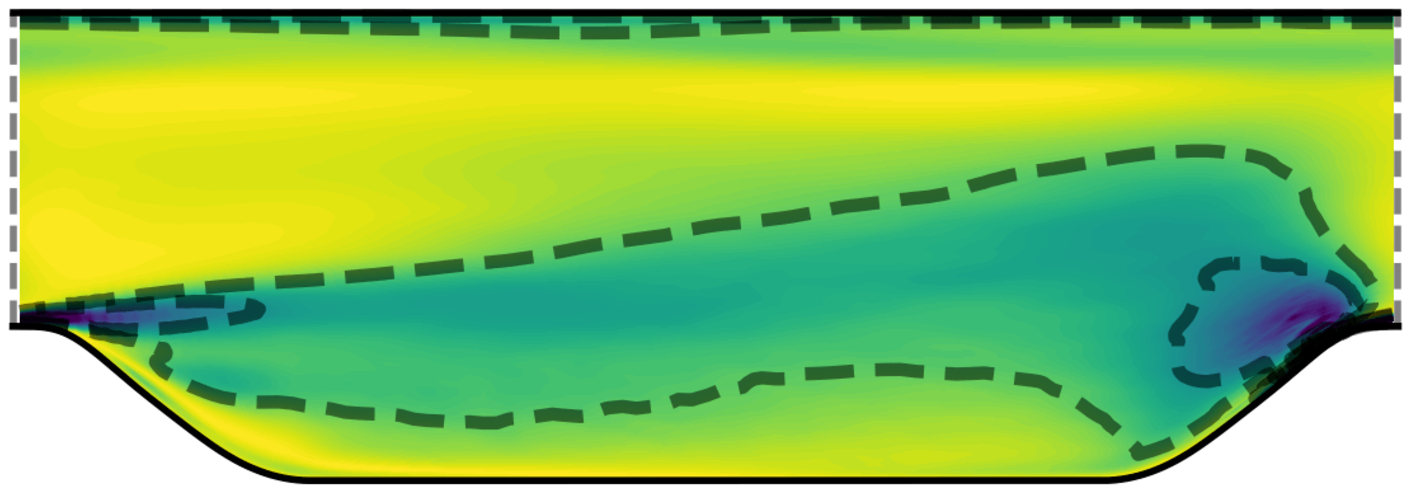}} &
        \raisebox{-.5\height}{\includegraphics[scale=0.18]{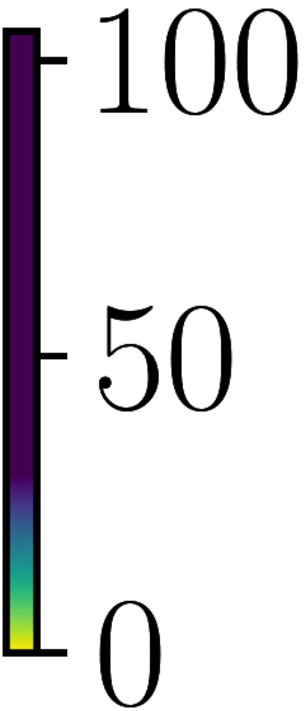}}
        \\
        \rotatebox[origin=c]{90}{$\theta_2$} & 
        \raisebox{-.5\height}{\includegraphics[scale=0.25]{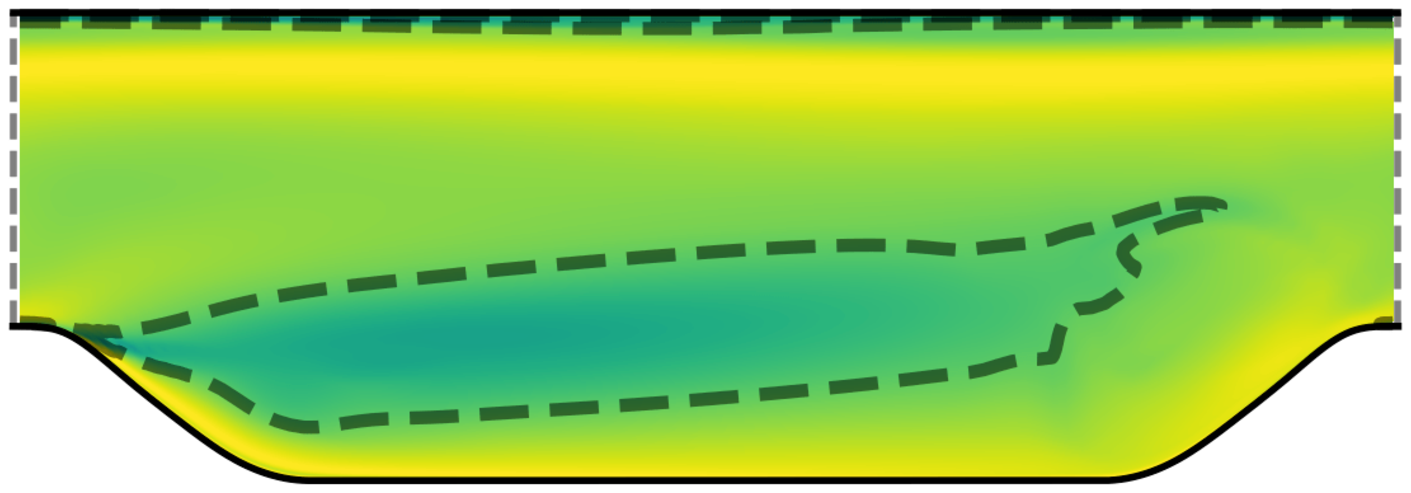}} &
        \raisebox{-.5\height}{\includegraphics[scale=0.25]{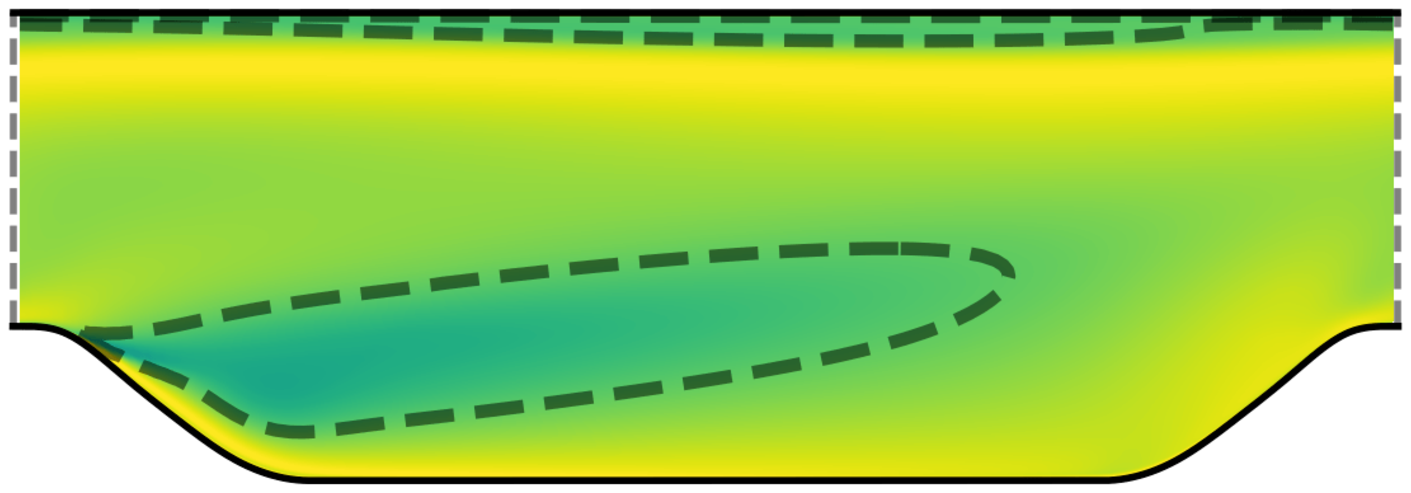}} &
        \raisebox{-.5\height}{\includegraphics[scale=0.25]{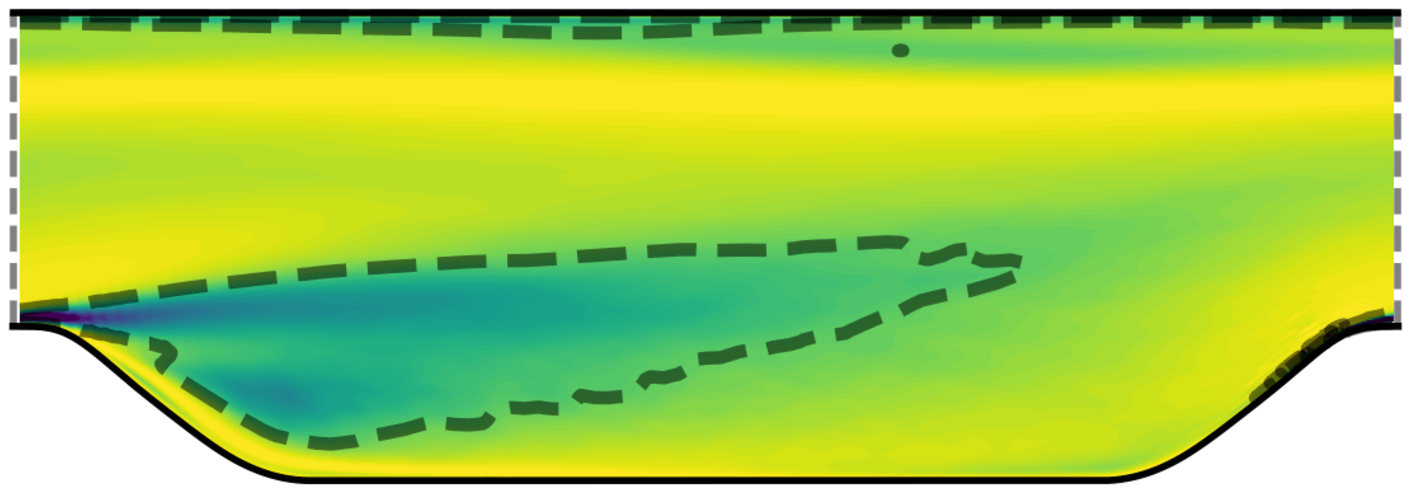}} & 
        \raisebox{-.5\height}{\includegraphics[scale=0.18]{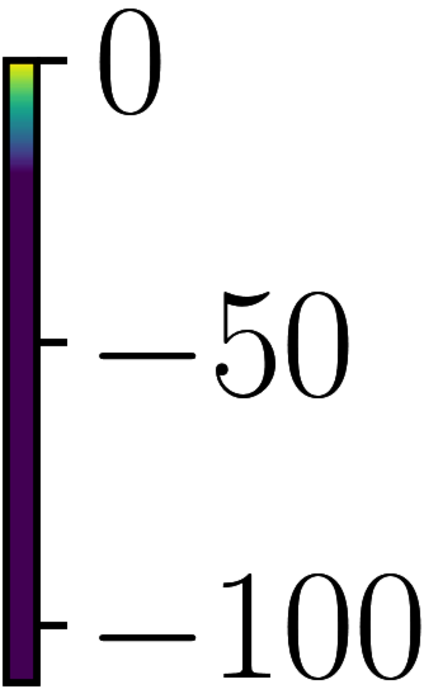}}
        \\
        \rotatebox[origin=c]{90}{$|\theta_1|-|\theta_2|$} & 
        \raisebox{-.5\height}{\includegraphics[scale=0.25]{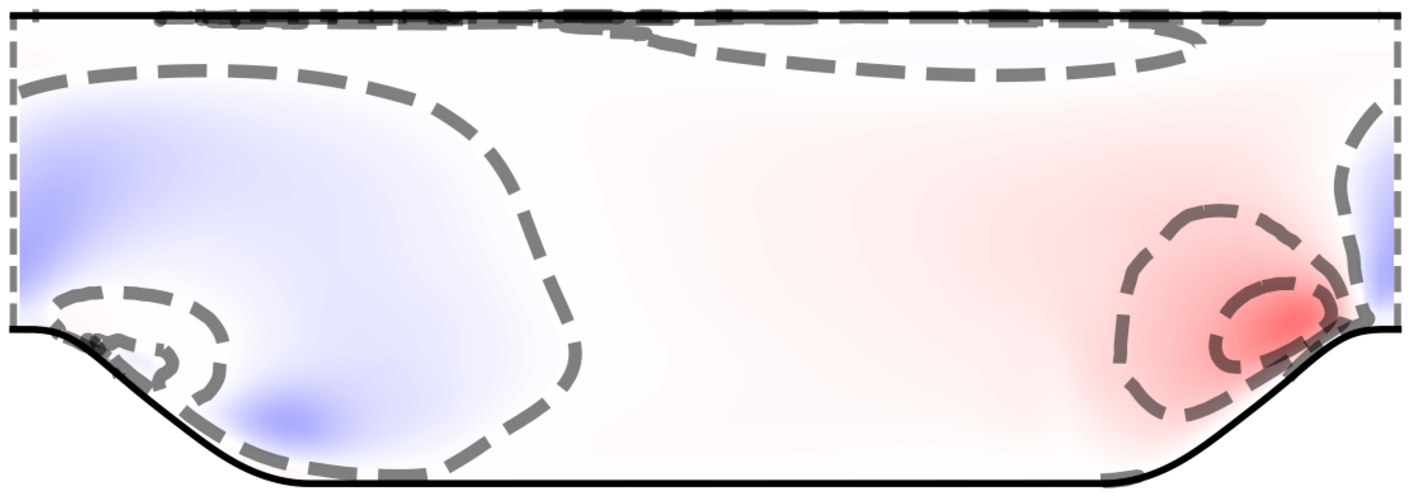}} &
        \raisebox{-.5\height}{\includegraphics[scale=0.25]{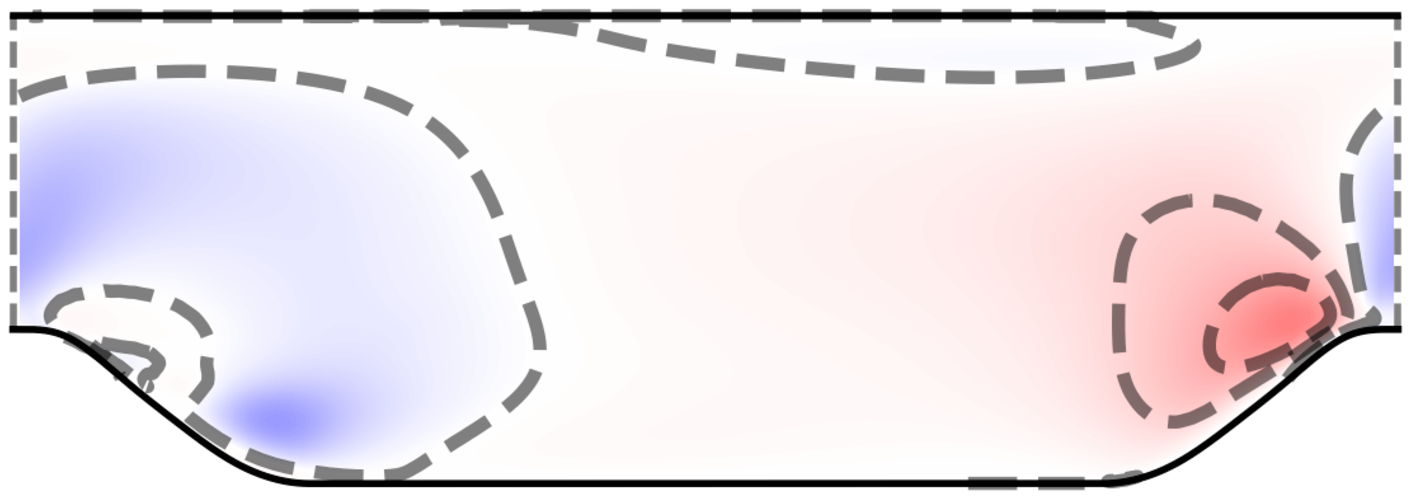}} &
        \raisebox{-.5\height}{\includegraphics[scale=0.25]{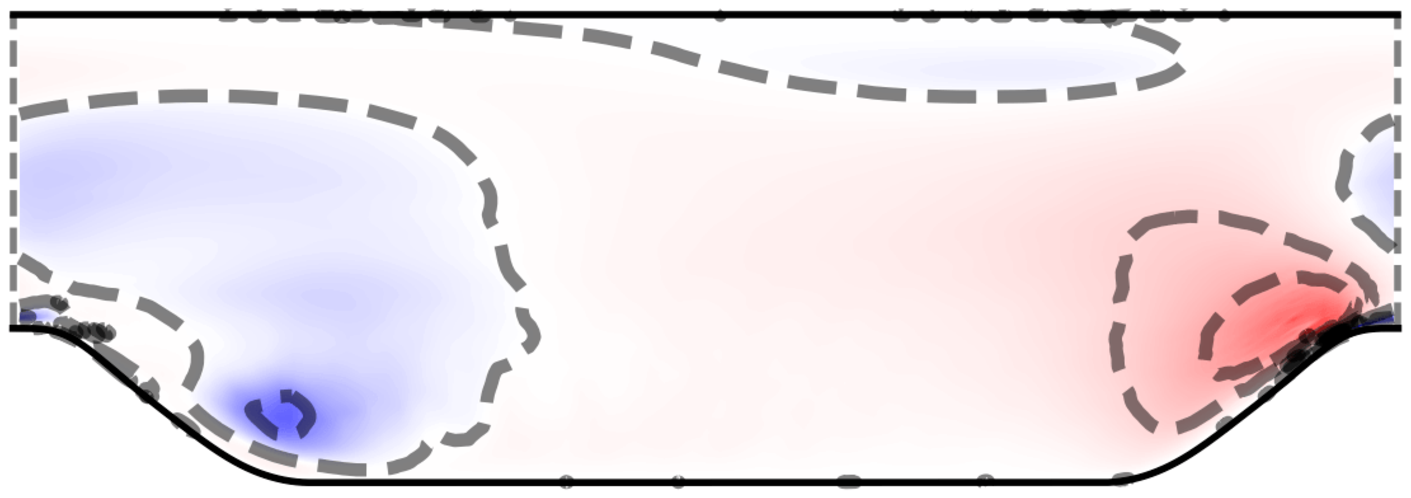}} & 
        \raisebox{-.5\height}{\includegraphics[scale=0.18]{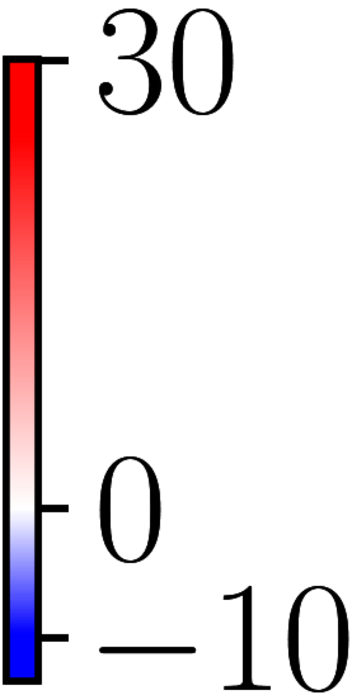}}
        \\
    \end{tabular}
    }\\
    \subfloat[kernel density of $\theta_1$]{\includegraphics[width=0.43\textwidth]{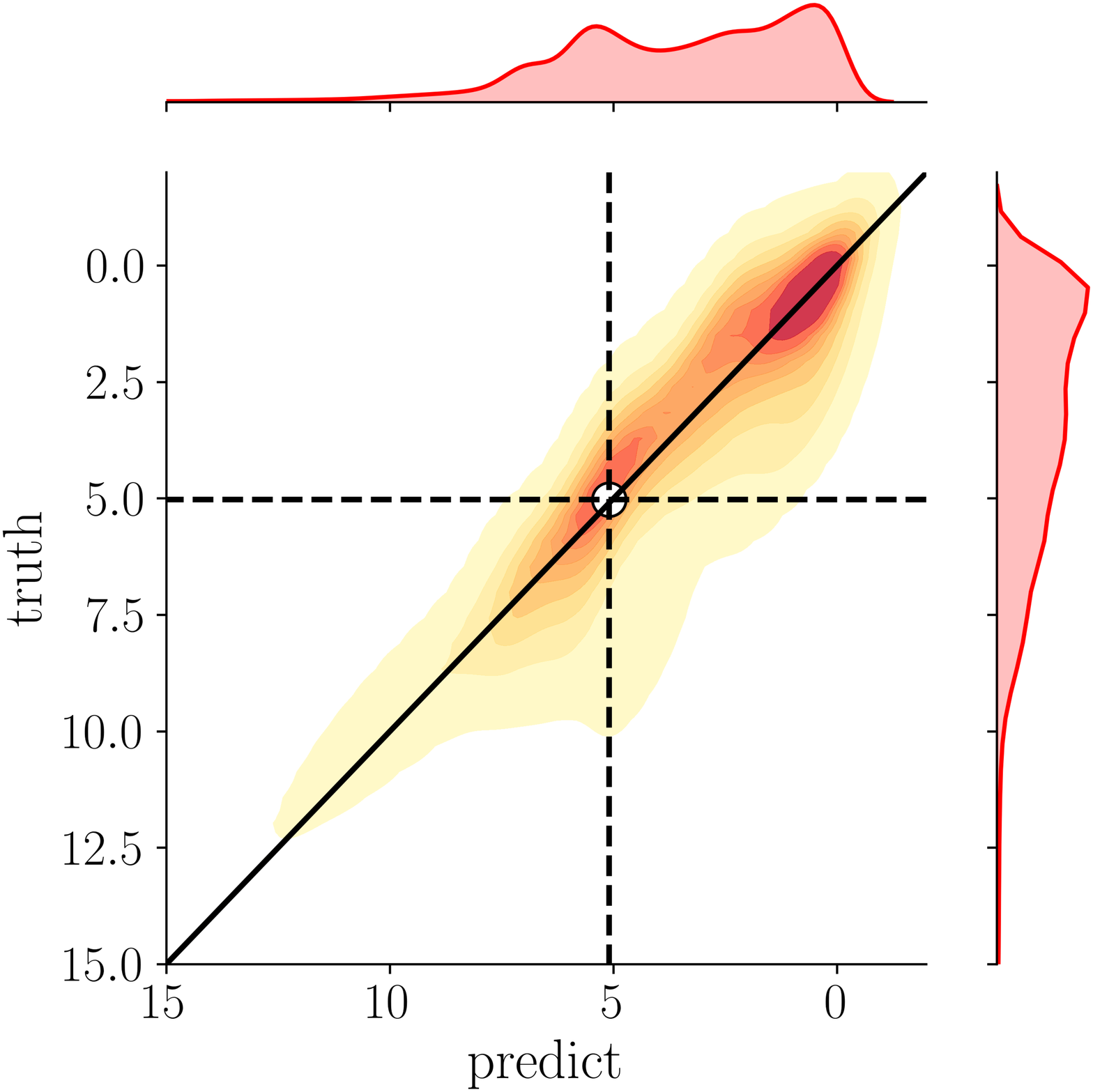}}
    \subfloat[kernel density of $\theta_2$]{\includegraphics[width=0.43\textwidth]{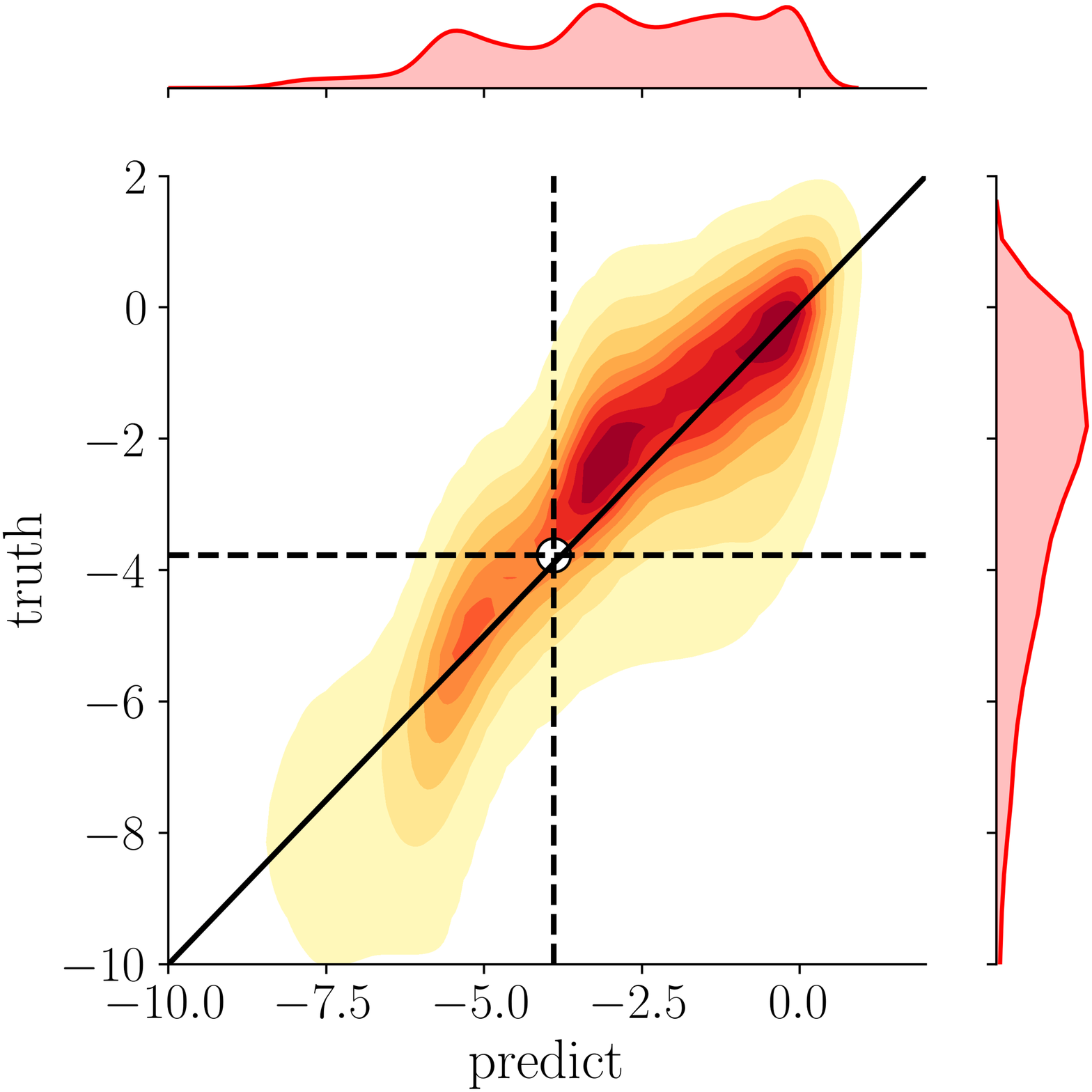}}
    \caption{Panel (a) shows contour plots of $\theta_1$, $\theta_2$, and $|\theta_1|-|\theta_2|$ with comparison among the direct learned model, the ensemble-based learned model and the DNS; Panels (b) and (c) show kernel density plot of $\theta_1$ and $\theta_2$ from the truth and the model estimation for periodic hill case, respectively. 
    The round circles in panels (b) and (c) indicate the values of the $30 \%$ quantiles (i.e., $30\%$ of the cells have $\bm{\theta}$ larger than this value in the magnitude). The probability densities of the truth and the model estimation are plotted on the margins.}
    \label{fig:hill_theta_pdf}
\end{figure}

The learned model has good predictive performance in capturing flow features for $\theta_1$ and $\theta_2$ at intermediate or small magnitudes.
This is illustrated in Figure~\ref{fig:hill_theta_pdf}(a) where the comparison of scalar invariants among the direct learned model, the ensemble-based learned model, and the DNS data is presented.
The scalar invariants $\bm{\theta}$ from  both the direct learned model and the ensemble-based learned model exhibit patterns similar to those of the DNS data, while the noticeable difference exists mainly in the $\bm{\theta}$ with large magnitudes around the separation point. 
This difference is likely due to the fact that the velocity data near the separation point is not used to train the model.
Specifically, in this case the used data is distributed along four profiles, i.e., $x/H=1, 3, 5,$ and $7$, which are away from the separation point.
It can be seen from Figure~\ref{fig:hill_theta_pdf}(a) that the magnitude of scalar invariants around the separation point significantly exceeds that of the training data.
Hence, the learned model with these data only achieves limited improvements in that region.
To further improve the model estimation, additional data around the separation point should be used for training. 
It is noted that the training data, if positioned close spatially, may lead to poor training performance because the correlation among the observation errors is neglected.
Specifically, the observation error includes the measurement and process errors in the ensemble Kalman method.
The measurement error can be negligible for the DNS data, while the process error is significant in this case since it includes the intrinsic discrepancy between the RANS simulation and DNS.
Such error correlation is difficult to estimate and often neglected as in this work.
However, the error of spatially close data would have relatively strong correlations with each other, particularly in the stream-wise direction due to the advection effects. 
As such, the neglected correlation information could deteriorate the training performance.
Alternatively, one can place sparse data at particular positions, e.g., the separation point in this case, but the specific position is often not known as \textit{a priori}.
Hence we suggest positioning training data evenly with a distance of more than one correlation length over the computational domain.
The correlation length in the periodic hill case is approximated as the height of hill crest, i.e., $l_c/H=1$, and without loss of generality, we choose the velocity along profiles at $x/H=1, 3, 5,$ and $7$ as the training data.
The comparisons of $\theta_1$ and $\theta_2$ between the ensemble-based learned model and the truth are presented in Figures~\ref{fig:hill_theta_pdf}(b) and~\ref{fig:hill_theta_pdf}(c), respectively. 
The plots of kernel densities in Figures~\ref{fig:hill_theta_pdf}(b) and~\ref{fig:hill_theta_pdf}(c) indicate that there are relatively small number of cells with input features $\theta_1$ and $\theta_2$ with large magnitudes.
Specifically, only $30\%$ of the cells in the domain have magnitudes of $\theta_1$ and $\theta_2$ larger than $5.0$ and $3.8$, respectively.
This is the probable cause of the deteriorated estimation in the regions with large $\theta$ magnitudes (Figure~\ref{fig:hill_theta_pdf}a).

\begin{figure}
    \centering
    \includegraphics[width=0.9\textwidth]{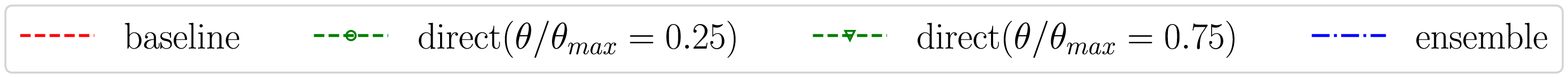}\\
    \subfloat[$g^{(1)}$ vs $\bm{\theta}$]{\includegraphics[width=0.38\textwidth]{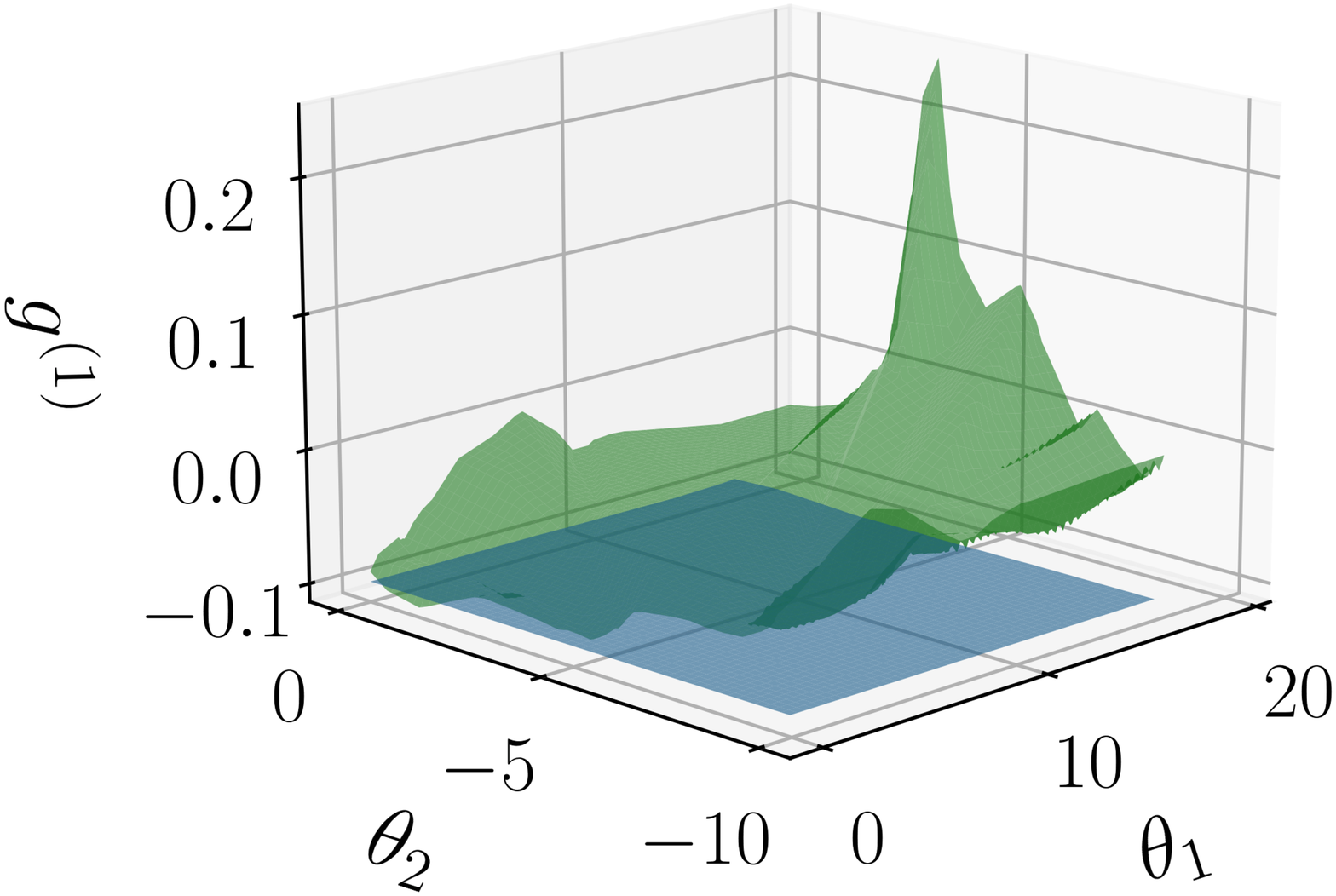}} 
    \subfloat[$g^{(1)}$ vs $\theta_1$]{\includegraphics[width=0.3\textwidth]{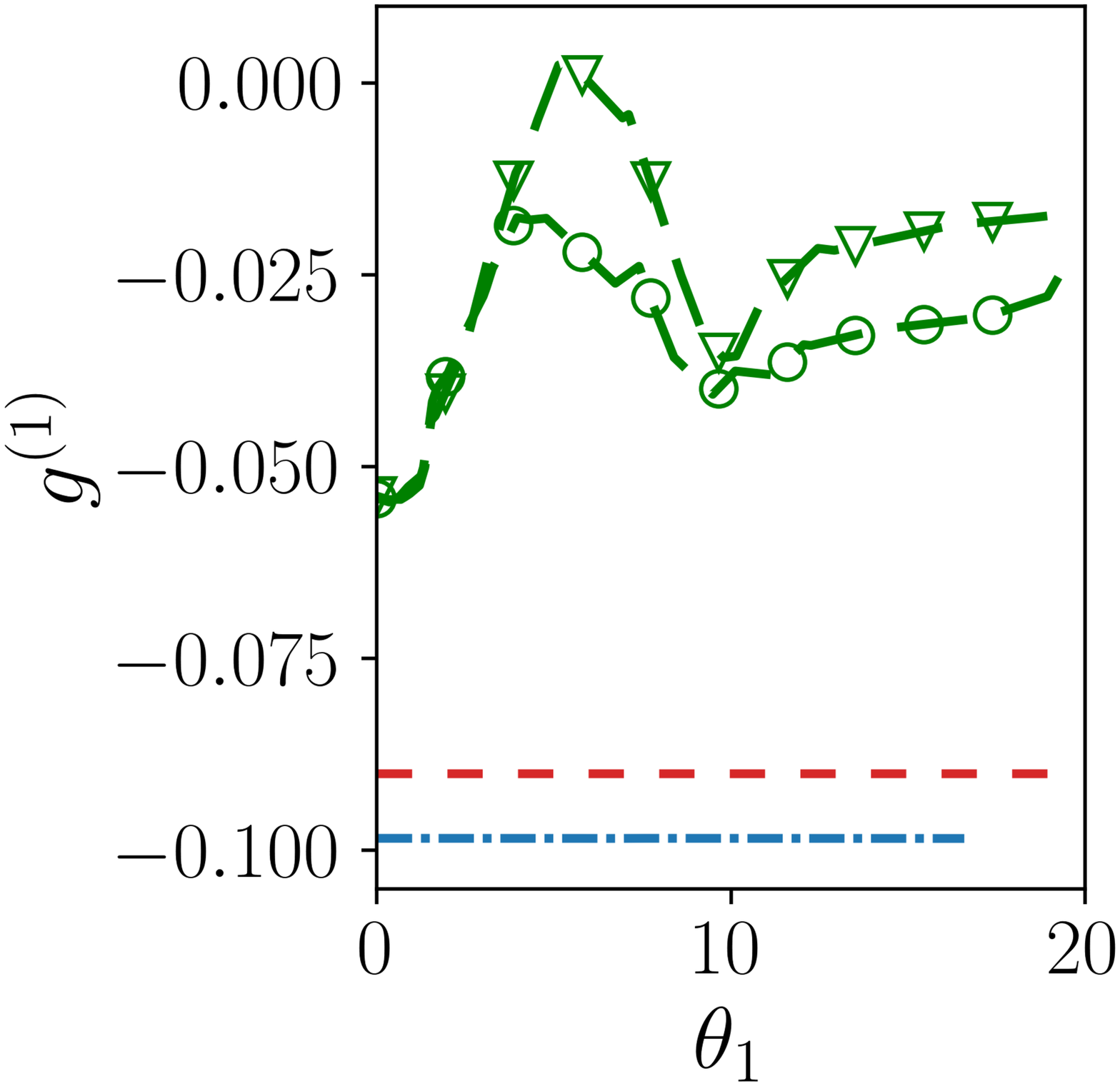}} 
    \subfloat[$g^{(1)}$ vs $\theta_2$]{\includegraphics[width=0.3\textwidth]{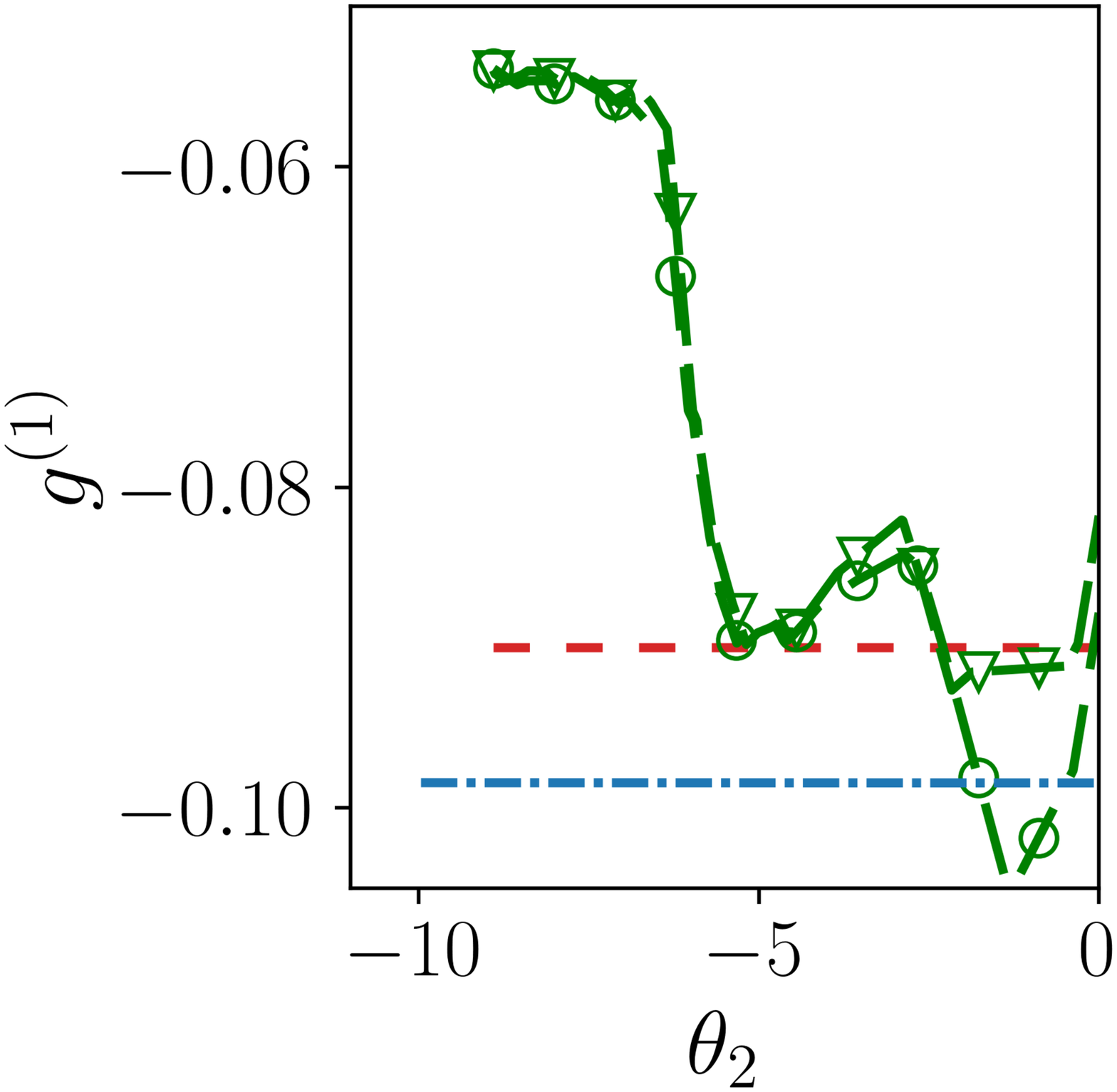}} \\
    \subfloat[$g^{(2)}$ vs $\bm{\theta}$]{\includegraphics[width=0.38\textwidth]{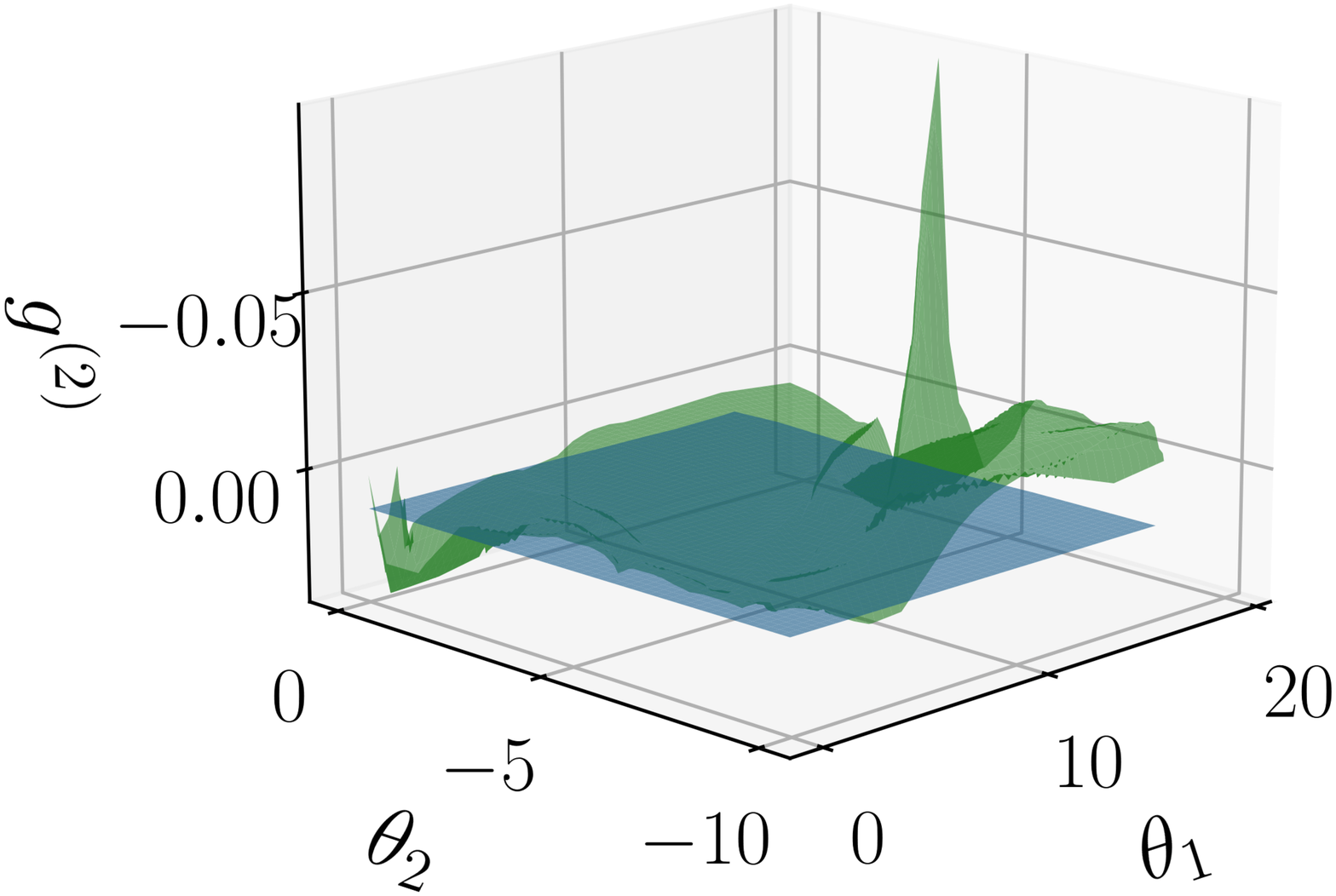}} 
    \subfloat[$g^{(2)}$ vs $\theta_1$]{\includegraphics[width=0.3\textwidth]{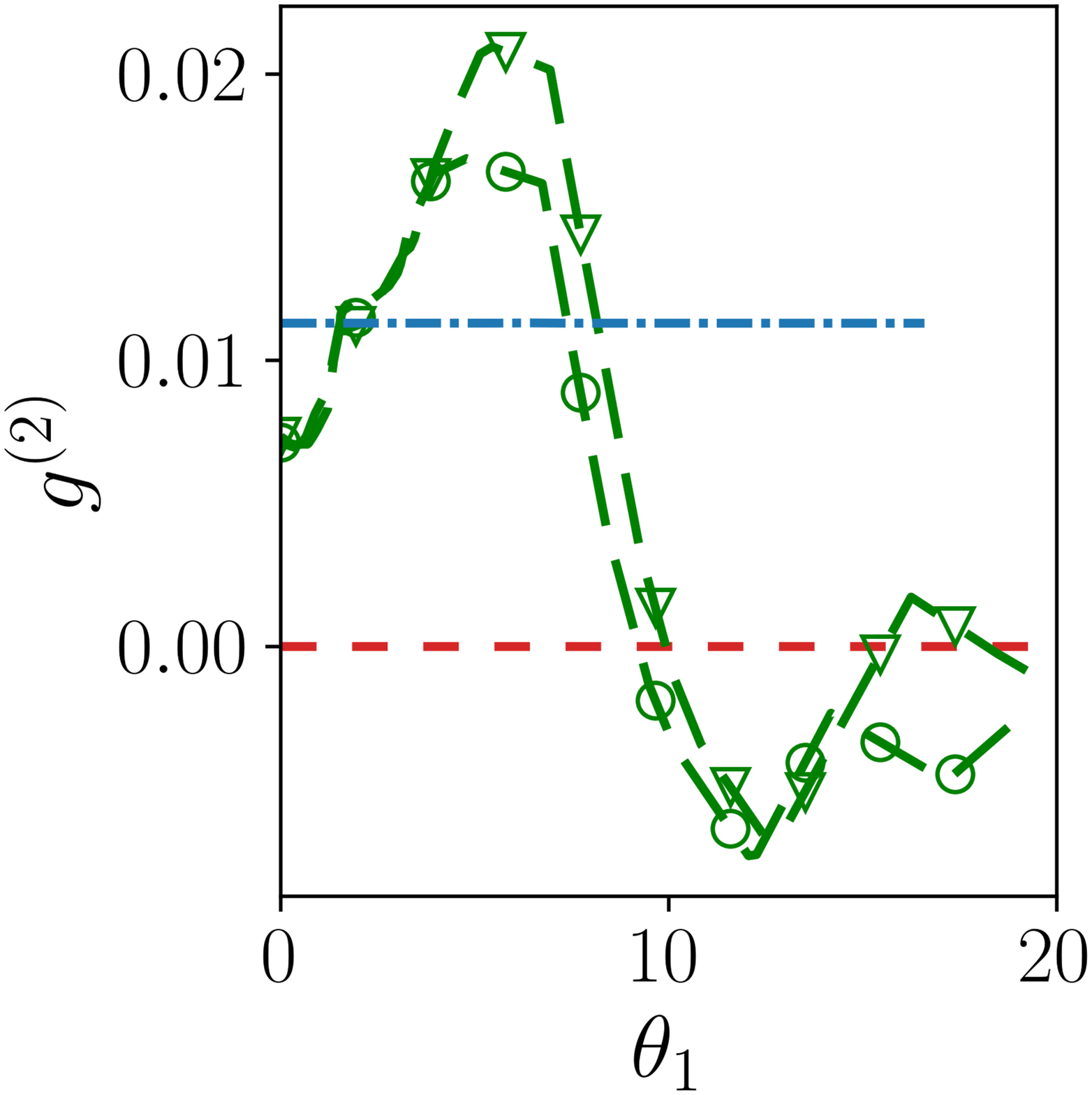}} 
    \subfloat[$g^{(2)}$ vs $\theta_2$]{\includegraphics[width=0.3\textwidth]{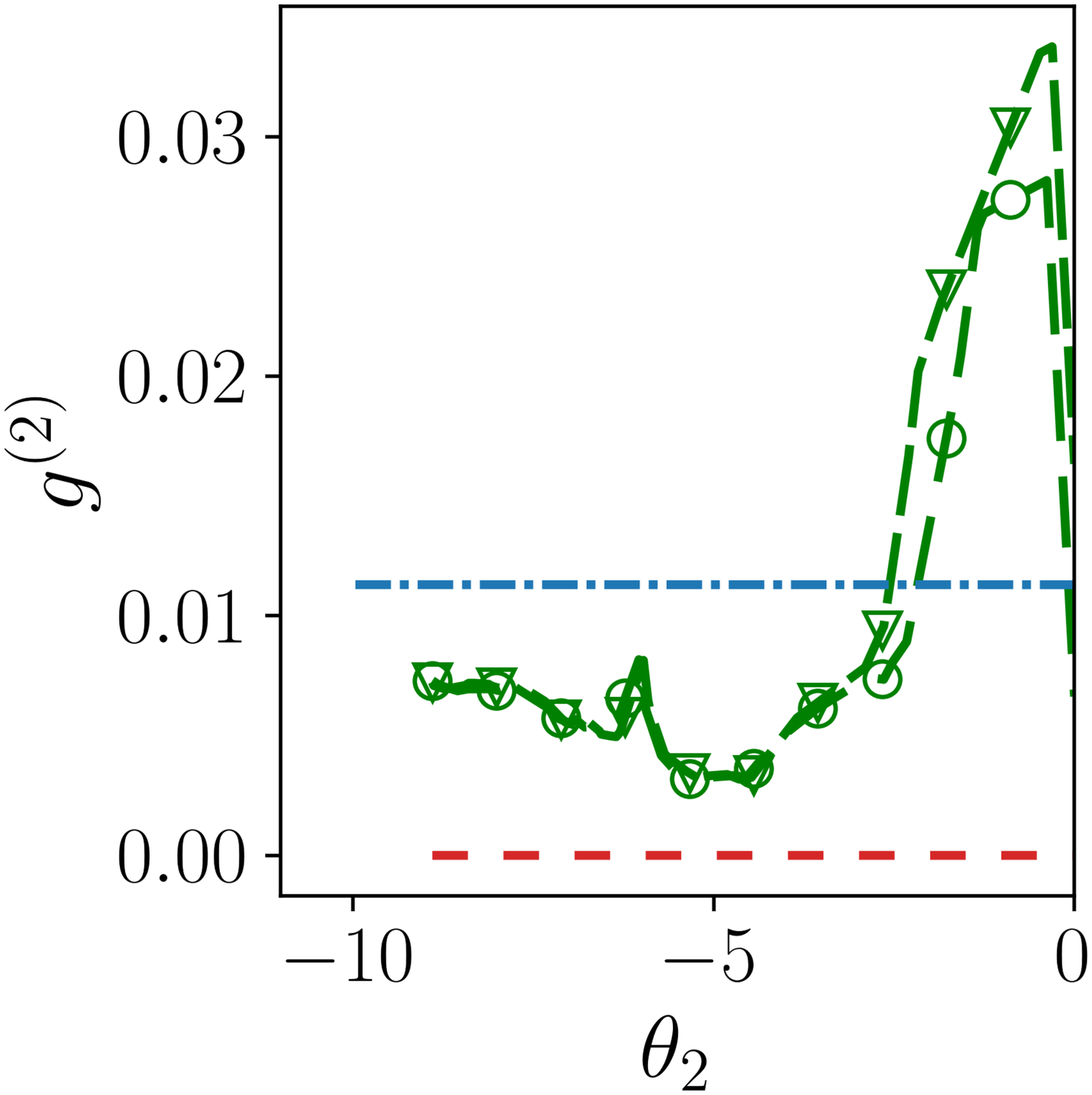}}
    \caption{Plots of the mapping between the scalar invariants~$\bm{\theta}$ and the tensor coefficient~$\bm{g}$ with comparison among the baseline model, the direct learned model, and the ensemble-based learned model for the periodic hill case.
    Panels (a) and (d) show the model function of $g^{(1)}$ and $g^{(2)}$, respectively, where the blue surface represents the ensemble-based learned model and the green surface represents the direct learned model.
    Panels (b), (c), (e), and (f) show the curve plots of the model function at specific planes.
    For the direct learned model, the plots indicate the learned function at $\theta/\theta_\text{max}=0.25$ and $0.75$.
    For the baseline and the ensemble-based learned model, the plots only show the learned function at $\theta/\theta_\text{max}=0.25$, since they are almost constant in the entire function space.
    }
    \label{fig:hill_g}
\end{figure}

The nonlinear mapping between the scalar invariant~$\bm{\theta}$ and the $g$ function is learned from the training data.
The functional mappings with the direct learning method and the ensemble method are shown in Figure~\ref{fig:hill_g}.
In this case no ground truth of the mapping $\bm{\theta} \mapsto g$ exists for validation.
Here we show the baseline mapping from the linear eddy viscosity, i.e., $g^{(1)} = -0.09$ and $g^{(2)}=0$.
The direct learned function has relatively strong non-linearity, while the ensemble-based learned function is almost constant at about $-0.098$ for $g^{(1)}$ and $0.01$ for $g^{(2)}$.
The $g$ function varies slightly for the large invariant $\theta_1$ and the small invariant $\theta_2$, mainly in the uphill region with large strain rates.

\begin{figure}
    \centering
    \includegraphics[width=0.8\textwidth]{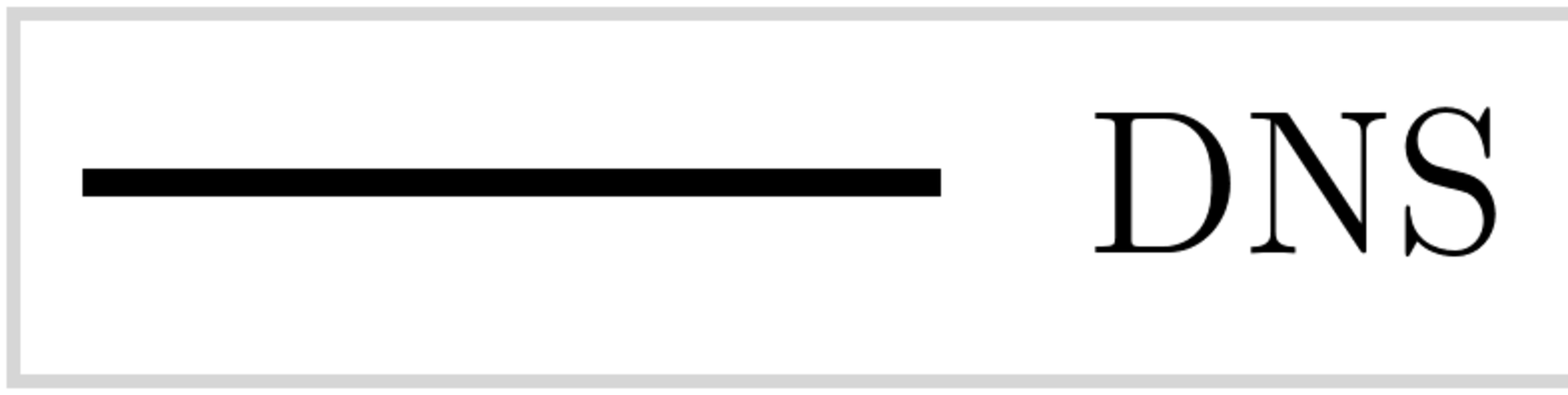} \\
    \subfloat[TKE $k$]{\includegraphics[width=0.49\textwidth, trim=0 0 0 30, clip]{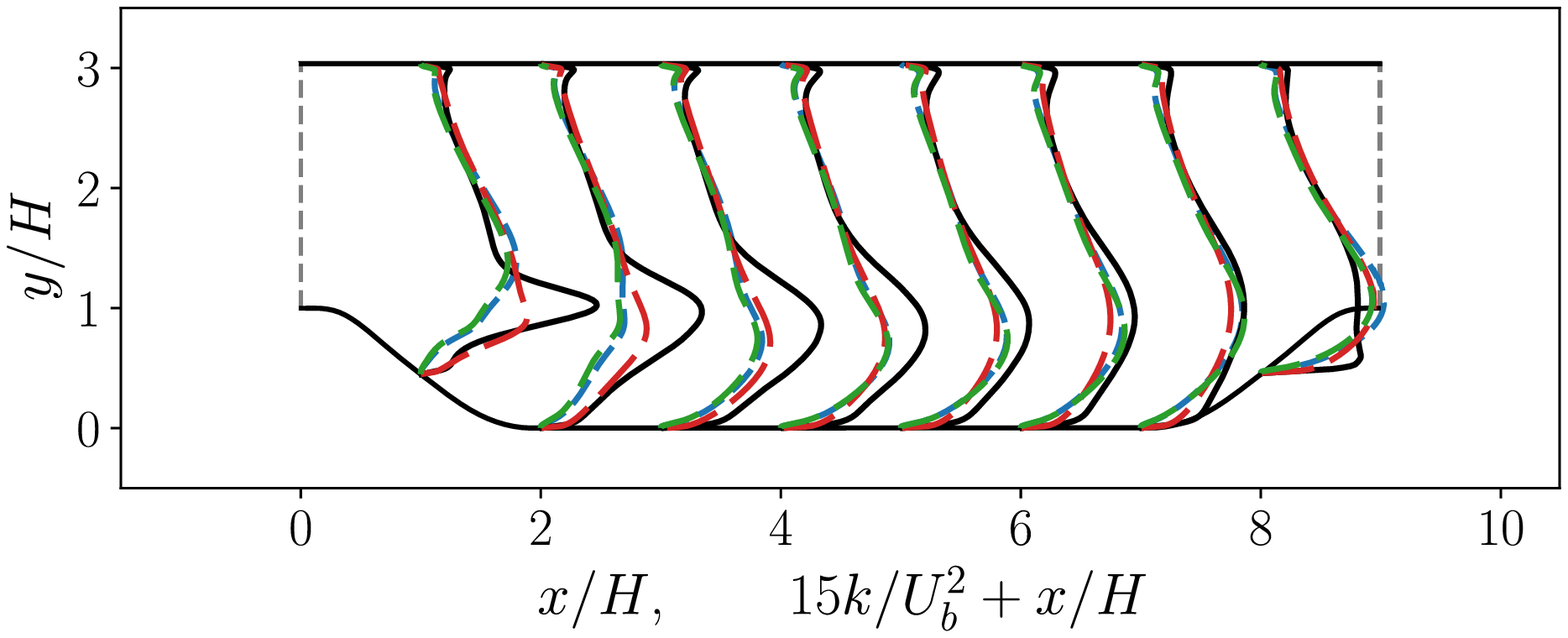}}
    \subfloat[ $b_{xy}$]{\includegraphics[width=0.49\textwidth, trim=0 0 0 30, clip]{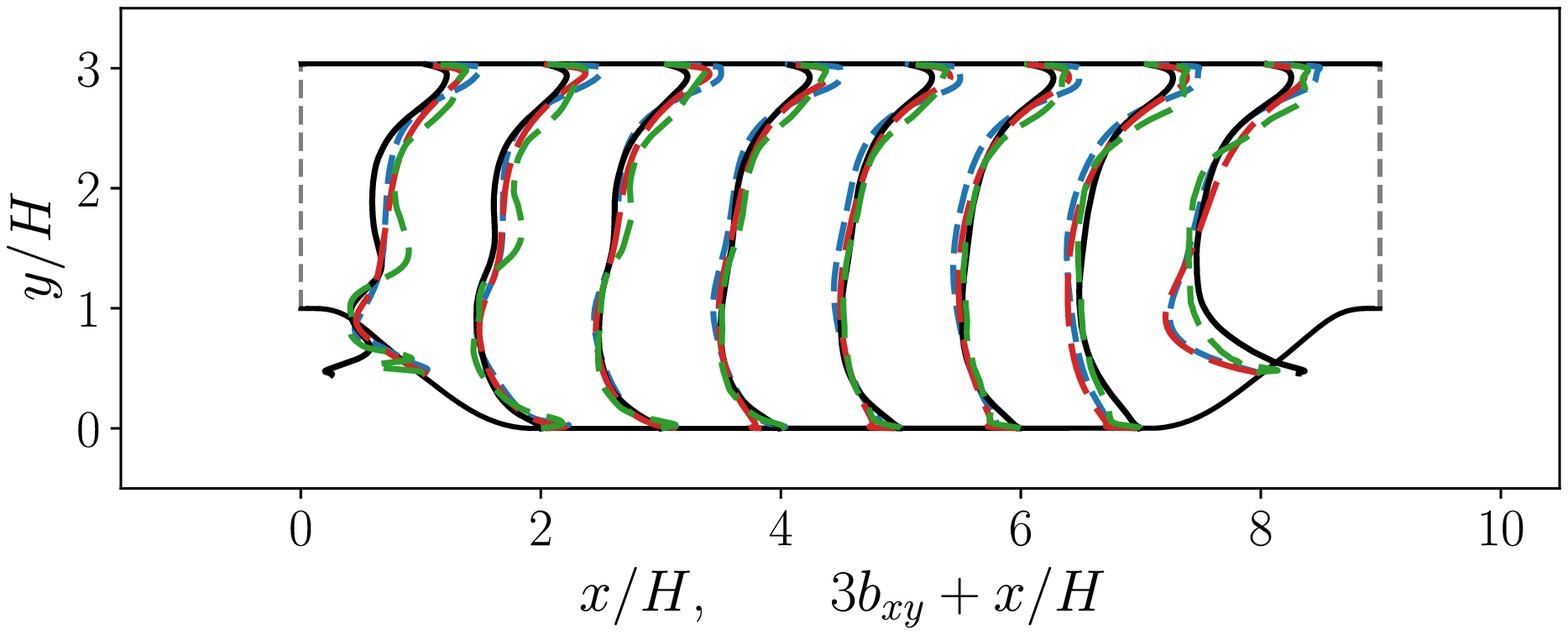}}
    \caption{Comparison of turbulent kinetic energy and deviatoric part of Reynolds stress~$b_{xy}$ along profiles among the $k$--$\varepsilon$ model, the direct learned model, the ensemble-based learned model, and the DNS for the periodic hill case.}
    \label{fig:pehills_2order_profile}
\end{figure}

The estimation of turbulent kinetic energy (TKE) and the deviatoric part of Reynolds stress deviates noticeably from the DNS data. 
The results are presented in Figure~\ref{fig:pehills_2order_profile}.
The direct learned model can estimate the Reynolds stress in relatively better agreement with DNS results than the ensemble-based learned model, due to the use of direct data.
As for the turbulence kinetic energy, both the direct learning and ensemble-based methods lead to significant discrepancies compared to the DNS results.
The discrepancies can be due to the fact that
the divergence-free part of the Reynolds stress has no effects on velocity~\citep{perot1999turbulence}, which poses difficulties in reconstructing the Reynolds stress from the velocity data accurately.
One can use velocity field from high fidelity data to obtain the divergence of Reynolds stress tensor by balancing the momentum equation and further regard the divergence of Reynolds stress tensor as the training target to avoid this issue~\citep{cruz2019use}.
Moreover, the large discrepancies in the TKE estimation can be caused by the deficiency of the model representation.
Specifically, the TKE transport equation can be derived rigorously from the Navier–Stokes equation, and thus it is an exact equation if all terms are modelled correctly. 
Two terms in the equations are not exact, i.e., the production term $\mathcal{P} = \bm{\tau} : \mathbf{S} = 2k (\mathbf{b} + \frac{1}{3} \mathbf{I} ) : \mathbf{S}$, and the dissipation term. 
The turbulence modelling addresses the modelling of the deviatoric tensor $\mathbf{b}$, which if modelled correctly would yield the correct TKE production. 
However, the dissipation rate is modelled by another transport equation that is much less rigorously derived than the TKE transport equation. 
The present work focuses on addressing the deficiency of the turbulence closure by learning a nonlinear algebraic Reynolds stress model, i.e., a nonlinear function $\mathbf{b} = f(\mathbf{S}, \mathbf{W})$. 
This work does not address the shortcomings of the dissipation modelling, which is present in both algebraic models and Reynolds stress transport models (i.e., differential stress models). 
Nor does it make the stress–-strain-strain rate function nonlocal (e.g., as in~\citet{zhou2021learning, zhou2022frame}), which is needed for non-equilibrium turbulence requiring Reynolds stress transport models.

Previous efforts of data-driven, Reynolds stress-based models~\citep{schmelzer2020discovery,waschkowski2022multi} have introduced a corrective production term~$\delta \mathcal{P}(\mathbf{S}, \mathbf{W})$ to the TKE transport equation, which improves the TKE prediction. 
This correction goes beyond the turbulence constitutive modelling and addresses the structure of turbulence quantity transport models. 
It has the same effects as the multiplicative factor $\beta$ applied to the production term~\citep{singh2016using}. 
However, note that the latter operates in the realm of Boussinesq assumption (stress–-strain-rate relation), and thus data-driven Reynolds stress models (e.g.,~\citet{schmelzer2020discovery, waschkowski2022multi} and the present work) would introduce more degrees of freedom in the correction if such a production correction term is used.

Our investigation suggests that the nonlinearity of the stress--strain-rate relation may not be the dominant deficiency in the flows studied here. This is evident from Figure~\ref{fig:pehills_2order_profile}, which shows that the learned model does not significantly improve the TKE estimation. 
One can also consider $k$ an ``operation variable" described by the TKE transport equation, which is consistent with the widely accepted interpretation that the dissipation rate $\varepsilon$ is an operation variable~\citep{pope2001turbulent}. 
They are not physical variables and do not necessarily need to be compared directly to their DNS counterpart. 
The purpose of operation variables is to make good predictions of the velocities and their derived field.

\begin{figure}
    \centering
    \includegraphics[width=0.6\textwidth]{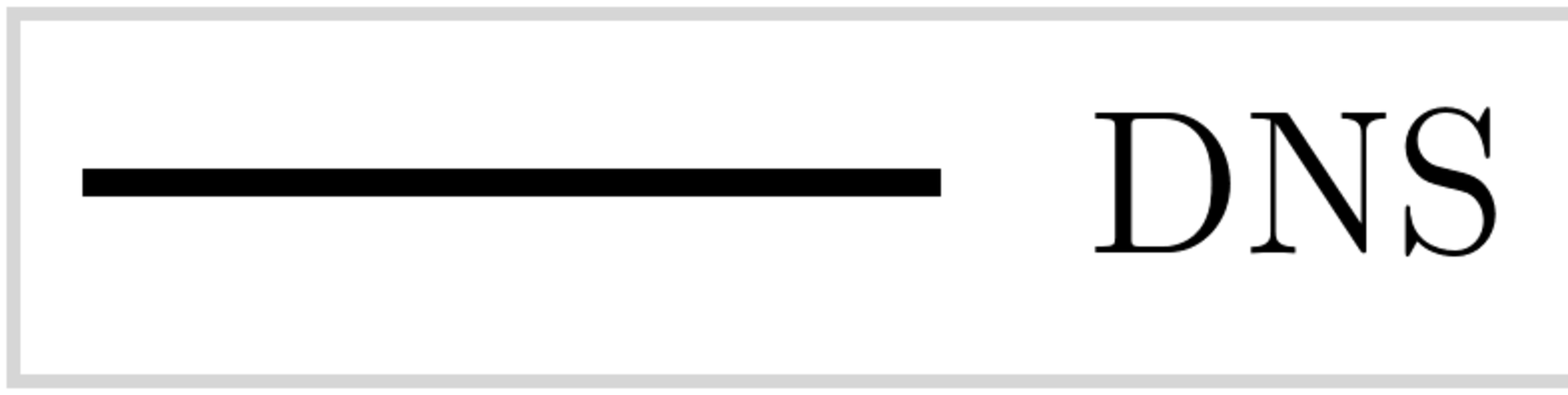}
    \subfloat[$\alpha=0.5$]{\includegraphics[width=0.48\textwidth]{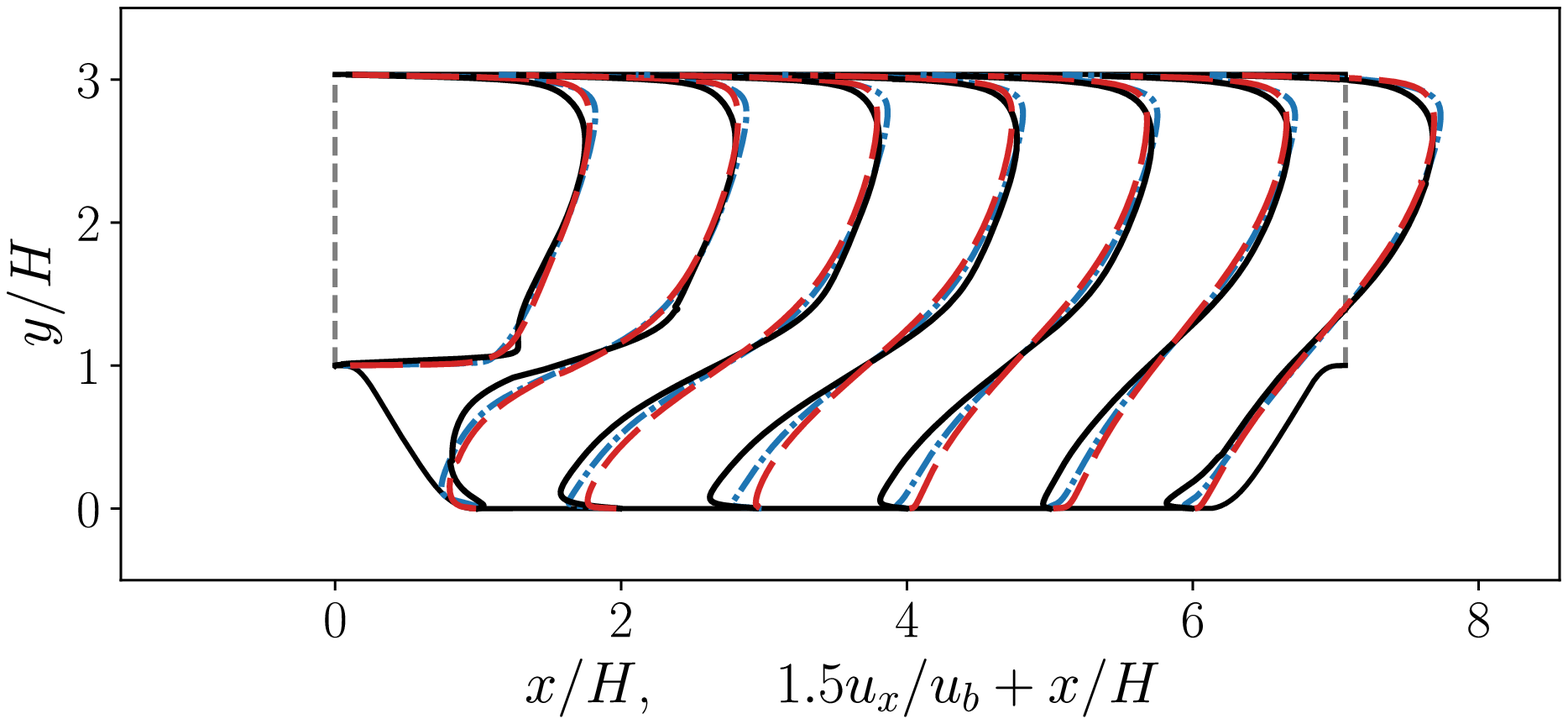}}
    \subfloat[$\alpha=0.8$]{\includegraphics[width=0.48\textwidth]{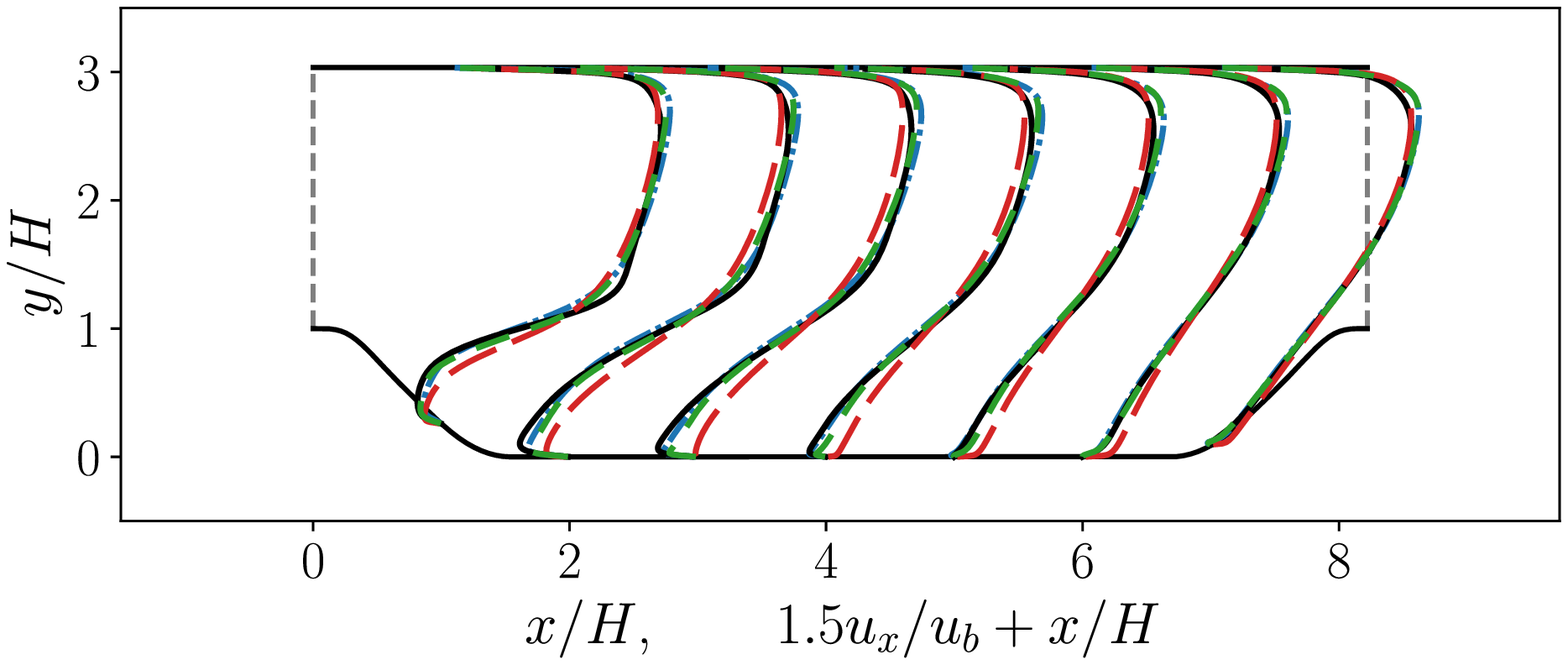}} \\
    \subfloat[$\alpha=1.2$]{\includegraphics[width=0.48\textwidth]{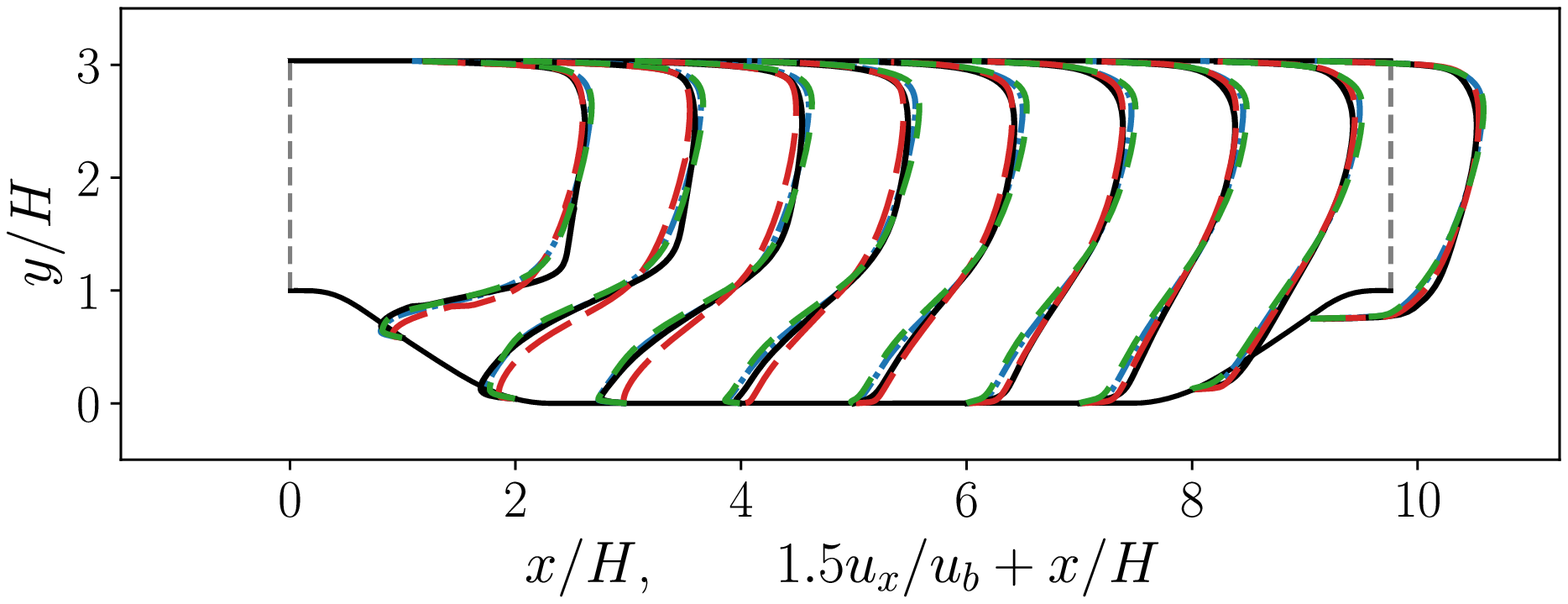}}
    \subfloat[$\alpha=1.5$]{\includegraphics[width=0.48\textwidth]{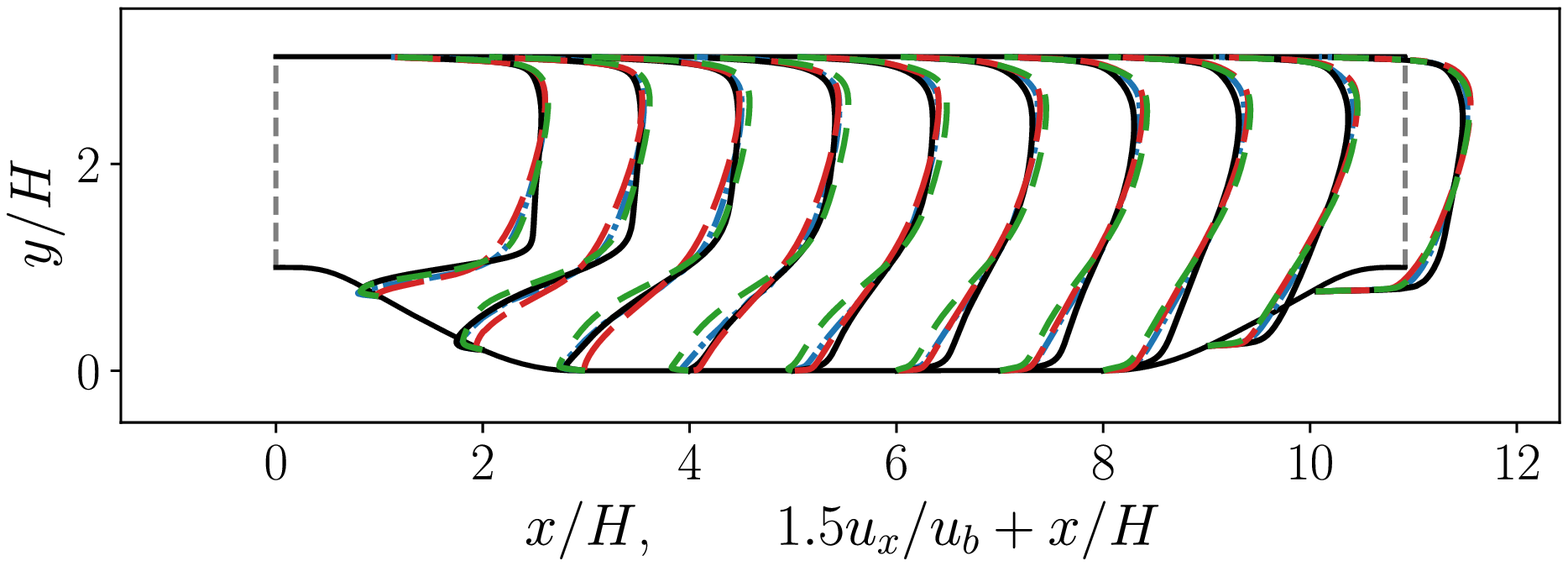}} \\
    \subfloat[error over entire field]{\includegraphics[width=0.48\textwidth]{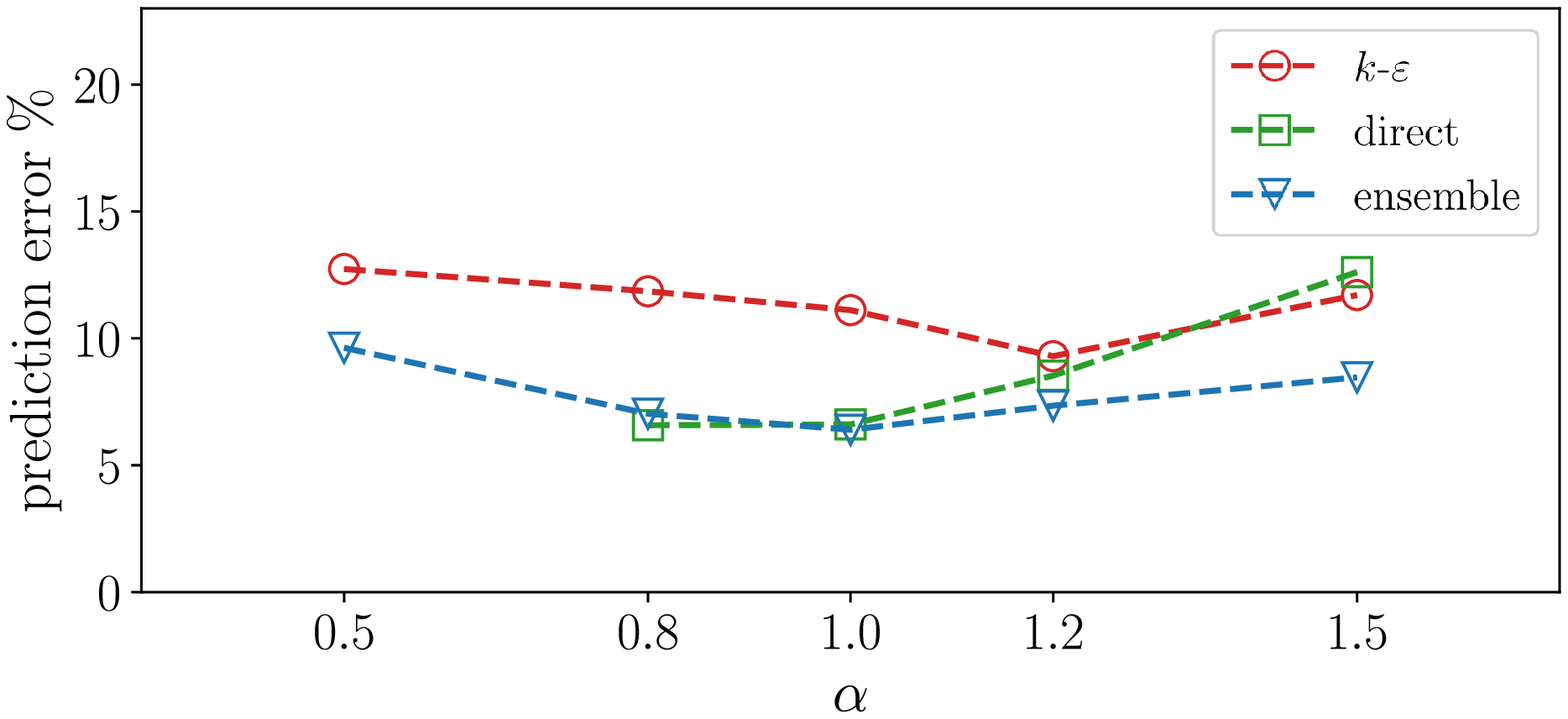}}
    \subfloat[error over recirculation region]{\includegraphics[width=0.48\textwidth]{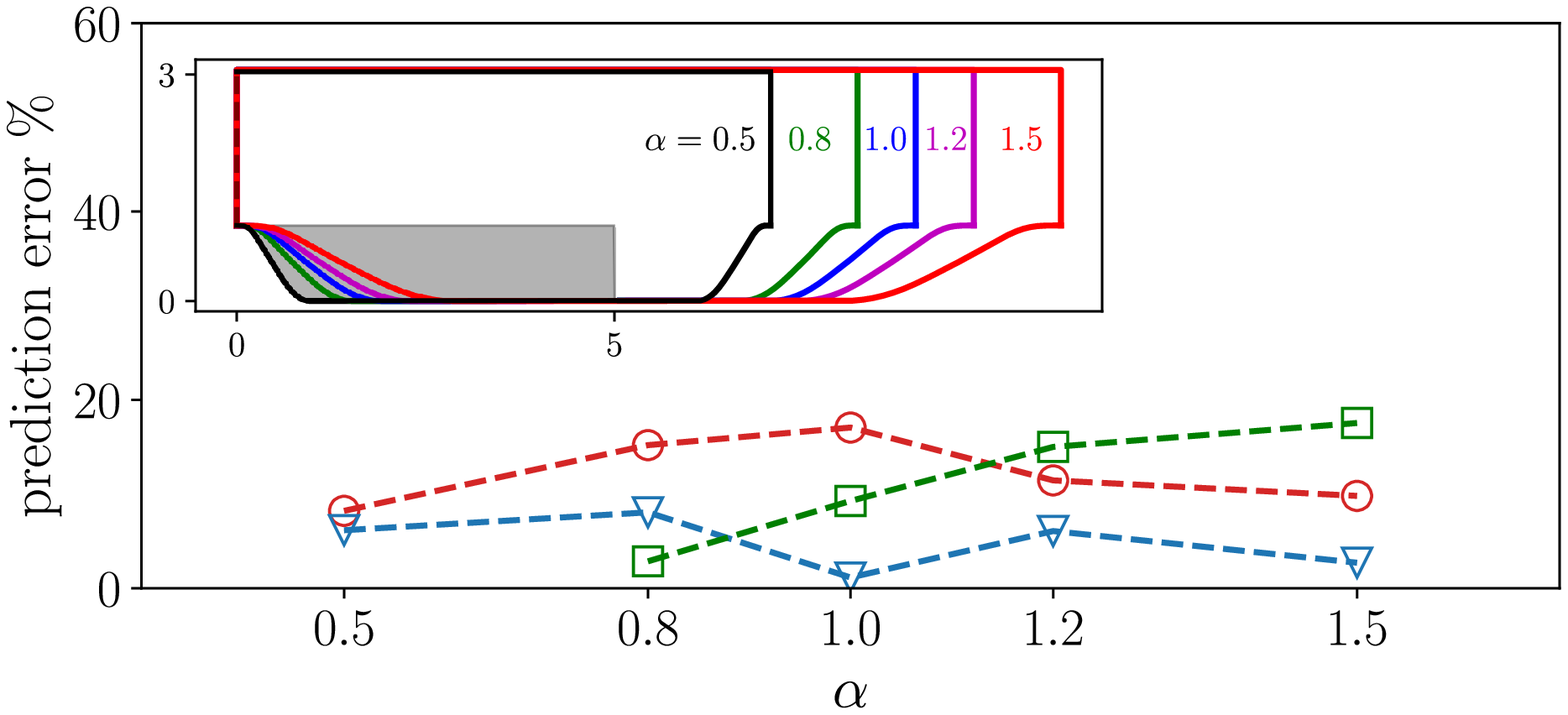}}
    \caption{Results of generalizability tests on configurations with different slopes ($\alpha=0.5, 0.8, 1.2, 1.5$). The panels (a)--(d) show the velocity profiles with comparison among the $k$--$\varepsilon$ model, the direct learned model, the ensemble-based learned model, and the DNS. The panels (e) and (f) show the plots of prediction error over the entire field and recirculation region, respectively. The shadow in panel (f) indicate the recirculation region for error calculations. The results of the direct learned model are not shown for the configuration with $\alpha=0.5$, since the solver diverges in that case.}
    \label{fig:pehill_generalizability}
\end{figure}

Our generalizability test suggests that the learned model with the ensemble method is able to generalize to cases that are similar (in terms of feature space) to the trained cases but perform less well in cases with large differences from the trained cases.
This test suggests that a wide range of input features should be embedded in order to obtain a practical model.
The results of the predicted velocity~$u_x$ for different slopes~$\alpha$ are shown in Figure~\ref{fig:pehill_generalizability}.
The prediction with the direct learned model is also presented for comparison. 
For the case of $\alpha=0.5$, the solver diverges with the direct learned model, and hence no results are presented.
In the case of $\alpha=0.8$ and $1.2$, the learned model can generally improve the prediction compared to the $k$--$\varepsilon$ model.
However, in the case of $\alpha=1.5$, the model leads to significant discrepancies.
In contrast, all the cases show that the learned model with the ensemble method can noticeably improve the mean flow estimation in terms of the velocity compared to the $k$--$\varepsilon$ model.
Particularly, for the case of $\alpha=0.8$ and $\alpha=1.2$, the velocity profiles~$u_x$ have a remarkable agreement with the DNS data.
That is probably due to the similar input features of these two cases to the training case of $\alpha=1$.
Additionally, the error between the prediction and the DNS data over the entire field and the recirculation region ($0 < x/H < 5$ and $0 < y/H < 1$) is shown in Figures~\ref{fig:pehill_generalizability}(e) and \ref{fig:pehill_generalizability}(f), respectively, where the error from the training case of $\alpha=1$ is also indicated based on the propagated velocity. 
It is obvious that the ensemble-based learned models provide better prediction than the $k$--$\varepsilon$ model in all the test cases and the direct learned model in all the cases except for $\alpha=0.8$.
For the training case ($\alpha=1$), the learned model provides the lowest prediction error, which is reasonable since the prediction is directly informed by the training data.
The model prediction error increases as the extrapolation case is further away from the training case.
Particularly there exhibit noticeable discrepancies in the case of $\alpha=1.5$.
The maximum value of the input feature is provided in Table~\ref{tab:max_theta_vs_alpha} to show the feature difference among these cases.
It can be seen that the range of the input feature for $\alpha =0.8$ and $1.2$ is relatively close to the training case in contrast to the cases of $\alpha=0.5$ and $1.5$.
This confirms that the consistency of the input features between the training case and the test cases is essential for the generalizability of the data-driven model.
For the flow with similar input features, the trained model is able to provide satisfactory predictions.
This suggests that a wide range of input features should be included in the training case to obtain a practical model.

\begin{table}
    \centering
    \begin{tabular}{cccccc}
        Geometry (slope parameter $\alpha$)  & 0.5 & 0.8 & 1.0 & 1.2 & 1.5  \\ \\
        Max.~of input feature $\theta_1$ & 161 & 115 & 105 & 91 & 69 \\ \\
        Exceeds training case ($\alpha=1$) by 
         & $53.3\%$ & $9.52\%$ & 0 & $13.3\%$ & $34.3\%$ \\
    \end{tabular}
    \caption{Comparison of the maximum value of input features in flow configurations with different slopes~$\alpha$.}
    \label{tab:max_theta_vs_alpha}
\end{table}

\section{Discussion}
\label{sec:discuss}

\subsection{Parallelization}
To enhance the generalizability of the learned model, training data should embed various flow features from different configurations, e.g., the square duct, the periodic hills, and airfoils.
To handle a large data set, the conventional machine learning training algorithms need to randomly split the data into multiple batches.
Further, the stochastic gradient descent (SGD) is employed to train the model by looping over the entire data set sequentially~\citep{kovachki2019ensemble}.
This makes it inefficient to handle the large data set.
The ensemble-based framework is able to learn the model from a large data set in a parallelizable manner.
The ensemble-based method is inherently parallelizable and can handle the data with random noise so as to avoid data overfitting.
This achieves the same goal as SGD for machine learning.
Furthermore, the model-consistent training framework can train with the data from different configurations simultaneously as noted by~\citet{waschkowski2022multi}.
These training cases do not need communication (e.g., embarrassingly parallel workload), such that the wall time is not significantly increased when the number of used CPU cores is equal to the number of configurations.

\subsection{Flexibility in learning from different observation data}
The ensemble-based framework is extremely flexible in terms of the loss function, specific applications, and observation data, due to its derivative-free nature.
Specifically, the loss function can even be non-differentiable, e.g., when learning dynamic model parameters with statistical observation data. 
In such a scenario, the adjoint-based method would be difficult to deploy, while the ensemble method only needs to evaluate the cost function and approximate the corresponding gradient based on the model input and output.
Moreover, the framework here is used for the turbulence closure problem.
Other physical systems where the adjoint solver is not readily available can apply the proposed method to learn the underlying closure model based on the measurable observations. 
Besides, in specific cases, e.g., the RANS modelling, the available data are often collected from different configurations with varying physical quantities and dimensionality.
It is difficult for the conventional methods to use these disparate data, as they need to develop specific adjoint solvers for different measurable quantities, which is a challenging task for complex CFD solvers.
The proposed model-consistent learning framework can approximate the sensitivity of the model prediction to the model parameters based on model inputs and outputs.
With the non-intrusive and derivative-free nature, the ensemble-based model-consistent learning is naturally flexible for different loss functions, physical systems, and disparate data.

\section{Conclusion}
\label{sec:conclusion}

This work proposes an ensemble-based framework to learn nonlinear eddy viscosity turbulence models from indirect data.
Earlier works~\citep{duraisamy2021perspectives} have noted that there exists an inconsistency issue between training and prediction environments when learning turbulence models from direct data.
In this work, we observe that this inconsistency leads to poor generalizability of the learned model.
The proposed framework can ensure the consistency and thus improve the generalizability of the learned model. 
Furthermore, the training method is non-intrusive and does not need an adjoint solver.
Moreover, the ensemble-based method has been shown to learn a turbulence model from indirect observation data more efficiently than the adjoint-based method based on our particular implementation of adjoint solver and test cases (both of which are steady-state flows).

The capability of the proposed framework is demonstrated on two flows, the flow in a square duct and the flow over periodic hills. The duct flow demonstrated the capability of the proposed method in learning underlying closure relationships from velocity observation data, and the periodic hill case showed the generalizability of the learned model to flows in~similar configurations with varying slopes.
Both cases highlight the straightforward implementation of the ensemble-based learning method.
It runs in parallel and can learn from large sets of training flows simultaneously.
Moreover, the non-intrusive nature of the ensemble-based method makes it convenient to handle different types of observations without developing an adjoint solver for each new objective function.

The main limitations and perspectives of the present ensemble-based learning method are discussed below.
First, the present algorithm would be difficult to be applied for scenarios having large data sets such as unsteady three-dimensional flow data, due to the prohibitive computational cost of large matrix inversion in the update scheme.
Dimension reduction techniques such as the truncated singular value decomposition would be incorporated in the ensemble method to address this issue~\citep{luo2015iterative}. 
Second, the position of observation data is critical to the training performance.
The strategy to select the optimal position of these training data needs to be further investigated.
Third, the present framework is based on the nonlinear eddy viscosity model, which is under the weak equilibrium assumption. 
It would be worthy of investigation to use a neural network to emulate the Reynolds stress transport equation~\citep{zhou2022frame} with embedded non-equilibrium effects.
Besides, future works will focus on training with different classes of flows to enhance the generalizability of the learned model, which is a step towards representing a universal or unified turbulence model.

\appendix

\section{Practical implementation}
\label{sec:implementation}

\begin{figure}
    \centering
    \includegraphics[width=\textwidth]{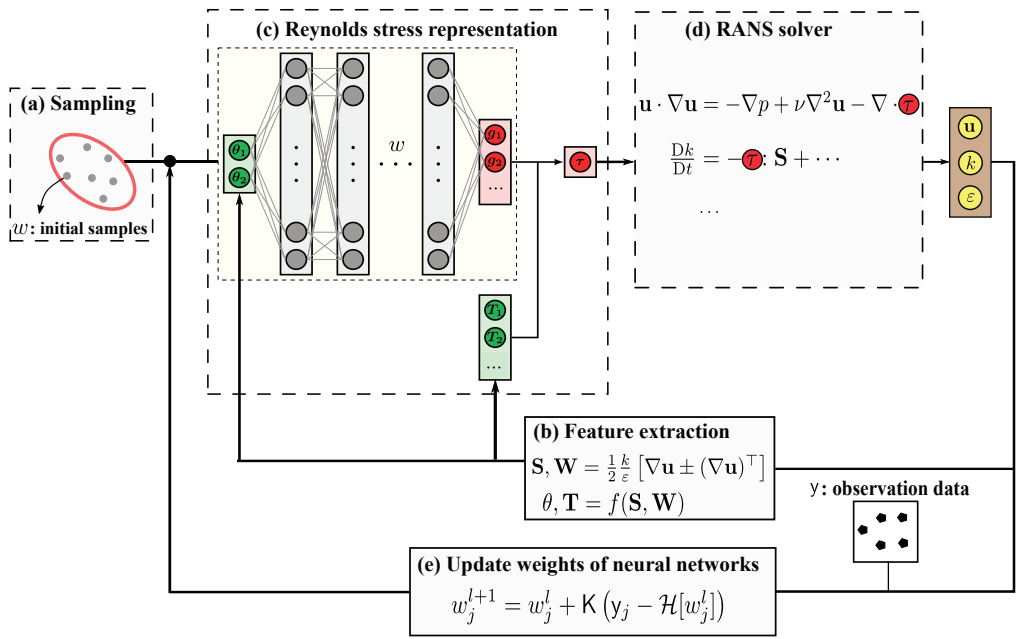}
    \caption{Detailed schematic of ensemble-based model-consistent training of the tensor basis neural network. (a) generate samples of neural network weights; (b) extract input features; (c) evaluate Reynolds stress based on tensor basis neural network; (d) propagate the Reynolds stress to velocities; (e) update weights of neural networks based on ensemble Kalman method.}
    \label{fig:scheme}
\end{figure}

The practical implementation of the proposed ensemble-based model-consistent turbulence modelling framework is detailed in this section and illustrated schematically in Figure~\ref{fig:scheme}.
Given the observation error~$\mathsf{R}$, the data set~$\mathsf{y}$, and the sample variance~$\sigma$, the procedure for the ensemble-based model learning is summarized below:
\begin{enumerate}
   \item Pre-training: 
    To obtain the initial weight $\bm{w}^0$ of the neural network, we pre-train the network to be an equivalent linear eddy viscosity model such that $g^{(1)}=-0.09$ and $g^{(i)}=0$ (for $i=2$ to $10$). The weights so obtained, $\bm{w}^0$, are set as the initial value for optimization~\citep{strofer2021end}. The pre-training is necessary because conventional initialization methods (e.g., random initialization) may lead to nonphysical values such as the positive $g_1$ (negative eddy viscosity), which would cause divergence of the RANS solver. Pre-training is needed to address this difficulty and accelerate model learning.
    \item  Initial sampling: We assume that the weights are independent and identically distributed (i.i.d.)~Gaussian random variables with mean $\bm{w}^0$ and variance $\sigma^2$.
    We draw random samples of the weights (Fig.~\ref{fig:scheme}a) through the formula $\bm{w}_j = \bm{w}^0 + \bm{\epsilon}_j$, where $\bm{\epsilon} \sim \mathcal{N}(0, \sigma^2)$. 

    \item Feature extraction: the velocity field~$\boldsymbol{u}$ and turbulence time scale~$\frac{k}{\varepsilon}$ are used to compute the scalar invariants~$\bm{\theta}$ and the tensor  bases~$\boldsymbol{\mathbf{T}}$ (Fig.~\ref{fig:scheme}b) based on the equations~\eqref{eq:tensor_basis} and~\eqref{eq:scalar_invariant}. 
    The scalar invariants are then adopted as the inputs of the neural network function $\bm{g}$, while the tensor bases are employed to construct the Reynolds stress by combining with the outputs of the neural network as illustrated in step (iv) below. The input features of the neural network are scaled into the range of $[0, 1]$ with the maximum and minimum values. In posterior tests, the maximum and minimum values in the training case need to be used to scale the input features in test cases.
   	Note that this strategy uses global quantities, i.e., the maximum and minimum values of the scalar invariants, to normalize the input features.
   	Such global normalization may lead to feature clustering along certain directions within the feature space and further cause convergence issues. 
   	Particularly when jointly training with different classes of flows, the input features can have a wide diversity of the scalar invariants~$\bm{\theta}$.
   	For example, the flow with shock waves can provide significant extreme values of scalar invariants.
   	Using these values to scale the input features of other flows would lead to feature clustering in a narrow range near $0$ in posterior tests or jointing training. 
   	In such scenarios, it is necessary to normalize the input features based on local quantities~\citep{ling2015evaluation,wang2017physics,wu2018physics}, e.g.,~$\hat{\theta}=\theta / (| \theta| + | \theta^* |)$, where $\theta^*$ is local normalization.
   	Such a normalization can ensure that the normalized quantity $\hat{\theta}$ falls within the range $[-1, 1]$. 
   	The local normalization can avoid the feature clustering issue with appropriate choices of the normalization factor~$\theta^*$, which is worthy of further investigation in future studies.
    \item Evaluation of Reynolds stress: input features~$\bm{\theta}$ are propagated to the basis coefficient $g$ with each realization of the weights~$\bm{w}$, and then the Reynolds stress can be constructed (Fig.~\ref{fig:scheme}c) through combining the coefficient $g$ and the tensor basis~$\boldsymbol{\mathbf{T}}$, i.e., $\bm{\tau} = 2k\sum_{i} g^{(i)} \boldsymbol{\mathbf{T}}^{(i)} + \frac{2}{3} k \mathbf{I}$.
    
    \item Propagation to velocity: 
    the velocity is obtained by solving the RANS equations for each constructed Reynolds stress.
    Moreover, the turbulence kinetic energy and the dissipation rate are obtained by solving the turbulence transport equations (Fig.~\ref{fig:scheme}d).
    \item Computation of Kalman gain from samples. To this end, we first compute the square root matrices at iteration step $l$ as follows:
    \begin{subequations}
    \label{eq:sqrt_root}
    \begin{align}
    \mathsf{S}_w^l &= \dfrac{1}{\sqrt{N_e -1}} \left[\bm{w}_1^l - \overline{\bm{w}}^l, \bm{w}_2^l - \overline{\bm{w}}^l, \dotsb, \bm{w}_{N_e}^l - \overline{\bm{w}}^l\right], \\
    \mathsf{S}_y^l &= \dfrac{1}{\sqrt{N_e -1}} \left[\mathcal{H}[\bm{w}_1^l] - \mathcal{H}[\overline{\bm{w}}^l], \mathcal{H}[\bm{w}_2^l] - \mathcal{H}[\overline{\bm{w}}^l], \dotsb, \mathcal{H}[\bm{w}_{N_e}^l] - \mathcal{H}[\overline{\bm{w}}^l]\right] , \label{eq:sqrt_root_y} \\
    \overline{\bm{w}}^l &= \dfrac{1}{N_e} \sum_{j=1}^{N_e} \bm{w}_j^l \text{,}
    \end{align}
\end{subequations}
where $N_e$ is the sample size. The Kalman gain matrix is then computed as:
\begin{equation*}
\mathsf{K}  = \mathsf{S}_w \mathsf{S}_y^\top \left(\mathsf{S}_y \mathsf{S}_y^\top + \gamma^l \mathsf{R} \right)^{-1} .
\end{equation*}
    \item Update weights of neural networks: use the iterative ensemble Kalman method to update the weights of the neural network (Fig.~\ref{fig:scheme}e), i.e.,
    \begin{equation*}
    \bm{w}_j^{l+1} = \bm{w}_j^l + \mathsf{K} \left(\mathsf{y}_j - \mathcal{H}[\bm{w}_j^l]\right) \text{.}
    \end{equation*}

    In steps (vi) and (vii), the parameter $\gamma$ is adjusted in an inner loop.
    This inner loop adaptively adjusts the update step length by inflating the observation error covariance with the parameter~$\gamma$.
    Specifically, we let 
    $\gamma^{\upsilon} = \beta^{\upsilon} \{ \mathsf{S}_y^{\upsilon} (\mathsf{S}_y^{\upsilon})^\top \} / \{ \mathsf{R} \}$
    where $\beta^{\upsilon}$ is a scalar coefficient whose value changes at each subiteration index $\upsilon$. 
    Specifically, at each iteration, an initial value (i.e., at sub-iteration step $\upsilon=0$) is set to be $\beta^{0}=1$.
    If at the $\upsilon$-th sub-iteration step, the average data misfit (over the ensemble of model predictions) is reduced, then at the next sub-iteration step, we set $\beta^{\upsilon+1} = 0.8 \beta^{\upsilon}$ and break out of the inner loop; otherwise we set $\beta^{\upsilon+1} = 1.2 \beta^{\upsilon}$ and repeat step (vi).
    We allow up to five sub-iterations in this inner loop.
    
    \item If the ensemble variance is smaller than the observation error, consider the iteration converged and end the iteration; otherwise, continue to step (iii) until the convergence criterion above is met or the maximum number of iterations is reached.
    
\end{enumerate}

\section{Hessian matrix in the ensemble Kalman method}
\label{sec:theory}

In this section, we illustrate how the approximated Hessian matrix, as well as the gradient (Jacobian), are implicitly incorporated in the ensemble Kalman method, which leads to accelerated learning and improved robustness. This is a crucial difference compared to the stochastic gradient descent optimization used for neural network training in deep learning.

The weight update scheme of the iterative ensemble Kalman method is formulated as in Equation~\eqref{eq:ies_weight}:
\begin{equation*} 
    \bm{w}_j^{l+1}  = \bm{w}_j^l + \mathsf{K} \left(\mathsf{y}_j - \mathcal{H}[\bm{w}_j^l]\right) \quad
    \textrm{with} \quad \mathsf{K} = \mathsf{S}_w \mathsf{S}_y^\top \left(\mathsf{S}_y \mathsf{S}_y^\top + \gamma^l \mathsf{R} \right)^{-1}.
     \tag{\ref{eq:ies_weight}}
\end{equation*}
We first establish its connection to the common form of the Kalman gain matrix $\mathsf{K} = \mathsf{P}\mathsf{H}^\top\left(\mathsf{H}\mathsf{P}\mathsf{H}^\top + \gamma^l \mathsf{R} \right)^{-1}$. To this end, we write the model error covariance matrix $\mathsf{P}$ and other associated quantities in terms of the square root matrix~$\mathsf{S}_w$ and its projection~$\mathsf{S}_y$ to the observation space, i.e.,
\begin{equation}
	\mathsf{P} = \mathsf{S}_w \mathsf{S}_w^\top
	\quad \textrm{and} \quad 
	\mathsf{S}_y = \mathsf{H} \mathsf{S}_w .
\end{equation}
Consequently, the cross-covariance $\mathsf{P}\mathsf{H}^\top$ between the weights $\bm{w}$ and the predictions~$\mathcal{H}[\bm{w}]$ and the projection of $\mathsf{P}$ to the observation space are:
\begin{equation*}
	\mathsf{P}\mathsf{H}^\top = \mathsf{S}_w \mathsf{S}_y^\top
	\quad \textrm{and} \quad
	\mathsf{H}\mathsf{P}\mathsf{H}^\top = \mathsf{S}_y \mathsf{S}_y^\top ,
\end{equation*}
respectively. The two forms of the Kalman gain matrix are thus established.

Next, we show that the Kalman gain matrix $\mathsf{K}$ in the update scheme implicitly contains the inverse of an approximated Hessian matrix of the cost function. To see this point, let~$\mathsf{H}$ be the local gradient of the observation operator $\mathcal{H}$ (with respect to the parameter $\bm{w}$; same for all gradient and Hessian mentioned hereafter). After dropping the iteration index, it can be shown that the gradient of the data misfit term 
$J' = \| \mathsf{y} - \mathcal{H}[\bm{w}]  \|_{\gamma \mathsf{R}}^2$ in Equation~\eqref{eq:cost_iter} is given by (neglecting a constant factor 2)
\begin{equation} \label{eq:local_misfit_gradient}
\frac{\partial J'}{\partial \bm{w}}= - \mathsf{H}^\top (\gamma\mathsf{R})^{-1} \left(\mathsf{y}_j - \mathcal{H}[\bm{w}_j]\right) \text{,}
\end{equation}
and the local Hessian matrix of the entire objective function is given by (neglecting a constant factor 2)
\begin{equation} \label{eq:local_hessian}
\frac{\partial^2 J}{\partial \bm{w}^2}= \mathsf{P}^{-1} +  \mathsf{H}^\top (\gamma\mathsf{R})^{-1} \mathsf{H} \text{.}
\end{equation}
We will utilize the following matrix identity:
\begin{equation} \label{eq:mtx_id}
\mathsf{P} \mathsf{H}^\top \left( \mathsf{H} \mathsf{P} \mathsf{H}^\top + \gamma\mathsf{R} \right)^{-1} =
\left(\mathsf{P}^{-1} +  \mathsf{H}^\top (\gamma\mathsf{R})^{-1} \mathsf{H} \right)^{-1} \mathsf{H}^\top (\gamma\mathsf{R})^{-1} \text{.}
\end{equation}
See Equation~(49) in~\citet{luo2021novel} for detailed derivations of the identify above. In general, the observation operator $\mathcal{H}$ is nonlinear, in which case the square root matrix $\mathsf{S}_y$ as estimated in Equation~\eqref{eq:sqrt_root} provides a derivative-free approximation to the projected square root matrix $\mathsf{H} \mathsf{S}_w$. Accordingly, one can see that the term $\mathsf{K} \left(\mathsf{y}_j - \mathcal{H}[\bm{w}_j^l]\right)$ in Equation~\eqref{eq:ies_weight} is an ensemble-based derivative-free approximation to the product between the inverse of the local Hessian matrix in Equation~\eqref{eq:local_hessian} and the (negative) local gradient in Equation~\eqref{eq:local_misfit_gradient}. In other words, the weight update formula Equation~\eqref{eq:ies_weight} implicitly utilizes the information of both approximated gradient and Hessian matrices.

\section{Sensitivity study of network architecture and observation data}
\label{sec:sensitivity}

Neural networks with different architectures are used in the model consistent training of the square duct case to show the sensitivity of the framework.
Three network architectures are tested: (1) two hidden layers with $5$ neurons per layer (baseline), (2) two hidden layers with $10$ neurons per layer, and (3) ten hidden layers and $10$ neurons per layer.
The results of errors in the velocity and Reynolds stress over the entire field are summarized in Table~\ref{tab:sensitivity_study}.
It can be seen that the results are not very sensitive to the neural network architecture for the square duct case.
The errors in the velocity and the Reynolds stress among the three cases are very similar.
It is noted that the case with $2$ layers and $5$ neuron per layer is able to predict well the flow fields in both velocities and the Reynolds stresses.
This is likely due to the narrow range of the input features in this case.
The maximum of the input features is approximately $7$, which can be sufficiently captured with $69$ parameters in the neural network.
Moreover, we test the setting of using the velocity observation along the anti-diagonal line of the computational domain.
The results in both the velocity and the Reynolds stress field become slightly inferior compared to the case with the full field.

\begin{table}
    \centering
    \begin{tabular}{c c c c c}
     \shortstack{Network architecture \\ (neurons/layer $\times$ layers )} & $5 \times 2$ & $10 \times 2$ & $10 \times 10$ & \shortstack{$5 \times 2$ \\ (less data)} \\ \\
    Number of weights & 69 & 184 & 1064 & 69 \\ \\
    Number of data points & 2500 & 2500 & 2500 & 50 \\ \\
    Error in mean velocities $\mathcal{E}(\boldsymbol{u})$ & $0.47\%$  & $0.91\%$ & $0.52\%$ & $2.0\%$ \\ \\
     Error in Reynolds stresses $\mathcal{E}(\boldsymbol{\tau})$ & $5.8\%$ & $6.9\%$ & $6.0\%$ & $9.4\%$ \\
    \end{tabular}
    \caption{Sensitivity of predictive performance to network architecture and observation data for the square duct case.}
    \label{tab:sensitivity_study}
\end{table}

We perform the sensitivity analysis on the neural network for the periodic hill case. 
The neural network with ten hidden layers and $10$ neurons per layer is regarded as the baseline since it has been used in the work of~\citep{ling2016reynolds}.
Moreover, we test the neural network with two hidden layers and $5$ neurons per layer and the neural network with two hidden layers and $10$ neurons per layer.
The results are summarized in the Table.~\ref{tab:sensitivity_study_pehills}.
In general, the velocity and Reynolds stress results do not vary significantly with different neural networks.
In other words, the training performance is not sensitive to the neural network architecture. 
Hence, we choose the baseline network with ten hidden layers and $10$ neurons per layer in this case.

In summary, the neural network architecture has no significant effects on the training accuracy based on the sensitivity study for both the square duct and periodic hill cases.
Hence, we choose the baseline networks for both cases.
That is, the network with two hidden layers and $5$ neurons per layer is used for the square duct case, and the network with ten hidden layers and $10$ neurons per layer is adopted for the periodic hill case.
Additionally, this choice keeps consistency to the previous works~\citep{strofer2021end, strofer2021ensemble} such that the training performance with different approaches shown in Table~\ref{tab:performance_sum} are compared in a consistent manner.

\begin{table}
    \centering
    \begin{tabular}{c c c c}
    \shortstack{Network architecture \\ (neurons/layer $\times$ layers )} & $5 \times 2$ & $10 \times 2$ & $10 \times 10$ \\ \\
        Number of weights & 57 & 162 & 1042  \\ \\
        Error in mean velocities $\mathcal{E}(\boldsymbol{u})$ & $6.5\%$  & $6.6\%$ & $6.4\%$  \\ \\
        Error in Reynolds stresses $\mathcal{E}(\boldsymbol{\tau})$ & $43.0\%$ & $45.0\%$ & $44.0\%$ \\
    \end{tabular}
    \caption{Sensitivity of predictive performance to network architecture for the periodic hill case.}
    \label{tab:sensitivity_study_pehills}
\end{table}

\section{Implementation of direct learning method}
\label{sec:direct_learning}

This section briefly presents the implementation of the direct learning method~\citep{ling2016reynolds}.
The deviatoric part of Reynolds stress tensor~$\mathbf{b}$ from DNS is used as the training data to optimize the tensor-basis neural network.
The input features of scalar invariants are from the baseline RANS prediction.
The cost function can be written as
\begin{equation}
     J = \| \mathbf{g}[\bm{w}]\mathbf{T} - \bm{b}^\text{DNS} \|^2 + \lambda \| \mathbf{g}[\bm{w}] - \mathbf{g}_0 \|^2 \text{,}
\end{equation}
where the tensor bases~$\mathbf{T}$ are obtained from the baseline RANS results, and $\mathbf{g}_0$ represents the prior value, i.e., $g_0^{(1)}=-0.09$ and $g_0^{(2)-(10)}=0$.
The model training amounts to finding optimal weights of the neural network such that the cost function~$J$ is minimized.
In the region where the tensor basis $\mathbf{T}$ is small, the coefficients $\bm{g}$ can have extremely large, nonphysical values, which leads to negative eddy viscosity.
Hence, the regularization is required to alleviate the ill-conditioning by penalizing the deviation from the prior value.
The regularization parameter $\lambda$ is adjusted to achieve good data fit and avoid extreme value of $\mathbf{g}$ simultaneously.
It is taken as $10$ in this work.
To solve the least-squares problem, the gradient of the cost function is obtained based on auto differentiation, and the Adam algorithm is adapted to update the weights of the neural network.

\section*{Acknowledgment}
XLZ and GH are supported by the NSFC Basic Science Center Program for ``Multiscale Problems in Nonlinear Mechanics'' (No. 11988102). XLZ also acknowledges supports from the National Natural Science Foundation
of China (No.~12102435) and the China Postdoctoral Science Foundation (No.~2021M690154). XL acknowledges partial financial supports from the National Centre for Sustainable Subsurface Utilization of the Norwegian Continental Shelf (NCS2030), Norway. HX is not funded when performing this work. Figure 1 in this manuscript was prepared with the help of Mr.~Xuhui Zhou, whose efforts are greatly appreciated by the authors. 
Finally, the authors thank the reviewers for their constructive and valuable comments, which greatly improved the quality and clarity of this paper.

\section*{Declaration of interests}
The authors report no conflict of interest.


\begin{thebibliography}{61}
	\expandafter\ifx\csname natexlab\endcsname\relax\def\natexlab#1{#1}\fi
	\def\au#1{#1} \def\ed#1{#1} \def\yr#1{#1}\def\at#1{#1}\def\jt#1{\textit{#1}}
	\def\bt#1{#1}\def\bvol#1{\textbf{#1}} \def\vol#1{#1} \def\pg#1{#1}
	\def\publ#1{#1}\def\arxiv#1{#1}\def\org#1{#1}\def\st#1{\textit{#1}}
	
	\bibitem[Abadi {\em et~al.\/}(2015)Abadi, Agarwal, Barham, Brevdo, Chen, Citro,
	Corrado, Davis, Dean, Devin, Ghemawat, Goodfellow, Harp, Irving, Isard, Jia,
	Jozefowicz, Kaiser, Kudlur, Levenberg, Man\'{e}, Monga, Moore, Murray, Olah,
	Schuster, Shlens, Steiner, Sutskever, Talwar, Tucker, Vanhoucke, Vasudevan,
	Vi\'{e}gas, Vinyals, Warden, Wattenberg, Wicke, Yu \&
	Zheng]{abadi2015tensorflow}
	{\sc \au{Abadi, M.}, \au{Agarwal, A.}, \au{Barham, P.}, \au{Brevdo, E.},
		\au{Chen, Z.}, \au{Citro, C.}, \au{Corrado, G.~S.}, \au{Davis, A.}, \au{Dean,
			J.}, \au{Devin, M.}, \au{Ghemawat, S.}, \au{Goodfellow, I.}, \au{Harp, A.},
		\au{Irving, G.}, \au{Isard, M.}, \au{Jia, Y.}, \au{Jozefowicz, R.},
		\au{Kaiser, L.}, \au{Kudlur, M.}, \au{Levenberg, J.}, \au{Man\'{e}, D.},
		\au{Monga, R.}, \au{Moore, S.}, \au{Murray, D.}, \au{Olah, C.}, \au{Schuster,
			M.}, \au{Shlens, J.}, \au{Steiner, B.}, \au{Sutskever, I.}, \au{Talwar, K.},
		\au{Tucker, P.}, \au{Vanhoucke, V.}, \au{Vasudevan, V.}, \au{Vi\'{e}gas, F.},
		\au{Vinyals, O.}, \au{Warden, P.}, \au{Wattenberg, M.}, \au{Wicke, M.},
		\au{Yu, Y.} \& \au{Zheng, X.}} \yr{2015} {TensorFlow}: Large-scale machine
	learning on heterogeneous systems. Software available from tensorflow.org.
	
	\bibitem[Bae \& Koumoutsakos(2022)]{bae2022scientific}
	{\sc \au{Bae, H.~J.} \& \au{Koumoutsakos, P.}} \yr{2022}  \at{Scientific
		multi-agent reinforcement learning for wall-models of turbulent flows}.
	\jt{Nature Communications}  \bvol{13}~(1),  \pg{1--9}.
	
	\bibitem[Brener {\em et~al.\/}(2021)Brener, Cruz, Thompson \&
	Anjos]{brener2021conditioning}
	{\sc \au{Brener, B.~P.}, \au{Cruz, M.~A.}, \au{Thompson, R.~L.} \& \au{Anjos,
			R.~P.}} \yr{2021}  \at{Conditioning and accurate solutions of {Reynolds}
		average {Navier--Stokes} equations with data-driven turbulence closures}.
	\jt{Journal of Fluid Mechanics}  \bvol{915},  \pg{1--20}.
	
	\bibitem[Chen {\em et~al.\/}(2019)Chen, Chang, Meng \& Zhang]{chen2019ensemble}
	{\sc \au{Chen, Y.}, \au{Chang, H.}, \au{Meng, J.} \& \au{Zhang, D.}} \yr{2019}
	\at{{Ensemble Neural Networks (ENN)}: A gradient-free stochastic method}.
	\jt{Neural Networks}  \bvol{110},  \pg{170--185}.
	
	\bibitem[Chen \& Oliver(2013)]{chen2013levenberg}
	{\sc \au{Chen, Y.} \& \au{Oliver, D.~S.}} \yr{2013}  \at{Levenberg--marquardt
		forms of the iterative ensemble smoother for efficient history matching and
		uncertainty quantification}.  \jt{Computational Geosciences}  \bvol{17}~(4),
	\pg{689--703}.
	
	\bibitem[Cruz {\em et~al.\/}(2019)Cruz, Thompson, Sampaio \&
	Bacchi]{cruz2019use}
	{\sc \au{Cruz, M.~A.}, \au{Thompson, R.~L.}, \au{Sampaio, L. E.~B.} \&
		\au{Bacchi, R. D.~A.}} \yr{2019}  \at{The use of the {Reynolds} force vector
		in a physics informed machine learning approach for predictive turbulence
		modeling}.  \jt{Computers \& Fluids}  \bvol{192},  \pg{104258}.
	
	\bibitem[Duraisamy(2021)]{duraisamy2021perspectives}
	{\sc \au{Duraisamy, K.}} \yr{2021}  \at{Perspectives on machine
		learning-augmented {Reynolds-averaged} and large eddy simulation models of
		turbulence}.  \jt{Physical Review Fluids}  \bvol{6}~(5),  \pg{050504}.
	
	\bibitem[Eisfeld {\em et~al.\/}(2016)Eisfeld, Rumsey \&
	Togiti]{eisfeld2016verification}
	{\sc \au{Eisfeld, B.}, \au{Rumsey, C.} \& \au{Togiti, V.}} \yr{2016}
	\at{Verification and validation of a second-moment-closure model}.  \jt{AIAA
		Journal}  \bvol{54}~(5),  \pg{1524--1541}.
	
	\bibitem[Evensen(2009)]{evensen2009data}
	{\sc \au{Evensen, G.}} \yr{2009} {\em Data assimilation: the ensemble {Kalman}
		filter\/}.  \publ{Springer}.
	
	\bibitem[Evensen(2018)]{evensen2018analysis}
	{\sc \au{Evensen, G.}} \yr{2018}  \at{Analysis of iterative ensemble smoothers
		for solving inverse problems}.  \jt{Computational Geosciences}
	\bvol{22}~(3),  \pg{885--908}.
	
	\bibitem[Han {\em et~al.\/}(2022)Han, Zhou \& Xiao]{han2022vcnn}
	{\sc \au{Han, J.}, \au{Zhou, X.-H.} \& \au{Xiao, H}} \yr{2022}  \at{{VCNN-e}: A
		vector-cloud neural network with equivariance for emulating {Reynolds} stress
		transport equations}.  \jt{arXiv preprint 2201.01287} .
	
	\bibitem[Holland {\em et~al.\/}(2019)Holland, Baeder \&
	Duraisamy]{holland2019field}
	{\sc \au{Holland, J.~R.}, \au{Baeder, J.~D.} \& \au{Duraisamy, K.}} \yr{2019}
	Field inversion and machine learning with embedded neural networks:
	Physics-consistent neural network training.  \bt{In {\em AIAA Aviation 2019
			Forum\/}},  \pg{p. 3200}.
	
	\bibitem[Kovachki \& Stuart(2019)]{kovachki2019ensemble}
	{\sc \au{Kovachki, N.~B.} \& \au{Stuart, A.~M.}} \yr{2019}  \at{Ensemble
		{Kalman} inversion: a derivative-free technique for machine learning tasks}.
	\jt{Inverse Problems}  \bvol{35}~(9),  \pg{095005}.
	
	\bibitem[Launder {\em et~al.\/}(1975)Launder, Reece \&
	Rodi]{launder1975progress}
	{\sc \au{Launder, B.~E.}, \au{Reece, G.~J.} \& \au{Rodi, W.}} \yr{1975}
	\at{Progress in the development of a {Reynolds-stress} turbulence closure}.
	\jt{Journal of Fluid Mechanics}  \bvol{68}~(3),  \pg{537--566}.
	
	\bibitem[Launder \& Sandham(2002)]{launder2002closure}
	{\sc \au{Launder, B.~E.} \& \au{Sandham, N.~D.}} \yr{2002} {\em Closure
		strategies for turbulent and transitional flows\/}.  \publ{Cambridge
		University Press}.
	
	\bibitem[Launder \& Sharma(1974)]{launder1974application}
	{\sc \au{Launder, B.~E.} \& \au{Sharma, B.~I.}} \yr{1974}  \at{Application of
		the energy-dissipation model of turbulence to the calculation of flow near a
		spinning disc}.  \jt{Letters in Heat and Mass Transfer}  \bvol{1}~(2),
	\pg{131--137}.
	
	\bibitem[Ling {\em et~al.\/}(2016)Ling, Kurzawski \&
	Templeton]{ling2016reynolds}
	{\sc \au{Ling, J.}, \au{Kurzawski, A.} \& \au{Templeton, J.}} \yr{2016}
	\at{Reynolds averaged turbulence modelling using deep neural networks with
		embedded invariance}.  \jt{Journal of Fluid Mechanics}  \bvol{807},
	\pg{155--166}.
	
	\bibitem[Ling \& Templeton(2015)]{ling2015evaluation}
	{\sc \au{Ling, J.} \& \au{Templeton, J.}} \yr{2015}  \at{Evaluation of machine
		learning algorithms for prediction of regions of high {Reynolds averaged
			Navier Stokes} uncertainty}.  \jt{Physics of Fluids}  \bvol{27}~(8),
	\pg{085103}.
	
	\bibitem[Luo(2021)]{luo2021novel}
	{\sc \au{Luo, X.}} \yr{2021}  \at{Novel iterative ensemble smoothers derived
		from a class of generalized cost functions}.  \jt{Computational Geosciences}
	\bvol{25}~(3),  \pg{1159--1189}.
	
	\bibitem[Luo {\em et~al.\/}(2018)Luo, Bhakta \& Naevdal]{luo2018correlation}
	{\sc \au{Luo, X.}, \au{Bhakta, T.} \& \au{Naevdal, G.}} \yr{2018}
	\at{Correlation-based adaptive localization with applications to
		ensemble-based {4D-seismic} history matching}.  \jt{SPE Journal}
	\bvol{23}~(02),  \pg{396--427}.
	
	\bibitem[Luo {\em et~al.\/}(2015)Luo, Stordal, Lorentzen \&
	N{\ae}vdal]{luo2015iterative}
	{\sc \au{Luo, X.}, \au{Stordal, A.~S.}, \au{Lorentzen, R.~J.} \&
		\au{N{\ae}vdal, G.}} \yr{2015}  \at{Iterative ensemble smoother as an
		approximate solution to a regularized minimum-average-cost problem: theory
		and applications}.  \jt{SPE Journal}  \bvol{20}~(05),  \pg{962--982}.
	
	\bibitem[MacArt {\em et~al.\/}(2021)MacArt, Sirignano \&
	Freund]{macart2021embedded}
	{\sc \au{MacArt, J.~F.}, \au{Sirignano, J.} \& \au{Freund, J.~B.}} \yr{2021}
	\at{Embedded training of neural-network subgrid-scale turbulence models}.
	\jt{Physical Review Fluids}  \bvol{6}~(5),  \pg{050502}.
	
	\bibitem[Michel{\'e}n-Str{\"o}fer \& Xiao(2021)]{strofer2021end}
	{\sc \au{Michel{\'e}n-Str{\"o}fer, C.~A.} \& \au{Xiao, H.}} \yr{2021}
	\at{End-to-end differentiable learning of turbulence models from indirect
		observations}.  \jt{Theoretical and Applied Mechanics Letters}
	\bvol{11}~(4),  \pg{100280}.
	
	\bibitem[Michel{\'e}n-Str{\"o}fer {\em et~al.\/}(2021{\natexlab{{\em
				a\/}}})Michel{\'e}n-Str{\"o}fer, Zhang \& Xiao]{strofer2021dafi}
	{\sc \au{Michel{\'e}n-Str{\"o}fer, C.~A.}, \au{Zhang, X.-L.} \& \au{Xiao, H.}}
	\yr{2021{\natexlab{{\em a\/}}}}  \at{{DAFI}: An open-source framework for
		ensemble-based data assimilation and field inversion}.  \jt{Communications in
		Computational Physics}  \bvol{29}~(5),  \pg{1583--1622}.
	
	\bibitem[Michel{\'e}n-Str{\"o}fer {\em et~al.\/}(2021{\natexlab{{\em
				b\/}}})Michel{\'e}n-Str{\"o}fer, Zhang \& Xiao]{strofer2021ensemble}
	{\sc \au{Michel{\'e}n-Str{\"o}fer, C.~A.}, \au{Zhang, X.-L.} \& \au{Xiao, H.}}
	\yr{2021{\natexlab{{\em b\/}}}}  \at{Ensemble gradient for learning
		turbulence models from indirect observations}.  \jt{Communications in
		Computational Physics}  \bvol{30}~(5),  \pg{1269--1289}.
	
	\bibitem[Nocedal \& Wright(2006)]{nocedal2006numerical}
	{\sc \au{Nocedal, J.} \& \au{Wright, S.}} \yr{2006} {\em Numerical
		Optimization\/}.  \publ{Springer Science \& Business Media}.
	
	\bibitem[Novati {\em et~al.\/}(2021)Novati, de~Laroussilhe \&
	Koumoutsakos]{novati2021automating}
	{\sc \au{Novati, G.}, \au{de~Laroussilhe, H.~L.} \& \au{Koumoutsakos, P.}}
	\yr{2021}  \at{Automating turbulence modelling by multi-agent reinforcement
		learning}.  \jt{Nature Machine Intelligence}  \bvol{3}~(1),  \pg{87--96}.
	
	\bibitem[Othmer(2008)]{othmer2008continuous}
	{\sc \au{Othmer, C.}} \yr{2008}  \at{A continuous adjoint formulation for the
		computation of topological and surface sensitivities of ducted flows}.
	\jt{International journal for numerical methods in fluids}  \bvol{58}~(8),
	\pg{861--877}.
	
	\bibitem[Park \& Choi(2021)]{park2021toward}
	{\sc \au{Park, J.} \& \au{Choi, H.}} \yr{2021}  \at{Toward neural-network-based
		large eddy simulation: application to turbulent channel flow}.  \jt{Journal
		of Fluid Mechanics}  \bvol{914}.
	
	\bibitem[Perot(1999)]{perot1999turbulence}
	{\sc \au{Perot, B.}} \yr{1999}  \at{Turbulence modeling using body force
		potentials}.  \jt{Physics of Fluids}  \bvol{11}~(9),  \pg{2645--2656}.
	
	\bibitem[Pope(1975)]{pope1975more}
	{\sc \au{Pope, S.~B.}} \yr{1975}  \at{A more general effective-viscosity
		hypothesis}.  \jt{Journal of Fluid Mechanics}  \bvol{72}~(2),  \pg{331--340}.
	
	\bibitem[Pope(2000)]{pope2001turbulent}
	{\sc \au{Pope, S.~B.}} \yr{2000} {\em Turbulent flows\/}.  \publ{Cambridge
		University Press}.
	
	\bibitem[Sa{\"\i}di {\em et~al.\/}(2022)Sa{\"\i}di, Schmelzer, Cinnella \&
	Grasso]{saidi2021cfddriven}
	{\sc \au{Sa{\"\i}di, I. B.~H.}, \au{Schmelzer, M.}, \au{Cinnella, P.} \&
		\au{Grasso, F.}} \yr{2022}  \at{{CFD}-driven symbolic identification of
		algebraic {Reynolds}-stress models}.  \jt{Journal of Computational Physics}
	\bvol{457},  \pg{111037}.
	
	\bibitem[Schmelzer {\em et~al.\/}(2020)Schmelzer, Dwight \&
	Cinnella]{schmelzer2020discovery}
	{\sc \au{Schmelzer, M.}, \au{Dwight, R.~P.} \& \au{Cinnella, P.}} \yr{2020}
	\at{Discovery of algebraic {Reynolds-stress} models using sparse symbolic
		regression}.  \jt{Flow, Turbulence and Combustion}  \bvol{104}~(2),
	\pg{579--603}.
	
	\bibitem[Schneider {\em et~al.\/}(2020{\natexlab{{\em a\/}}})Schneider, Stuart
	\& Wu]{schneider2020ensemble}
	{\sc \au{Schneider, T.}, \au{Stuart, A.~M.} \& \au{Wu, J.-L.}}
	\yr{2020{\natexlab{{\em a\/}}}}  \at{Ensemble {Kalman} inversion for sparse
		learning of dynamical systems from time-averaged data}.  \jt{arXiv preprint
		2007.06175} .
	
	\bibitem[Schneider {\em et~al.\/}(2020{\natexlab{{\em b\/}}})Schneider, Stuart
	\& Wu]{schneider2020imposing}
	{\sc \au{Schneider, T.}, \au{Stuart, A.~M.} \& \au{Wu, J.-L.}}
	\yr{2020{\natexlab{{\em b\/}}}}  \at{Imposing sparsity within ensemble
		{Kalman} inversion}.  \jt{arXiv preprint arXiv:2007.06175} .
	
	\bibitem[Shih(1993)]{shih1993realizable}
	{\sc \au{Shih, T.-H.}} \yr{1993}  \bt{ \at{A realizable {Reynolds} stress
			algebraic equation model}}. ,  \vol{vol. 105993}.  \publ{National Aeronautics
		and Space Administration}.
	
	\bibitem[Singh \& Duraisamy(2016)]{singh2016using}
	{\sc \au{Singh, A.~P.} \& \au{Duraisamy, K.}} \yr{2016}  \at{Using field
		inversion to quantify functional errors in turbulence closures}.  \jt{Physics
		of Fluids}  \bvol{28}~(4),  \pg{045110}.
	
	\bibitem[Sirignano \& Spiliopoulos(2022)]{sirignano2021online}
	{\sc \au{Sirignano, J.} \& \au{Spiliopoulos, K.}} \yr{2022}  \at{Online adjoint
		methods for optimization of {PDEs}}.  \jt{Applied Mathematics $\&$
		Optimization} .
	
	\bibitem[Slotnick {\em et~al.\/}(2014)Slotnick, Khodadoust, Alonso, Darmofal,
	Gropp, Lurie \& Mavriplis]{pi2014cfd}
	{\sc \au{Slotnick, J.}, \au{Khodadoust, A.}, \au{Alonso, J.}, \au{Darmofal,
			D.}, \au{Gropp, W.}, \au{Lurie, E.} \& \au{Mavriplis, D.}} \yr{2014} {CFD}
	vision 2030 study: a path to revolutionary computational aerosciences.
	\bt{In {\em NASA CR-2014-218178\/}}.  \publ{Langley Research Center}.
	
	\bibitem[Spalart(2000)]{spalart2000strategies}
	{\sc \au{Spalart, P.~R.}} \yr{2000}  \at{Strategies for turbulence modelling
		and simulations}.  \jt{International Journal of Heat and Fluid Flow}
	\bvol{21}~(3),  \pg{252--263}.
	
	\bibitem[Spalart \& Allmaras(1992)]{spalart1992one-equation}
	{\sc \au{Spalart, P.~R.} \& \au{Allmaras, S.~R.}} \yr{1992} A one-equation
	turbulence model for aerodynamic flows. AIAA Paper 1992-439.
	
	\bibitem[Speziale {\em et~al.\/}(1991)Speziale, Sarkar \&
	Gatski]{speziale1991modelling}
	{\sc \au{Speziale, C.~G.}, \au{Sarkar, S.} \& \au{Gatski, T.~B.}} \yr{1991}
	\at{Modelling the pressure--strain correlation of turbulence: an invariant
		dynamical systems approach}.  \jt{Journal of Fluid Mechanics}  \bvol{227},
	\pg{245--272}.
	
	\bibitem[Sun \& Wang(2020)]{sun2020physics}
	{\sc \au{Sun, L.} \& \au{Wang, J.-X.}} \yr{2020}  \at{Physics-constrained
		{Bayesian} neural network for fluid flow reconstruction with sparse and noisy
		data}.  \jt{Theoretical and Applied Mechanics Letters}  \bvol{10}~(3),
	\pg{161--169}.
	
	\bibitem[{The OpenFOAM Foundation}(2021)]{opencfd21openfoam}
	{\sc \au{{The OpenFOAM Foundation}}} \yr{2021} {\em {OpenFOAM} User Guide\/}.
	
	\bibitem[Wallin \& Johansson(2000)]{wallin2000explicit}
	{\sc \au{Wallin, S.} \& \au{Johansson, A.~V.}} \yr{2000}  \at{An explicit
		algebraic {Reynolds} stress model for incompressible and compressible
		turbulent flows}.  \jt{Journal of Fluid Mechanics}  \bvol{403},
	\pg{89--132}.
	
	\bibitem[Wang {\em et~al.\/}(2017)Wang, Wu \& Xiao]{wang2017physics}
	{\sc \au{Wang, J.-X.}, \au{Wu, J.-L.} \& \au{Xiao, H.}} \yr{2017}
	\at{Physics-informed machine learning approach for reconstructing {Reynolds}
		stress modeling discrepancies based on {DNS} data}.  \jt{Physical Review
		Fluids}  \bvol{2}~(3),  \pg{034603}.
	
	\bibitem[Waschkowski {\em et~al.\/}(2022)Waschkowski, Zhao, Sandberg \&
	Klewicki]{waschkowski2022multi}
	{\sc \au{Waschkowski, F.}, \au{Zhao, Y.}, \au{Sandberg, R.} \& \au{Klewicki,
			J.}} \yr{2022}  \at{Multi-objective {CFD-driven} development of coupled
		turbulence closure models}.  \jt{Journal of Computational Physics}
	\bvol{452},  \pg{110922}.
	
	\bibitem[Weatheritt \& Sandberg(2016)]{weatheritt2016novel}
	{\sc \au{Weatheritt, J.} \& \au{Sandberg, R.}} \yr{2016}  \at{A novel
		evolutionary algorithm applied to algebraic modifications of the {RANS}
		stress--strain relationship}.  \jt{Journal of Computational Physics}
	\bvol{325},  \pg{22--37}.
	
	\bibitem[Wu {\em et~al.\/}(2019{\natexlab{{\em a\/}}})Wu,
	Michel{\'e}n-Str{\"o}fer \& Xiao]{wu2019physics}
	{\sc \au{Wu, J.-L.}, \au{Michel{\'e}n-Str{\"o}fer, C.~A.} \& \au{Xiao, H.}}
	\yr{2019{\natexlab{{\em a\/}}}}  \at{Physics-informed covariance kernel for
		model-form uncertainty quantification with application to turbulent flows}.
	\jt{Computers \& Fluids}  \bvol{193},  \pg{104292}.
	
	\bibitem[Wu {\em et~al.\/}(2018)Wu, Xiao \& Paterson]{wu2018physics}
	{\sc \au{Wu, J.-L.}, \au{Xiao, H.} \& \au{Paterson, E.}} \yr{2018}
	\at{Physics-informed machine learning approach for augmenting turbulence
		models: A comprehensive framework}.  \jt{Physical Review Fluids}
	\bvol{3}~(7),  \pg{074602}.
	
	\bibitem[Wu {\em et~al.\/}(2019{\natexlab{{\em b\/}}})Wu, Xiao, Sun \&
	Wang]{wu2019reynolds}
	{\sc \au{Wu, J.-L.}, \au{Xiao, H.}, \au{Sun, R.} \& \au{Wang, Q.}}
	\yr{2019{\natexlab{{\em b\/}}}}  \at{{Reynolds-averaged Navier--Stokes}
		equations with explicit data-driven {Reynolds} stress closure can be
		ill-conditioned}.  \jt{Journal of Fluid Mechanics}  \bvol{869},
	\pg{553--586}.
	
	\bibitem[Xiao \& Cinnella(2019)]{xiao2019quantification}
	{\sc \au{Xiao, H.} \& \au{Cinnella, P.}} \yr{2019}  \at{Quantification of model
		uncertainty in {RANS} simulations: {A} review}.  \jt{Progress in Aerospace
		Sciences}  \bvol{108},  \pg{1--31}.
	
	\bibitem[Xiao {\em et~al.\/}(2020)Xiao, Wu, Laizet \& Duan]{xiao2020flows}
	{\sc \au{Xiao, H.}, \au{Wu, J.-L.}, \au{Laizet, S.} \& \au{Duan, L.}} \yr{2020}
	\at{Flows over periodic hills of parameterized geometries: {A} dataset for
		data-driven turbulence modeling from direct simulations}.  \jt{Computers \&
		Fluids}  \bvol{200},  \pg{104431}.
	
	\bibitem[Yang \& Griffin(2021)]{yang2021grid}
	{\sc \au{Yang, X. I.~A.} \& \au{Griffin, K.~P.}} \yr{2021}  \at{Grid-point and
		time-step requirements for direct numerical simulation and large-eddy
		simulation}.  \jt{Physics of Fluids}  \bvol{33}~(1),  \pg{015108}.
	
	\bibitem[Zhang {\em et~al.\/}(2020{\natexlab{{\em a\/}}})Zhang,
	Michel{\'e}n-Str{\"o}fer \& Xiao]{zhang2020regularized}
	{\sc \au{Zhang, X.-L.}, \au{Michel{\'e}n-Str{\"o}fer, C.A.} \& \au{Xiao, H.}}
	\yr{2020{\natexlab{{\em a\/}}}}  \at{Regularized ensemble {Kalman} methods
		for inverse problems}.  \jt{Journal of Computational Physics}  \bvol{416},
	\pg{109517}.
	
	\bibitem[Zhang {\em et~al.\/}(2020{\natexlab{{\em b\/}}})Zhang, Xiao, Gomez \&
	Coutier-Delgosha]{zhang2020evaluation}
	{\sc \au{Zhang, X.-L.}, \au{Xiao, H.}, \au{Gomez, T.} \& \au{Coutier-Delgosha,
			O.}} \yr{2020{\natexlab{{\em b\/}}}}  \at{Evaluation of ensemble methods for
		quantifying uncertainties in steady-state {CFD} applications with small
		ensemble sizes}.  \jt{Computers \& Fluids}  \pg{p. 104530}.
	
	\bibitem[Zhang {\em et~al.\/}(2022)Zhang, Xiao, Luo \& He]{zhang2022dafi}
	{\sc \au{Zhang, X.-L.}, \au{Xiao, H.}, \au{Luo, X.} \& \au{He, G.}} \yr{2022}
	Ensemble-based learning of turbulence models. Software available from
	github.com/xiaoh/DAFI/ensemble-learning.
	
	\bibitem[Zhao {\em et~al.\/}(2020)Zhao, Akolekar, Weatheritt, Michelassi \&
	Sandberg]{zhao2020rans}
	{\sc \au{Zhao, Y.}, \au{Akolekar, H.~D.}, \au{Weatheritt, J.}, \au{Michelassi,
			V.} \& \au{Sandberg, R.~D.}} \yr{2020}  \at{{RANS} turbulence model
		development using {CFD-driven} machine learning}.  \jt{Journal of
		Computational Physics}  \bvol{411},  \pg{109413}.
	
	\bibitem[Zhou {\em et~al.\/}(2021)Zhou, Han \& Xiao]{zhou2021learning}
	{\sc \au{Zhou, X.-H.}, \au{Han, J.} \& \au{Xiao, H.}} \yr{2021}  \at{Learning
		nonlocal constitutive models with neural networks}.  \jt{Computer Methods in
		Applied Mechanics and Engineering}  \bvol{384},  \pg{113927}.
	
	\bibitem[Zhou {\em et~al.\/}(2022)Zhou, Han \& Xiao]{zhou2022frame}
	{\sc \au{Zhou, X.-H.}, \au{Han, J.} \& \au{Xiao, H.}} \yr{2022}
	\at{Frame-independent vector-cloud neural network for nonlocal constitutive
		modeling on arbitrary grids}.  \jt{Computer Methods in Applied Mechanics and
		Engineering}  \bvol{388},  \pg{114211}.
	
\end{thebibliography}
\end{document}